\documentclass[leqno,draft,11pt,a4paper]{article}
\usepackage{amssymb,amsmath,latexsym,theorem}
\usepackage[includeheadfoot,margin=1in]{geometry}
\usepackage[final]{graphicx}

\allowdisplaybreaks[2]

\newcommand\brak[1]{\textup(\nobreak\hskip0pt\relax#1\textup)}

\newcommand\clap{\mathpalette\doclap}
\newcommand\doclap[2]{\hbox to0pt{\hss$#1#2$\hss}}

\newcommand\et{\mathrel\&}
\newcommand\eq{\leftrightarrow}
\newcommand\LOR{\bigvee}
\newcommand\ET{\bigwedge}
\newcommand\model{\vDash}
\newcommand\nmodel{\nvDash}
\newcommand\dual{\mathrm d}
\newcommand\Var{\mathrm{Var}}
\newcommand\Form{\mathrm{Form}}
\newcommand\Ex{\mathrm E}
\newcommand\fii{\varphi}

\newcommand\p[1]{\langle#1\rangle}
\newcommand\lh[1]{\lvert#1\rvert}
\newcommand\bez{\smallsetminus}
\newcommand\sset{\subseteq}
\newcommand\nsset{\nsubseteq}
\newcommand\Sset{\supseteq}
\newcommand\sSset{\supsetneq}
\newcommand\bigcupd{\mathop{\dot\bigcup}}
\newcommand\pw[1]{\mathcal P(#1)}
\newcommand\nul{\varnothing}
\newcommand\res{\mathbin\restriction}
\newcommand\two{\mathbf2}
\newcommand\id{\mathrm{id}}

\newcommand\dia{\Diamond}
\newcommand\diadot{{\centdot\dia}}
\newcommand\centdot{\mathpalette\docentdot}
\ifx\symlasy\undefined
  
  \renewcommand\boxdot{{\origboxdot}}
  \newcommand\docentdot[2]{%
    \setbox0\hbox{$#1\mathop{#2}$}\dimen0 \ht0
    \setbox0\hbox{$#1#2$}\advance\dimen0 -\ht0
    \setbox2\hbox to\wd0{\hss$#1\mathop{\cdot}$\hss}\wd2=0pt
    \lower\dimen0\box2\box0 }
\else
  \renewcommand\boxdot{{\centdot\Box}}
  \newcommand\docentdot[2]{%
     \setbox0\hbox{$#1#2$}%
     \raise0.206\ht0\hbox to\wd0{\hss$#1\cdot$\hss}%
     \kern-\wd0 \box0 }
\fi
\newcommand\T{\mathsf T}
\newcommand\ru{\mathrel/}
\newcommand\Ru{\Bigm/}
\newcommand\up{\mathord\uparrow}
\newcommand\down{\mathord\downarrow}
\newcommand\Up{{\setbox0\hbox{$\uparrow$}%
         \lower\dp0\hbox to\wd0{\hss\vrule width4pt height.4pt\hss}%
         \kern-\wd0\box0}}
\newcommand\Down{{\setbox0\hbox{$\downarrow$}%
         \raise\ht0\hbox to\wd0{\hss\vrule width4pt depth.4pt\hss}%
         \kern-\wd0\box0}}
\newcommand\nr[1]{%
  \setbox0\hbox{$\scriptstyle\bigcirc$}%
  \vcenter{\hbox to\wd0{\hss$\scriptscriptstyle#1$\hss}}%
  \kern-\wd0\box0 }

\DeclareMathOperator\dom{dom}
\DeclareMathOperator\cls{cl}
\DeclareMathOperator\Sub{Sub}
\DeclareMathOperator\decp{Dec}
\DeclareMathOperator\consp{Cons}
\DeclareMathOperator\poly{poly}

\newcommand\cxt[1]{\mathrm{#1}}
\newcommand\np{\cxt{NP}}
\newcommand\conp{\cxt{coNP}}
\newcommand\ptime{\cxt P}
\newcommand\fp{\cxt{FP}}
\newcommand\psp{\cxt{PSPACE}}

\newcommand\lgc[1]{\mathbf{#1}}

\newcommand\pfsys[2]{\ifx\relax#2\relax\else#2\text-\fi\mathrm{#1}}
\newcommand\EF{\pfsys{EF}}
\newcommand\SF{\pfsys{SF}}
\newcommand\CF{\pfsys{CF}}
\newcommand\Fr{\pfsys{F}}
\newcommand\SCF{\pfsys{SCF}}
\newcommand\TL[2]{#1{#2}^*}
\newcommand\CPC{\lgc{CPC}}
\newcommand\IPC{\lgc{IPC}}
\newcommand\kiv{\lgc{K4}}
\newcommand\I{{\bullet}}
\newcommand\R{{\circ}}
\newcommand\RI{{*}}
\newcommand\DP{\mathrm{DP}}
\newcommand\Ext{\mathrm{Ext}}
\newcommand\VR{\mathrm V}
\newcommand\rrule{\mathrm R}
\newcommand\itp{\mathrm{Itp}}

\newcommand\ob[1]{\overline{#1}}
\newcommand\txto{${}\to{}$}
\newcommand\php{\mathrm{PHP}}
\newcommand\bintr{\mathrm{BT}}

\newcommand\noproof{\relax\ifmmode\eqno\qed\else\leavevmode\unskip\bme\vadjust{}\nobreak\hfill$\qed$\par\fi}
\newcommand\qed{\Box}
\newcommand\bme{\hskip.75em\relax}
\newenvironment{Pf}
  {\par\noindent\textit{Proof:}\bme\ignorespaces}
  {\noproof\pagebreak[2]\vskip\medskipamount\ignorespacesafterend}

\theoremstyle{plain}
\newtheorem{Thm}{Theorem}[section]
\newtheorem{Cor}[Thm]{Corollary}
\newtheorem{Lem}[Thm]{Lemma}
\newtheorem{Obs}[Thm]{Observation}
\newtheorem{Que}[Thm]{Question}
\newtheorem{Cl}{Claim}[Thm]

\newenvironment{Pf*}{\let\qed\qedCl\Pf}\endPf
\theorembodyfont\upshape
\newtheorem{Def}[Thm]{Definition}
\newtheorem{Rem}[Thm]{Remark}
\newtheorem{Exm}[Thm]{Example}

\author{Emil Je\v r\'abek\\[\medskipamount]
Institute of Mathematics, Czech Academy of Sciences\\
\small \v Zitn\'a 25,
115\:67 Praha 1,
Czech Republic,
email: \texttt{jerabek@math.cas.cz}
}

\title{On the proof complexity of logics of bounded branching}

\begin{document}
\maketitle

\begin{abstract}
We investigate the proof complexity of extended Frege ($\EF{}$) systems for basic transitive modal logics ($\kiv$,
$\lgc{S4}$, $\lgc{GL}$, \dots) augmented with the bounded branching axioms $\lgc{BB}_k$. First, we study feasibility of
the disjunction property and more general extension rules in $\EF{}$ systems for these logics: we show that the
corresponding decision problems reduce to total $\conp$~search problems (or equivalently, disjoint $\np$~pairs, in the
binary case); more precisely, the decision problem for extension rules is equivalent to a certain special case of
interpolation for the classical $\EF{}$~system. Next, we use this characterization to prove superpolynomial (or even
exponential, with stronger hypotheses) separations between $\EF{}$ and substitution Frege ($\SF{}$) systems for all
transitive logics contained in $\lgc{S4.2GrzBB_2}$ or $\lgc{GL.2BB_2}$ under some assumptions weaker than $\psp\ne\np$.
We also prove analogous results for superintuitionistic logics: we characterize the decision complexity of
multi-conclusion Visser's rules in $\EF{}$ systems for Gabbay--de Jongh logics~$\lgc T_k$, and we show conditional
separations between $\EF{}$ and~$\SF{}$ for all intermediate logics contained in~$\lgc T_2+\lgc{KC}$.


\end{abstract}

\section{Introduction}\label{sec:introduction}

The primary focus of proof complexity is on questions about lengths of derivations or refutations in proof systems for
classical propositional logic~$\CPC$ (including algebraic proof systems dealing with polynomial equations or
inequalities, into which Boolean tautologies can be easily translated). While lower bounds on systems such as
resolution exhibit limitations of SAT-solving technology, the original motivation comes from computational complexity,
as the fundamental problem $\np\ne\conp$ is equivalent to superpolynomial lower bounds on all proof systems for~$\CPC$.
Despite years of effort, we can currently only prove lower bounds on relatively weak systems such as constant-depth
Frege. The unrestricted Frege system (the simplest textbook proof system for~$\CPC$, also p-equivalent to
sequent or natural deduction calculi) is well out of reach.

The situation is rather different in proof complexity of nonclassical propositional logics such as modal logics or
intuitionistic logic, where Frege and related systems are the main objects of study. First, unlike the plethora of
classical proof systems, there are not many alternatives to variants of Frege systems (or equivalent sequent calculi)
in nonclassical logics, though \emph{extended Frege} ($\EF{}$) systems are perhaps even more natural, or at least more
robust: on the one hand, extension axioms formalize the intuitive practice of naming longer formulas so that they can
be referred to succinctly in the proof; on the other hand, bounds on the size of $\EF{}$ proofs are essentially
equivalent to bounds on the \emph{number of lines} in Frege (or $\EF{}$) proofs, which is a measure easier to work
with than size, and $\EF{}$ systems can be thought of as Frege systems operating with \emph{circuits} instead of
formulas, which makes many arguments go through more smoothly.

Crucially, there are a number of nontrivial results on the complexity of Frege and $\EF{}$ systems in
various nonclassical logics, in contrast to~$\CPC$. The underlying theme in many works on the proof complexity of modal
or (super)intuitionistic logics is that of \emph{feasibility of the disjunction property} ($\DP$): given a proof of
$\Box\fii_0\lor\Box\fii_1$ (or just $\fii_0\lor\fii_1$ in the intuitionistic case), can we efficiently decide which
$\fii_u$ is provable, or better yet, can we construct its proof?

Buss and Mints~\cite{buss-mints} proved the feasibility of $\DP$ in the natural deduction system for intuitionistic
logic ($\IPC$); Buss and Pudl\'ak~\cite{buss-pud} extended this result, and made the important connection that it
implies \emph{conditional lower bounds} in a similar way as feasible interpolation does in classical proof systems.
Feasibility of $\DP$ for some modal proof systems was shown by Ferrari et al.~\cite{fff}. Mints and
Kojevnikov~\cite{min-koj} generalized feasible $\DP$ in $\IPC$ to feasibility of \emph{Visser's rules}, and used it to
show that all Frege systems for~$\IPC$ are p-equivalent, even if allowed to include inference rules that are not valid,
but merely admissible. A similar result was proved for a certain family of transitive modal logics by Je\v
r\'abek~\cite{ej:modfrege}, using feasibility of modal \emph{extension rules} generalizing~$\DP$.

A breakthrough was achieved by Hrube\v s~\cite{hru:lbmod,hru:lbint,hru:nonclas} who proved \emph{unconditional}
exponential lower bounds on (effectively) $\EF{}$ proofs in some modal logics and~$\IPC$, using a modified version of
feasible~$\DP$ as a form of monotone interpolation. Building on his results, Je\v r\'abek~\cite{ej:sfef} proved
exponential separation between $\EF{}$ and \emph{substitution Frege} ($\SF{}$) systems for a class of transitive
modal and superintuitionistic logics, while $\EF{}$ and~$\SF{}$ systems are equivalent for some other classes of
logics (this equivalence was well known for classical $\EF{}$ and $\SF{}$ systems).

More specifically, it was shown in~\cite{ej:sfef} that the proof complexity of modal and superintuitionistic logics is
connected to their model-theoretic properties, in particular frame measures such as \emph{width} (maximum size of
finite antichains) and \emph{branching} (maximum number of immediate successors): on the one hand, $\SF L$ has
exponential speed-up over $\EF L$ for all transitive modal or superintuitionistic logics~$L$ of unbounded branching. On
the other hand, $\EF L$ and $\SF L$ are p-equivalent (and, in a suitable sense, p-equivalent to $\EF\CPC$) for many
logics of bounded width: basic logics of bounded width such as $\lgc{K4BW}_k$, $\lgc{S4BW}_k$, $\lgc{GLBW}_k$,
and~$\lgc{LC}$, all logics of bounded width and depth, and---for a restricted class of tautologies---all
cofinal-subframe logics of bounded width. Note that branching is upper bounded by width, hence all logics of bounded
width have bounded branching, but the converse is not true---there are logics of branching~$2$ and unbounded width.

Although these results reveal considerable information about modal $\EF{}$ systems, they do not precisely delimit the
boundary between logics for which we have unconditional $\EF{}$ lower bounds and separations from $\SF{}$, and logics
where $\EF{}$ and $\SF{}$ are equivalent and lower bounds on them imply classical $\EF{}$ lower bounds; nor do they
establish that such a sharp boundary exists in the first place. Can we say something about the proof complexity of
$\EF{}$ for logics of bounded branching and unbounded width? (Cf.~\cite[Prob.~7.1]{ej:sfef}.)

This is the question we take up in the present paper. We look at basic logics~$L$ of bounded branching such as $\lgc{K4BB}_k$,
$\lgc{S4BB}_k$, and $\lgc{GLBB}_k$ (more generally, extensible logics as in~\cite{ej:modfrege} augmented with the
bounded branching axioms~$\lgc{BB}_k$). First, we study the feasibility of $\DP$ and extension rules for $\EF L$: while
they are (probably) no longer decidable in polynomial time as was the case for extensible logics, we will show that
they are decidable by total $\conp$ search problems (or equivalently, disjoint $\np$~pairs, for two-conclusion rules),
which is still much smaller complexity than the trivial $\psp$ upper bound. As a consequence, we prove a
superpolynomial separation between $\EF L$ and $\SF L$ unless $\psp=\np=\conp$; in fact, this holds not just for the
basic logics of bounded branching, but for all logics included in $\lgc{GLBB_2}$ or $\lgc{S4GrzBB_2}$. (Note that
logics with the~$\DP$ are $\psp$-hard, hence $\psp\ne\np$ implies superpolynomial lower bounds on \emph{all} proof
systems for these logics; however, such trivial arguments cannot separate $\EF L$ from $\SF L$.) The speed-up of
$\SF L$ over $\EF L$ can be improved to exponential if we assume $\psp\nsset\cxt{NSUBEXP}$.

We elaborate our basic argument by internalizing parts of it in the $\EF{}$ system itself. In this way, we
can characterize the complexity of extension rules for $\EF{}$ systems of basic logics of bounded branching exactly:
they are equivalent to certain special cases of \emph{interpolation} for $\EF\CPC$. We also extend the argument to
cover monotone interpolation in the style of Hrube\v s~\cite{hru:lbmod,hru:nonclas}. This leads to separations of
$\EF L$ from $\SF L$ under weaker hypotheses than $\psp\ne\np$, but unfortunately we still do not obtain unconditional
separations or lower bounds.

We extend the scope of our results in two ways. First, by using positive ($\bot$-free) tautologies, we show (under the
same hypotheses) that $\SF L$ has a superpolynomial speed-up over $\EF L$ for a class of logics~$L$ that includes all
logics contained in $\lgc{S4.2GrzBB_2}$ or $\lgc{GL.2BB_2}$. Second, we adapt our results to
\emph{superintuitionistic logics}: we characterize the complexity of Visser's rules (which generalize the
intuitionistic~$\DP$) for $\EF{}$ systems of the Gabbay--de Jongh logics $\lgc T_k$, and we prove a conditional
superpolynomial speed-up of $\SF L$ over $\EF L$ for all logics $L\sset\lgc T_2+\lgc{KC}$.

The paper is organized as follows. In Section~\ref{sec:preliminaries}, we review the necessary background on modal
logics, their proof complexity, and extension rules. Section~\ref{sec:summary-results} presents an overview of the main
results. Section~\ref{sec:disj-prop-bb} presents the reduction of extension
rules for $\EF{}$ systems of our logics to $\conp$ search problems, and the ensuing separation between $\EF{}$
and~$\SF{}$ conditional on $\psp\ne\np$. In Section~\ref{sec:intern} we internalize the argument inside~$\EF{}$,
leading to separation under weaker assumptions, and in Section~\ref{sec:hrubes-style} we extend it to Hrube\v s-style
monotone interpolation, leading to further weakening of the assumptions. The separations between $\EF{}$ and~$\SF{}$
are generalized to a larger class of logics using positive tautologies in Section~\ref{sec:lower-bounds-separ}, and
parallel results for superintuitionistic logics are proved in Section~\ref{sec:super-logics}. We conclude the paper
with a few remarks and open problems in Section~\ref{sec:conclusion}.

\section{Preliminaries}\label{sec:preliminaries}

As a general notational convention, we denote the set of natural numbers (including~$0$) by~$\omega$, and unless
stated otherwise, our indices and similar integer variables start from~$0$, so that, e.g., $\{\fii_i:i<n\}$ means
$\{\fii_0,\dots,\fii_{n-1}\}$, and $\LOR_{i<n}\fii_i$ is $\fii_0\lor\dots\lor\fii_{n-1}$. If $n=0$, we understand
$\LOR_{i<n}\fii_i$ as~$\bot$, and $\ET_{i<n}\fii_i$ as~$\top$.

\subsection{Modal logic}\label{sec:modal-logic}

We refer the reader to Chagrov and Zakharyaschev~\cite{cha-zax} for background on modal logic.

We consider monomodal propositional modal logics in a language using countably infinitely many propositional variables
$p_i$, $i\in\omega$ (often denoted also by other letters such as $q$, $r$, \dots\ for convenience), a complete set of
Boolean connectives (say, $\{\land,\lor,\to,\neg,\top,\bot\}$, but for the most part the choice will not matter), and a
unary modal connective~$\Box$. Let $\Var$ denote the set of variables, and $\Form$ the set of formulas. We define the abbreviations
$\dia\fii=\neg\Box\neg\fii$, $\boxdot\fii=\fii\land\Box\fii$, and $\diadot\fii=\neg\boxdot\neg\fii$. We will generally
denote formulas by lower-case Greek letters $\fii$, $\psi$, \dots, or upper-case Latin letters $A$, $B$, $C$, \dots. If
$X$ is a formula or a set of formulas, then $\Sub(X)$ denotes the set of subformulas of (formulas from)~$X$.

A \emph{normal modal logic} is a set of formulas~$L$ that contains all classical (Boolean) tautologies and the schema
\begin{equation}
\label{eq:8}\tag{$\lgc K$}\Box(\fii\to\psi)\to(\Box\fii\to\Box\psi),
\end{equation}
and it is closed under substitution and the rules of modus ponens and necessitation:
\begin{align}
\label{eq:90}\tag{MP}\fii,\fii\to\psi&\ru\psi,\\
\label{eq:91}\tag{Nec}\fii&\ru\Box\fii.
\end{align}
Elements of~$L$ are also more explicitly called \emph{$L$-tautologies}.
The \emph{consequence relation~$\vdash_L$} of~$L$ is defined such that for any set of formulas $\Gamma\cup\{\fii\}$,
$\Gamma\vdash_L\fii$ iff $\fii$ is in the closure of $L\cup\Gamma$ under \eqref{eq:90} and~\eqref{eq:91}. The least
normal modal logic is denoted~$\lgc K$.

If $L$ is a normal modal logic and $X$ a formula or a set of formulas, let $L\oplus X$ be the least normal modal logic
containing $L\cup X$, i.e., the closure of $L$ and substitution instances of~$X$ under \eqref{eq:90} and~\eqref{eq:91}.
A logic is \emph{finitely axiomatizable} if can be written as $\lgc K\oplus\fii$ for some formula~$\fii$ (or
equivalently, $\lgc K\oplus X$ for a finite set~$X$).

A \emph{transitive modal logic} is a normal modal logic that also includes the schema
\begin{equation}
\label{eq:4}\tag{$\lgc4$}
\Box\fii\to\Box\Box\fii.
\end{equation}
The least transitive modal logic is denoted~$\kiv$. Unless stated otherwise, all logics in this paper are
\emph{finitely axiomatizable transitive modal logics}; we will also write $\kiv\sset L$ as a shorthand for $L$
being a (finitely axiomatizable transitive modal) logic.

A (transitive) \emph{Kripke frame} is a pair $\p{W,{<}}$ where $<$ is a transitive relation on a set~$W$. (Such notation
is not meant to imply that $<$ is irreflexive.) We will write $x\le y$ for $x<y\lor x=y$, $x\sim y$ for $x\le y\land
y\le x$, and $x\lnsim y$ for $x<y\land y\nless x$. Equivalence classes of~$\sim$ are called \emph{clusters}, and the
quotient partial order $\p{W,{\le}}/{\sim}$ is called the \emph{skeleton} of $\p{W,{<}}$. The cluster of a point~$x$ is
denoted $\cls(x)$. If $X\sset W$, let
\begin{align*}
X\down=\{y\in W:\exists x\in X\:(y<x)\},\\
X\Up=\{y\in W:\exists x\in X\:(x\le y)\},
\end{align*}
and similarly for $X\up$, $X\Down$. A frame $\p{W,{<}}$ is called \emph{rooted} if $W=\{x\}\Up$ for some $x\in W$; any
such $x$ is called the \emph{root} of~$W$. A point $x\in W$ is called \emph{reflexive} if $x<x$, and \emph{irreflexive}
otherwise. As a general notational convention, we will denote irreflexive points and related objects with~$\I$,
and reflexive points with~$\R$.

A \emph{valuation} in a Kripke frame $\p{W,{<}}$ is a mapping $v\colon\Var\to\pw W$. A \emph{Kripke model} is
$M=\p{W,{<},v}$, where $F=\p{W,{<}}$ is a Kripke frame, and $v$ a valuation in~$F$. The valuation uniquely defines a
satisfaction relation for all formulas:
\begin{alignat*}{2}
M,x&\model p_i&&\iff x\in v(p_i),\\
M,x&\model c(\fii_0,\dots,\fii_{d-1})
&&\iff c\bigl((M,x\model\fii_0),\dots,(M,x\model\fii_{d-1})\bigr),\quad c\in\{\land,\lor,\to,\neg,\top,\bot\},\\
M,x&\model\Box\fii&&\iff\forall y\in W\,(x<y\implies M,y\model\fii).
\end{alignat*}
Instead of $M,x\model\fii$, we may write $F,x\model\fii$ or just $x\model\fii$ if the model or frame is understood from
the context. We define
\begin{align*}
M\model\fii&\iff\forall x\in W\:M,x\model\fii,\\
F\model\fii&\iff\forall v\colon\Var\to\pw W\:\p{W,{<},v}\model\fii.
\end{align*}

A \emph{(general) frame} is $F=\p{W,{<},A}$, where $\p{W,{<}}$ is a Kripke frame, and $A\sset\pw W$ is a Boolean
algebra of sets, closed under the operation $X\mapsto\Box X=\{x\in W:\forall y>x\,(y\in X)\}$, or equivalently, under
$X\mapsto X\down$. An \emph{admissible valuation} in the frame~$F$ is a map $v\colon\Var\to A$; the closure conditions on~$A$
ensure that the resulting model $\p{W,{<},v}$ (which is said to be \emph{based on~$F$}) satisfies
$\{x\in W:x\model\fii\}\in A$ for all formulas~$\fii$. We put
\[F\model\fii\iff\forall v\colon\Var\to A\:\p{W,{<},v}\model\fii.\]
If $F\model\fii$, we say that $\fii$ is \emph{valid} in~$F$. We will identify a Kripke frame $\p{W,{<}}$ with the frame
$\p{W,{<},\pw W}$. If $L$ is a logic, an \emph{$L$-frame} is a frame $F$ such that $F\model\fii$ for all $\fii\in L$,
and an \emph{$L$-model} is a model based on an $L$-frame. A frame $\p{W,{<},A}$ is \emph{refined} if
\begin{align*}
x<y&\iff\forall X\in A\:(x\in\Box X\implies y\in X),\\
x=y&\iff\forall X\in A\:(x\in\phantom\Box X\implies y\in X),
\end{align*}
for all $x,y\in W$, and a refined frame is \emph{descriptive} if $A$ is compact: every $S\sset A$ with the finite
intersection property has a nonempty intersection. Kripke frames are refined. Every logic~$L$ is complete w.r.t.\ a
class of descriptive frames (whereas some logics are not complete w.r.t.\ Kripke frames): i.e., if $\nvdash_L\fii$,
there exists a descriptive $L$-frame $F$ such that $F\nmodel\fii$. If a frame~$F=\p{W,{<},A}$ is finite, the atoms
of~$A$ define a partition of~$W$, and the quotient of $F$ by the corresponding equivalence relation is a Kripke frame
that validates the same formulas as~$F$. For this reason, there is no loss of generality if we reserve the phrase
\emph{finite frame} to denote finite Kripke frames. A logic~$L$ has the \emph{finite model property} (FMP) if it is
complete w.r.t.\ a class of finite frames.

\begin{table}[t]
\centering
\def\arraystretch{1.2}
\begin{tabular}{lll}
\hline
logic&axiomatization over $\lgc{K4}$&finite rooted frames\\
\hline
$\lgc{S4}$&$\Box\fii\to\fii$&reflexive\\
$\lgc{D4}$&$\dia\top$&final clusters reflexive\\
$\lgc{GL}$&$\Box(\Box\fii\to\fii)\to\Box\fii$&irreflexive\\
$\lgc{K4Grz}$&$\Box\bigl(\Box(\fii\to\Box\fii)\to\fii\bigr)\to\Box\fii$&no proper clusters\\
$\lgc{K4.1}$&$\boxdot\dia\fii\to\dia\Box\fii$&no proper final clusters\\
$\lgc{K4.2}$&$\dia\boxdot\fii\to\Box\diadot\fii$&unique final cluster\\
$\lgc{K4.3}$&$\Box(\boxdot\fii\to\psi)\lor\Box(\Box\psi\to\fii)$&linear (width $1$)\\
$\lgc{K4B}$&$\fii\to\Box\dia\fii$&single cluster\\
$\lgc{S5}$&$\lgc{S4}\oplus\lgc{K4B}$&single reflexive cluster\\
$\lgc{K4BW}_{\!k}$&$\displaystyle\LOR_{i\le k}\Box\Bigl(\ET_{j\ne i}\boxdot\fii_j\to\fii_i\Bigr)$&width at most $k$\\
$\lgc{K4BD}_k$&(see below)&depth at most $k$\\
$\lgc{K4BC}_k$
&$\displaystyle\Box\Bigl[\LOR_{i\le k}\Box\Bigl(\ET_{j<i}\fii_j\to\fii_i\Bigr)\to\ET_{i\le k}\fii_i\Bigr]\to\Box\fii_0$
&cluster size at most $k$\\
$\lgc{K4BB}_k$&(see below)&branching at most $k$\\
$\lgc{S4.1.4}$
&$\Box\bigl(\Box(\fii\to\Box\fii)\to\fii\bigr)\to(\Box\dia\Box\fii\to\fii)$
&reflexive,\\
&&no inner proper clusters\\
\hline
\end{tabular}
\caption{Some transitive modal logics}
\label{tab:axi}
\end{table}
Several common (or otherwise interesting) transitive modal logics are listed in Table~\ref{tab:axi}, along with frame
conditions that characterize them on finite rooted frames. (A cluster is \emph{proper} if it has $\ge2$ elements. It is
\emph{final} if it has no successor clusters, otherwise it is \emph{inner}. Other semantic conditions are described
below.) Some of the entries are redundant: $\lgc{K4Grz}=\lgc{K4BC_1}$, $\lgc{K4.3}=\lgc{K4BW_{\!1}}$,
$\lgc{K4B}=\lgc{K4BD_1}$. We will generally form compound names of logics by stacking axiom names on a base logic
without $\oplus$~symbols, so that, e.g.,
$\lgc{S4.2GrzBB_2}=\lgc{S4}\oplus\lgc{K4.2}\oplus\lgc{K4Grz}\oplus\lgc{K4BB_2}$. An exception is $\lgc{S4.1.4}$, which
is not a systematic name, but a meaningless numerical label (see Zeman~\cite{zeman}).

If $F=\p{W,{<},A}$ is a frame, and $U\sset W$ is an upper subset (i.e., $U\Up=U$), then
$\p{U,{<}_U,A_U}$ is a \emph{generated subframe} of~$F$, where ${<}_U={<}\cap U^2$ and $A_U=\{X\cap U:X\in A\}$. The
\emph{disjoint sum} $\sum_{i\in I}F_i$ of a family of frames $F_i=\p{W_i,{<}_i,A_i}$, $i\in I$, is the frame
$\p{W,{<},A}$, where $W$ is the disjoint union $\bigcupd_iW_i$, ${<}=\bigcup_i{<}_i$, and $A=\{X\sset W:\forall i\in
I\,(X\cap W_i\in A_i)\}$. A \emph{subreduction} from a frame $F=\p{W,{<},A}$ to a frame $G=\p{V,{\prec},B}$ is a
partial mapping $f$ from $W$ onto~$V$ such that
\begin{enumerate}
\item[(S1)] $x<y\implies f(x)\prec f(y)$ for all $x,y\in\dom(f)$,
\item[(S2)] $f(x)\prec u\implies\exists y>x\,f(y)=v$ for all $x\in\dom(f)$ and $v\in V$, and
\item[(S3)] $f^{-1}[Y]\in A$ for all $Y\in B$ (which implies $\dom(f)\in A$).
\end{enumerate}
If $V\sset W$ and $f=\id_V$ (in which case the conditions reduce to ${\prec}={<}\cap V^2$ and $B\sset A$), then $G$ is
called a \emph{subframe} of~$F$. (Since this implies $V\in A$, generated subframes are \emph{not} necessarily
subframes.) A subreduction is called a \emph{p-morphism} (or \emph{reduction}) if it is total, i.e., $\dom(f)=W$.

For any logic~$L$, the class of $L$-frames is closed under generated subframes, disjoint sums, and p-morphic images;
that is, these frame operations preserve the validity of all formulas.

A subframe $\p{V,{<},B}$ of $\p{W,{<},A}$ is \emph{dense} if $V\Up\cap V\Down=V$, i.e., if $x<y<z$ and $x,z\in V$
imply $y\in V$. More generally, a subreduction $f$ from $F$ to~$G$ is dense if $\dom(f)$ is a dense subframe of~$F$.
Dense subreductions preserve the validity of \emph{positive formulas} (also called \emph{negation-free} or
\emph{$\bot$-free}): i.e., formulas built from propositional variables using $\{\Box,\land,\lor,\to,\top\}$,
disallowing $\neg$ and~$\bot$. (In general, a Boolean connective~$c$ is positive if $c(1,\dots,1)=1$.)

It will be also convenient to have a version of subreductions that is oblivious to reflexivity of points: we define a
\emph{weak subreduction} from $F=\p{W,{<},A}$ to $G=\p{V,{\prec},B}$ to be a partial mapping $f$ from $W$ onto~$V$
that satisfies
\begin{enumerate}
\item[(S1${}'$)] $x<y\implies f(x)\preceq f(y)$ for all $x,y\in\dom(f)$,
\item[(S2${}'$)] $f(x)\prec u\implies\exists y\ge x\,f(y)=v$ for all $x\in\dom(f)$ and $v\in V$,
\end{enumerate}
and (S3).

Let $k\ge1$. A rooted frame $\p{W,{<},A}$ has \emph{width~$\le k$} if it contains no antichain of size $k+1$, i.e.,
points $x_0,\dots,x_k\in W$ such that $x_i\nleq x_j$ for $i\ne j$. A logic~$L$ has \emph{width~$\le k$} if it is
complete w.r.t.\ a class of rooted frames of width~$\le k$, or equivalently, if all rooted refined $L$-frames have
width~$\le k$. We say that $L$ has \emph{bounded width} if it has width $\le k$ for some~$k$, and it has
\emph{unbounded width} otherwise.

A frame $\p{W,{<},A}$ has \emph{depth~$\le k$} if it contains no chain of length $k+1$, i.e., $x_0,\dots,x_k\in W$ such
that $x_0\lnsim x_1\lnsim\dots\lnsim x_k$. A frame~$F$ has \emph{cluster size $\le k$} if all clusters of~$F$ have at
most~$k$ elements. Similarly to width, we say a logic~$L$ has \emph{depth} (\emph{cluster size}) $\le k$ if it is
complete w.r.t.\ a class of frames of depth (cluster size, resp.) $\le k$, or equivalently, if all refined $L$-frames
have depth (cluster size) $\le k$; $L$ has \emph{bounded depth} (\emph{cluster size}) if it has depth (cluster size) $\le k$
for some~$k$, and it has \emph{unbounded depth} (\emph{cluster size}) otherwise.

These properties are modally definable: $L$ has width (depth, cluster size) $\le k$ iff it proves the
$\lgc{BW}_{\!k}$ ($\lgc{BD}_k$, $\lgc{BC}_k$, resp.) axioms, where $\lgc{BW}_{\!k}$ and $\lgc{BC}_k$ were given in
Table~\ref{tab:axi}, and $\lgc{BD}_k$ is the schema
\[\fii_0\lor\Box(\Box\fii_0\to\fii_1\lor\Box(\Box\fii_1\to\dots\to\fii_{k-1}\lor\Box(\Box\fii_{k-1}\to\bot)\cdots)).\]

A finite frame~$F$ has \emph{branching $\le k$} if every cluster of~$F$ has at most~$k$ immediate successor clusters.
If $L$ is a logic with FMP, then $L$ has \emph{branching $\le k$} if it is complete w.r.t.\ a class of finite frames of
branching~$\le k$, or equivalently, if all finite $L$-frames have branching~$\le k$. Again, $L$ has \emph{bounded
branching} if it has branching~$\le k$ for some~$k$, and \emph{unbounded branching} otherwise.

It is more complicated to extend the definition of branching to logics without FMP, as the concept of branching does
not make good sense for infinite frames: first, a non-leaf point in an infinite frame may have no immediate successors
at all, or its immediate successors may not lower bound all its other successors. Second, even in well-behaved frames
such as trees where immediate successors have reasonable graph-theoretic properties, a bound on their number does not
have the expected modal consequences: for example, it is not difficult to show that an arbitrary
finite rooted reflexive frame is a p-morphic image of the infinite complete \emph{binary} tree%
\footnote{See \cite[Thm.~2.21]{cha-zax} for the intuitionistic case; the only difference in the modal case is that
$f(x_0)$, $f(x_1)$, \dots\ will cycle through the root cluster of~$\mathfrak F$. One can also modify the
argument to apply to all countable rooted $\lgc{S4}$-frames.}%
, thus the logic of this tree is just~$\lgc{S4}$, which
has unbounded branching, even though the tree appears to have branching~$2$.

These issues are solved by showing that the logic of finite frames of branching~$\le k$ can be axiomatized by a
suitable axiom schema, namely
\[\tag{$\lgc{BB}_k$}
\Box\Bigl[\LOR_{i\le k}\Box\Bigl(\boxdot\fii_i\to\LOR_{\substack{j\le k\\j\ne i}}\boxdot\fii_j\Bigr)\to\LOR_{i\le k}\boxdot\fii_i\Bigr]
  \to\LOR_{i\le k}\Box\LOR_{\substack{j\le k\\j\ne i}}\boxdot\fii_j\]
(recall that we number indices from~$0$, hence $i\le k$ stands for $i=0,\dots,k$), and then we define a logic~$L$ to
have branching~$\le k$ iff it includes~$\lgc{K4BB}_k$. Since the $\lgc{BB}_k$ axioms are a central topic of this paper,
and in contrast to the well-known superintuitionistic Gabbay--de Jongh logics, this axiomatization is not commonly
found in modal logic literature, we provide more details. (Our $\lgc{BB}_k$ axioms are mentioned without proof in
\cite[Rem.~6.11]{ej:sfef}. The bounded branching logics as such appear in other sources, but they are defined
semantically: see e.g.\ Rybakov~\cite[p.~331]{ryb:bk}.)

Let $\Psi_k$ denote the \emph{$k$-prong fork}: the finite frame consisting of a root with $k$~immediate successors. (For
definiteness, let $\Psi_k$ be reflexive, but this does not matter.)
\begin{Lem}\label{lem:bb}
Let $k\ge1$.
\begin{enumerate}
\item\label{item:17}
A frame~$F$ validates~$\lgc{BB}_k$ iff there is no dense weak subreduction from~$F$ to~$\Psi_{k+1}$.
\item\label{item:18}
A finite frame~$F$ has branching $\le k$ iff there is no dense weak subreduction from~$F$ to~$\Psi_{k+1}$.
\item\label{item:19}
A formula~$\fii$ holds in all finite frames of branching~$\le k$ iff it is derivable in~$\lgc{K4BB}_k$.
\end{enumerate}
\end{Lem}
\begin{Pf}
Let us denote the root of~$\Psi_{k+1}$ as~$u$, and its leaves as $\{v_i:i\le k\}$.

\ref{item:17}: Let $f$ be a subreduction from~$F$ to~$\Psi_{k+1}$. We endow $F$ with an admissible valuation such that
\[F,x\model p_i\iff x\notin\dom(f)\text{ or }f(x)=v_i.\]
Clearly,
\begin{equation}\label{eq:92}
f(x)=v_i\implies F,x\model\boxdot p_i\land\neg\LOR_{j\ne i}\boxdot p_j,
\end{equation}
hence also
\[f(x)=u\implies F,x\model\neg\LOR_{i\le k}\Box\LOR_{j\ne i}\boxdot p_j.\]
We claim that
\[f(x)=u\implies F,x\model
   \Box\Bigl[\LOR_{i\le k}\Box\Bigl(\boxdot p_i\to\LOR_{j\ne i}\boxdot p_j\Bigr)\to\LOR_{i\le k}\boxdot p_i\Bigr],\]
hence $F\nmodel\lgc{BB}_k$. Indeed, if $f(x)=u$, and $x<y\model\neg\LOR_i\boxdot p_i$, let $z_i\ge y$ be such
that $z_i\nmodel p_i$ for each~$i\le k$, i.e., $z_i\in\dom(f)$ and $f(z_i)\ne v_i$. Since $f$ is dense, $x<y<z_0$
implies $y\in\dom(f)$. We cannot have $f(y)=v_i$, as $f(y)\le f(z_i)\ne v_i$. Thus, $f(y)=u$. But then $y$ sees points
in preimages of all~$v_i$, hence \eqref{eq:92} implies
\[F,y\model\neg\LOR_{i\le k}\Box\Bigl(\boxdot p_i\to\LOR_{j\ne i}\boxdot p_j\Bigr).\]

Conversely, assume that $F\nmodel\lgc{BB}_k$. Fix a model $M$ based on~$F$, and an instance of $\lgc{BB}_k$ using
$\{\fii_i:i\le k\}$ which is not true in~$M$. Notice that
\[\vdash_\kiv
  \Box\Bigl[\LOR_{i\le k}\Box\Bigl(\boxdot\fii_i\to\LOR_{j\ne i}\boxdot\fii_j\Bigr)\to\LOR_{i\le k}\boxdot\fii_i\Bigr]
       \to\ET_{i\le k}\Bigl[\Box\Bigl(\boxdot\fii_i\to\LOR_{j\ne i}\boxdot\fii_j\Bigr)
                 \to\Box\LOR_{j\ne i}\boxdot\fii_j\Bigr],\]
hence putting
\begin{align*}
\beta_i&=\boxdot\fii_i\land\ET_{j\ne i}\neg\boxdot\fii_j,\qquad i\le k,\\
\alpha&=\LOR_{i\le k}\Box\neg\beta_i\to\LOR_{i\le k}\boxdot\fii_i,
\end{align*}
we have $M\nmodel\Box\alpha\to\LOR_i\Box\neg\beta_i$. We define a partial (and a priori multi-valued) mapping $f$ from
$F$ to~$\Psi_{k+1}$ by
\[f(x)=\begin{cases}
  u&M,x\model\Box\alpha\land\ET_i\dia\beta_i,\\
  v_i&M,x\model\beta_i,\\
  \text{undefined}&\text{otherwise.}
\end{cases}\]
We claim that $f$ is a weak dense subreduction. The property (S3) is clear, and for (S2${}'$), it suffices to observe
that $f(x)=u$ implies $x\model\dia\beta_i$, hence $f(y_i)=v_i$ for some $y_i>x$. Since there exists $x$ such that
$f(x)=u$, this also implies that $f$ is onto.

For (S1${}'$), it is clear from the definition that $f(y_i)=v_i$ and $f(y_j)=v_j$ implies
$y_i\nleq y_j$ for $i\ne j$. Also, if $f(x)=u$ and $f(y_i)=v_i$, then $y_i\nleq x$: fixing $j\ne i$ (here we use
$k\ge1$), we already established that there exists $y_j>x$ such that $f(y_j)=v_j$, hence $y_i\nleq y_j$, and a fortiori
$y_i\nleq x$. This also ensures $f$ is single-valued.

It remains to prove that $f$ is dense. Assume $x<y<z$ and $x,z\in\dom(f)$. It is easy to see that $f(x)=f(z)$ implies
$f(y)=f(x)$. Otherwise $f(x)=u$ and $f(z)=v_i$ for some~$i\le k$. Then $y\model\boxdot\alpha$, thus either $f(y)=u$
and we are done, or $y\model\LOR_j\Box\neg\beta_j$, hence (in view of $y\model\alpha$) $y\model\boxdot\fii_{i'}$ for
some~$i'\le k$. Since $y<z$, we have $y\model\ET_{j\ne i}\neg\boxdot\fii_j$, hence $i'=i$ and $y\model\beta_i$, i.e.,
$f(y)=v_i$.

\ref{item:18}: If a point $x$ of~$F$ has immediate successors $y_0,\dots,y_k$, each belonging to a different cluster,
we can construct a weak dense subreduction from~$F$ to~$\Psi_{k+1}$ by mapping $\cls(x)$ to~$u$, and each $\cls(y_i)$
to~$v_i$.

On the other hand, if $f$ is such a weak dense subreduction, let $x$ be a $\lnsim$-maximal point of~$F$ mapped to~$u$.
For each $i\le k$, there exists $y_i\gnsim x$ such that $f(y_i)=v_i$. Let
$z_i$ be an immediate successor of~$x_i$ such that $z_i\le y_i$. Since $f$ is dense, $z_i\in\dom(f)$; by maximality
of~$x$, $u\ne f(z_i)\le f(y_i)$, hence $f(z_i)=v_i$. But then $\{z_i:i\le k\}$ are pairwise incomparable, i.e., they
belong to $k+1$ different clusters.

\ref{item:19}: The right-to-left implication follows from \ref{item:17} and~\ref{item:18}. Conversely, if
$\nvdash_{\lgc{K4BB}_k}\fii$, let us fix a $\lgc{K4BB}_k$-frame $F$ such that $F\nmodel\fii$. Then $F$ validates the
axioms $\alpha_{\I,k+1}$ and $\alpha_{\nr1,k+1}$ from \cite[Def.~4.30]{ej:modparami}: this follows from \ref{item:17}
and \cite[L.~4.31]{ej:modparami}, as any weak morphism to $F_{\I,k+1}$ or~$F_{\nr1,k+1}$ (as defined there) is a weak
dense subreduction to~$\Psi_{k+1}$. By \cite[L.~4.35]{ej:modparami}, there exists a finite frame
$F_0\model\kiv\oplus\alpha_{\I,k+1}\oplus\alpha_{\nr1,k+1}$ such that $F_0\nmodel\fii$. But then $F_0$ has
branching $\le k$ by \cite[L.~4.34]{ej:modparami}. (This argument also shows
$\lgc{K4BB}_k=\kiv\oplus\alpha_{\I,k+1}\oplus\alpha_{\nr1,k+1}$.)

Alternatively, a similar argument can be set up using \cite[L.~6.10]{ej:sfef} (note that the $\lgc{K4BB}_k$ appearing
in the statement of that lemma is \emph{defined} as the logic of all finite frames of branching~${\le k}$).
\end{Pf}

We remark that our definition of $\lgc{BB}_k$ does not have the correct semantics for $k=0$; in order to extend
Lemma~\ref{lem:bb} to $k=0$, we should redefine $\lgc{K4BB_0}$ as $\lgc{K4B}$.

We have $\lgc{K4BB_1}=\lgc{K4BW_{\!1}}=\lgc{K4.3}$. For $k\ge2$, all logics of width~$\le k$ also have
branching~$\le k$, but there exist logics of branching~$2$ and unbounded width such as $\lgc{K4BB_2}$ itself. We have
$\lgc{K4BB_1}\sSset\lgc{K4BB_2}\sSset\lgc{K4BB_3}\sSset\dots$, and $\bigcap_k\lgc{K4BB}_k=\lgc{K4}$.

We could drop the right-most $\boxdot$ in the definition of~$\lgc{BB}_k$, but for our purposes the definition above
will be more convenient to work with. Furthermore, the $\lgc{BB}_k$ axiom can be simplified to
\[\Box\LOR_{i\le k}\boxdot\fii_i\to\LOR_{i\le k}\Box\LOR_{j\ne i}\fii_j\]
over~$\lgc{GL}$.

\subsection{Proof complexity}\label{sec:proof-complexity}

An introduction to classical proof complexity can be found in Kraj\'\i\v cek~\cite{kra:pfcomp}; our setup for proof
complexity of modal logics is based on Je\v r\'abek~\cite{ej:sfef}.

A \emph{Frege rule} consists of all substitution instances of $\alpha_0,\dots,\alpha_{k-1}\ru\beta$, where $k\ge0$, and
$\alpha_i$ and~$\beta$ are formulas. A \emph{Frege system} is given by a finite set of Frege rules~$R$. A \emph{Frege
$R$-derivation} of a formula~$\fii$ from a set of formulas~$\Gamma$ is a sequence of formulas $\fii_0,\dots,\fii_m$
such that $\fii_m=\fii$, and for each~$i\le m$, $\fii_i\in\Gamma$, or $\fii_{j_0},\dots,\fii_{j_{k-1}}\ru\fii_i$ is an
instance of an $R$-rule for some $j_0,\dots,j_{k-1}<i$. A \emph{Frege $R$-proof} of~$\fii$ is a Frege $R$-derivation
of~$\fii$ from~$\nul$. The \emph{length} or \emph{size} of a derivation $\fii_0,\dots,\fii_m$ is $\sum_i\lh{\fii_i}$,
and the \emph{number of lines} is~$m+1$. A derivation is \emph{tree-like} if each formula is used at most once as a
premise of a Frege rule.

The associated consequence relation $\vdash_R$ is defined such that $\Gamma\vdash_R\fii$ iff there exists a Frege
$R$-derivation of~$\fii$ from~$\Gamma$. If $L$ is a logic, a Frege system using a set of rules~$R$ is a \emph{Frege
system for~$L$} if ${\vdash_R}={\vdash_L}$. (Note that this disallows the use of proper $L$-admissible rules as
in~\cite{min-koj,ej:modfrege}.)
\begin{Obs}\label{obs:frege-schema}
If $\fii_0,\dots,\fii_m$ is a Frege $R$-derivation of size~$s$ using variables $\{p_i:{i<n}\}$, and $\sigma$ is a
substitution, then $\sigma(\fii_0),\dots,\sigma(\fii_m)$ is a Frege $R$-derivation of size
$\le s\sum_{i<n}\lh{\sigma(p_i)}$ with the same number of lines.
\noproof\end{Obs}

A proof system~$P$ \emph{p-simulates} a proof system~$Q$, written as $Q\le_pP$, if there exists a poly-time
function~$f$ such that for any $Q$-proof $\pi$ of~$\fii$, $f(\pi)$ is a $P$-proof of~$\fii$. The systems $P$ and
$Q$ are \emph{p-equivalent}, written as $P\equiv_pQ$, if $P\le_pQ\le_pP$. The system~$P$ \emph{(weakly) simulates~$Q$}
if for any $Q$-proof $\pi$ of~$\fii$, there exists a $P$-proof of~$\fii$ of size polynomial in $\lh\pi$. If $P$ does
not weakly simulate~$Q$, we also say that $Q$ has \emph{superpolynomial speed-up} over~$P$; more generally, if
$S$ is a family of functions $s\colon\omega\to\omega$, then $Q$ has \emph{speed-up $S$} over~$P$ if there exist $s\in
S$, an infinite sequence of tautologies $\{\fii_n:n\in\omega\}$, and for each~$n$, a $Q$-proof $\pi_n$ of~$\fii_n$
such that all $P$-proofs of~$\fii_n$ have size at least $s\bigl(\lh{\pi_n}\bigr)$. (For example, for
$S=2^{n^{\Omega(1)}}$, we have \emph{exponential} speed-up.)

Observation~\ref{obs:frege-schema} implies that instances of a fixed Frege rule have linear-size proofs in any Frege system
where they are derivable at all, hence:
\begin{Cor}\label{cor:frege-sim}
For any logics $L\sset L'$, all Frege systems for~$L'$ p-simulate all Frege systems for~$L$. In particular, all
Frege-systems for~$L$ are p-equivalent.
\noproof\end{Cor}
(We rely here on all our proof systems having the same language. It is well known that in the classical case,
Corollary~\ref{cor:frege-sim} holds even if we allow Frege systems using different complete sets of connectives, but the
argument fails for modal logics.) In view of Corollary~\ref{cor:frege-sim}, we will speak of \emph{the} Frege system for a
logic~$L$, and we will denote it $\Fr L$. If $P$ is a line-based proof system such as $\Fr L$, we denote by $\TL{}P$ the
tree-like version of~$P$.

Let us fix an $\Fr L$ system using a set of rules~$R$. An \emph{extended Frege} derivation of~$\fii$ from~$\Gamma$ is a
sequence $\fii_0,\dots,\fii_m=\fii$ where each $\fii_i$ is either from~$\Gamma$, or derived by a Frege rule, or it is
an \emph{extension axiom} of the form $q\eq\psi$, where $q$ is a variable (an \emph{extension variable}) that does
not occur in $\fii$, $\Gamma$, $\psi$, or $\fii_j$ for any~$j<i$.

A \emph{substitution Frege} proof of~$\fii$ is a sequence $\fii_0,\dots,\fii_m=\fii$ such that each $\fii_i$ is derived
by a Frege rule, or by the \emph{substitution rule}: $\fii_i=\sigma(\fii_j)$ for some substitution~$\sigma$ and $j<i$.
($\SF{}$ \emph{derivations} from nonempty sets of premises do not make good sense.)

The extended Frege and substitution Frege systems for~$L$ are denoted $\EF L$ and~$\SF L$, respectively.
Corollary~\ref{cor:frege-sim} holds for $\EF{}$ systems, $\SF{}$ systems, as well as the
circuit-based systems below. It also holds for the tree-like systems $\TL\Fr L$, $\TL\EF L$, and
$\TL\CF L$ because of \cite[Prop.~3.17]{ej:sfef}, but for $\TL\SF L$, we need to assume that \eqref{eq:90} is included
among the Frege rules (or at least, that it has a tree-like Frege derivation in which one of the premises is used only
once).

For classical logic, $\EF{}$ and~$\SF{}$ are p-equivalent. The situation in modal logics is more complicated; the main
properties of the two systems are summarized below.
\begin{Thm}[\cite{ej:sfef}]\label{thm:sfef}
Let $L\Sset\kiv$.
\begin{enumerate}
\item $\Fr L\equiv_p\TL\Fr L$ and $\EF L\equiv_p\TL\EF L\equiv_p\TL\SF L$.
\item If $\fii$ has an $\EF L$~proof with $m$~lines, it has an $\Fr L$~proof with $O(m)$~lines.
If $\fii$ has an $\Fr L$~proof with $m$~lines, it has an $\EF L$~proof of size $O\bigl(m+\lh\fii^2\bigr)$.
\item If $\fii$ has an $\SF L$~proof of size~$s$ with~$m$ lines, it has an $\TL\Fr L$~proof of size $(s/m)^m<2^s$ with
$2^m$~lines.
\item If $L$ has unbounded branching, then $\SF L$ has exponential speed-up over $\EF L$.
\item If $L$ is a logic of bounded width and depth, or $L=\lgc{K4BW}_k$, $\lgc{S4BW}_k$, $\lgc{GLBW}_k$,
$\lgc{K4GrzBW}_k$, or $\lgc{S4GrzBW}_k$ for some~$k$, then $\SF L\equiv_p\EF L$.
\noproof
\end{enumerate}
\end{Thm}

Formulas (both Boolean and modal) can be represented more succinctly by \emph{circuits}: a circuit is a directed
acyclic graph (allowing multiple edges) with a unique node of out-degree~$0$ (the \emph{output} node); each node
of the circuit is labelled either with a variable, in which case it has in-degree~$0$, or with a $k$-ary connective, in
which case it has in-degree~$k$ (the incoming edges are ordered). Formulas can be identified with tree-like circuits
(i.e., each node other than the output has out-degree~$1$).

The \emph{circuit Frege system} $\CF L$ (introduced in~\cite{ej} for~$\CPC$) is defined essentially the same way as
$\Fr L$, except that it operates with circuits instead of formulas. There is an additional rule that allows to infer a
circuit from another circuit that represents the same formula (this property can be checked in polynomial time, or even
in~$\cxt{NL}$); alternatively, this rule may be replaced with several ``local'' transformation rules that only modify
the top part of the circuit.

When used for proving formulas (or deriving formulas from formulas), $\CF L$ is p-equivalent to $\EF L$. In fact, we
can in a sense simulate $\CF L$ by $\EF L$ even for proofs of circuits, but we need to translate them to formulas
first.

If $\fii$ is a circuit, we interpret $\Sub(\fii)$ as the set of subcircuits of~$\fii$. We fix distinct variables
$\{q_\psi:\psi\in\Sub(\fii)\}$ not occurring in~$\fii$, and define
\begin{align*}
\psi^*&=\begin{cases}
   \psi&\text{$\psi$ is a variable,}\\
   c(q_{\psi_0},\dots,q_{\psi_{k-1}})&\psi=c(\psi_0,\dots,\psi_{k-1})\text{ for a connective $c$,}
\end{cases}\\
\Ex_\fii&=\ET_{\psi\in\Sub(\fii)}\boxdot(q_\psi\eq\psi^*).
\end{align*}
\begin{Lem}\label{lem:cf-ef}
Let $L\Sset\kiv$. Given a modal circuit $\fii$, the following are polynomial-time constructible from each other:
\begin{enumerate}
\item\label{item:13}
An $\CF L$~proof of~$\fii$.
\item\label{item:14}
An $\CF L$~proof of\/ $\Ex_\fii\to q_\fii$.
\item\label{item:15}
An $\EF L$~proof of\/ $\Ex_\fii\to q_\fii$.
\end{enumerate}
\end{Lem}
\begin{Pf}
We can construct $\CF\kiv$~proofs of $\Ex_\fii\to\boxdot(q_\psi\eq\psi)$ for all $\psi\in\Sub(\fii)$ by induction on the
complexity of~$\psi$, which yields a $\CF\kiv$~proof of $\fii\to(\Ex_\fii\to q_\fii)$. Conversely, given an $\CF
L$~proof of $\Ex_\fii\to q_\fii$, we (simultaneously) substitute $\psi$ for~$q_\psi$ in the whole proof,
resulting in an $\CF L$~proof of $\ET_\psi\boxdot(\psi\eq\psi)\to\fii$, from which we can infer~$\fii$.

\ref{item:14} and~\ref{item:15} are mutually poly-time constructible by~\cite[Prop.~3.3]{ej:sfef}.
\end{Pf}

In view of Lemma~\ref{lem:cf-ef}, $\EF{}$ and $\CF{}$ are essentially identical proof systems. We find it much more
convenient to operate with circuits directly rather than by encoding them with extension axioms, hence we will work
almost exclusively with~$\CF{}$. We will still formulate lower bounds and similar results for~$\EF{}$ as it is the
better known of the two systems, but our results on feasibility of the disjunction property will be stated for~$\CF{}$
as it makes them more general (i.e., directly applicable to proofs of circuits rather than just formulas).

We would also like to work with circuits directly in~$\SF{}$. Let us define the \emph{substitution circuit Frege}
system $\SCF L$ as a version of the $\SF L$ system that operates with circuits in place of formulas, including the
$\CF L$ rules. Now, $\SF L$ is p-equivalent to~$\SCF L$ just like $\EF L$ is p-equivalent to $\CF L$:
\begin{Lem}\label{lem:scf-sf}
Let $L\Sset\kiv$. Given a modal circuit $\fii$, the following are polynomial-time constructible from each other:
\begin{enumerate}
\item\label{item:16}
An $\SCF L$~proof of~$\fii$.
\item\label{item:20}
An $\SCF L$~proof of\/ $\Ex_\fii\to q_\fii$.
\item\label{item:21}
An $\SF L$~proof of\/ $\Ex_\fii\to q_\fii$.
\end{enumerate}
\end{Lem}
\begin{Pf}
We can construct \ref{item:16} from~\ref{item:20} as in the proof of Lemma~\ref{lem:cf-ef}, and \ref{item:21} is trivially
an instance of~\ref{item:20}. Given an $\SCF L$~proof $\fii_0,\dots,\fii_m=\fii$, we consider the sequence of formulas
\[\Ex_{\fii_0}\to\boxdot q_{\fii_0},\dots,\Ex_{\fii_m}\to\boxdot q_{\fii_m},\]
and complete it to a valid $\SF L$~proof as follows.

If $\fii_i=\sigma(\fii_j)$ is derived by substitution from $\fii_j$, $j<i$, we use substitution to rename each $q_\psi$
from $\Ex_{\fii_j}$ to the corresponding $q_{\sigma(\psi)}$ from $\Ex_{\fii_i}$, and each original variable $p$ to
$q_{\sigma(p)}$. This turns $\Ex_{\fii_j}\to\boxdot q_{\fii_j}$ into $\Ex'_{\fii_i}\to\boxdot q_{\fii_i}$, where
$\Ex'_{\fii_i}$ is a conjunction of some conjuncts of $\Ex_{\fii_i}$ and the tautologies $\boxdot(q_{\sigma(x)}\eq
q_{\sigma(x)})$. We infer $\Ex_{\fii_i}\to\boxdot q_{\fii_i}$.

If $\fii_i$ is derived by an instance of a Frege rule
$\alpha_0,\dots,\alpha_{k-1}\ru\beta$, say $\fii_i=\beta(\vec\chi)$ and $\fii_{j_u}=\alpha_u(\vec\chi)$ with $j_u<i$, we
first apply the substitution rule on the premises $\Ex_{\fii_{j_u}}\to\boxdot q_{\fii_{j_u}}$ if necessary to rename the
extension variables $q_\psi$ so that they are used coherently in all $\Ex_{\fii_{j_u}}$ and~$\Ex_{\fii_i}$. We
unwind the top parts of the circuits to prove $\Ex_{\fii_{j_u}}\to\boxdot\bigl(q_{\fii_{j_u}}\eq\alpha_u(\vec
q_\chi)\bigr)$, and derive
\[\Ex_{\fii_{j_u}}\to\boxdot\alpha_u(\vec q_\chi).\]
We use an instance of the tautology $\ET_{u<k}\boxdot\alpha_u\to\boxdot\beta$ and
$\Ex_{\fii_i}\to\boxdot(q_{\fii_i}\eq\beta(\vec q_\chi))$ to derive
\[\Ex_{\fii_i}\land\ET_{u<k}\Ex_{\fii_{j_u}}\to\boxdot q_{\fii_i}.\]
Finally, we get rid of the conjuncts $\boxdot(q_\psi\eq\psi^*)$ of $\Ex_{\fii_{j_u}}$ not present in~$\Ex_{\fii_i}$ by
substituting $\psi^*$ for~$q_\psi$ and using the tautology $\boxdot(\psi^*\eq\psi^*)$. (We do this in a
top-down order, so that $q_\psi$ is not present elsewhere in the formula when it is being substituted for.)

If $\fii_i$ represents the same formula as $\fii_j$, $j<i$, we first use substitution to make sure the extension
variables $\{q_\psi:\psi\in\Sub(\fii_i)\}$ from $\Ex_{\fii_i}$ are disjoint from the extension variables
from~$\Ex_{\fii_j}$; let us denote the latter as $q'_\psi$. Then we prove bottom-up that whenever $\psi\in\Sub(\fii_i)$
and $\psi'\in\Sub(\fii_j)$ represent the same formula, we have
$\Ex_{\fii_j}\land\Ex_{\fii_i}\to\boxdot(q_\psi\eq q'_{\psi'})$. Using $\Ex_{\fii_j}\to\boxdot q'_{\fii_j}$, we infer
$\Ex_{\fii_j}\land\Ex_{\fii_i}\to\boxdot q_{\fii_i}$, and we discard $\Ex_{\fii_j}$ as in the case of Frege rules.
\end{Pf}

The upshot of Lemmas \ref{lem:cf-ef} and~\ref{lem:scf-sf} is not just that $\EF L\equiv_p\CF L$ and $\SF L\equiv_p\SCF L$ as proof
systems for formulas, but also that a speed-up of $\SCF L$ over $\CF L$ on circuit tautologies implies a speed-up
of $\SF L$ over $\EF L$: if $\{\fii_n:n\in\omega\}$ is a sequence of circuits that are easy for
$\SCF L$ and hard for $\CF L$, then the formulas $\{\Ex_{\fii_n}\to q_{\fii_n}:n\in\omega\}$ are easy for $\SF L$ and
hard for $\EF L$.

We remark that in a way, the term \emph{formulas} has a double meaning in the paper: formulas-1 are abstract entities
that may be $L$-tautologies, may be true or false in a given model, etc., and they are concretely represented by
syntactic objects such as circuits or formulas-2 (= tree-like circuits) that may be operated by proof systems.

Transitive modal logics have a deduction theorem in the form that $\Gamma\vdash_L\fii$ implies
$\vdash_L\ET\boxdot\Gamma\to\fii$. (Here, if $\Gamma$ is a sequence of formulas $\fii_0,\dots,\fii_{n-1}$, we write
$\Box\Gamma$ for $\Box\fii_0,\dots,\Box\fii_{n-1}$, and similarly for $\boxdot\Gamma$, $\neg\Gamma$, etc., while
$\ET\Gamma$ is $\fii_0\land\dots\land\fii_{n-1}$.) Frege systems and friends without an explicit substitution rule
satisfy a \emph{feasible deduction theorem}:
\begin{Lem}[{{\cite[Prop.~3.6]{ej:sfef}}}]\label{lem:feas-ded}
Let $L\Sset\kiv$, and $P$ be $\Fr L$, $\EF L$, or $\CF L$. Given a $P$-derivation of~$\fii$ from~$\Gamma$, we can
construct in polynomial time a $P$-proof of $\ET\boxdot\Gamma\to\fii$.
\noproof\end{Lem}

We also have feasible substitution of equivalence:
\begin{Lem}\label{lem:subset-eq}
Given modal circuits $\fii$, $\psi$, and $\chi(p)$ \brak{with other variables not shown}, we can construct in polynomial
time $\CF\kiv$~proofs of
\[\boxdot(\fii\eq\psi)\to\bigl(\chi(\fii)\eq\chi(\psi)\bigr).\]
\end{Lem}
\begin{Pf}
By induction on~$\chi$.
\end{Pf}

Let $\two=\{0,1\}$. A Boolean function $f\colon\two^n\to\two$ is \emph{monotone} if for all $a,b\in\two^n$, $a\le b$ (i.e.,
$a_i\le b_i$ for each $i<n$) implies $f(a)\le f(b)$. A \emph{monotone language} is $L\sset\two^*$ such that for all
$n\in\omega$, the characteristic function of $L_n=L\cap\two^n$ is monotone.

A Boolean formula or circuit is \emph{monotone} if it is built from variables using only the monotone connectives
$\{\land,\lor,\top,\bot\}$. More generally, $\fii$ is \emph{monotone in variables~$\vec p$} if it is built using
monotone connectives from the variables~$\vec p$, and from subformulas/subcircuits that do not contain~$\vec p$.
A Boolean formula or circuit is in \emph{negation normal form} if it has the form $\fii(\vec p,\neg\vec p)$, where
$\fii$ is monotone (i.e., it is built using monotone connectives from positive and negative literals).
\begin{Lem}\label{lem:mon}
Given a Boolean circuit~$\fii(p_0,\dots,p_{n-1})$ \brak{possibly using other variables} that is monotone in~$\vec p$, and
Boolean or modal circuits $\vec\psi$ and~$\vec\chi$, there is a polynomial-time constructible $\CF\CPC$~proof or
$\CF{\lgc K}$ proof (as appropriate) of
\begin{equation}\label{eq:76}
\ET_{i<n}(\psi_i\to\chi_i)\to\bigl(\fii(\vec\psi)\to\fii(\vec\chi)\bigr).
\end{equation}
\end{Lem}
\begin{Pf}
By induction on~$\fii$. (Note that \eqref{eq:76} is a substitution instance of the Boolean tautology
$\ET_i(p_i\to q_i)\to\bigl(\fii(\vec p)\to\fii(\vec q)\bigr)$, hence even in the modal case, the proof is essentially a
$\CF\CPC$ proof in modal language.)
\end{Pf}

\begin{Lem}\label{lem:mon-box}
Given a monotone Boolean circuit~$\fii(\vec p)$, and \brak{modal} circuits $\vec\psi$, there are poly-time constructible
$\CF{\lgc K}$~proofs of
\[\fii(\Box\vec\psi)\to\Box\fii(\vec\psi).\]
\end{Lem}
\begin{Pf}
By induction on the size of~$\fii$, using Lemma~\ref{lem:mon}, and the tautologies
$\Box\psi\land\Box\chi\to\Box(\psi\land\chi)$ and $\Box\psi\lor\Box\chi\to\Box(\psi\lor\chi)$.
\end{Pf}

\emph{Makinson's theorem} states that every consistent normal modal logic~$L$ is valid in a one-point Kripke frame
(irreflexive~$\I$, or reflexive~$\R$). In other words, $L$ is included in $L(\I)=\lgc K\oplus\Box\bot$ or in
$L(\R)=\lgc K\oplus(\fii\eq\Box\fii)$. In either case, we obtain a poly-time translation of $L$ into~$\CPC$: if
$\RI\in\{\I,\R\}$, we define a translation of modal formulas~$\fii$ to Boolean formulas $\fii^\RI$ such that it
preserves propositional variables, commutes with Boolean connectives, and
\begin{align*}
(\Box\fii)^\I&=\top,\\
(\Box\fii)^\R&=\fii^\R.
\end{align*}
Notice that $\fii^\RI=\fii$ for non-modal formulas~$\fii$, and $(\boxdot\fii)^\RI\equiv\fii^\RI$. Unwinding the
definition of satisfaction in one-point frames, we see that
\begin{equation}\label{eq:77}
{}\vdash_{L(\RI)}\fii\iff{}\vdash_\CPC\fii^\RI.
\end{equation}
Moreover, the translation acts efficiently on proofs:
\begin{Lem}\label{lem:one-point}
Let $\RI\in\{\I,\R\}$, and $L\sset L(\RI)$ be a normal modal logic. Given an $\CF L$ proof of
$\fii$, we can construct in polynomial time a $\CF\CPC$ proof of $\fii^\RI$.
\end{Lem}
\begin{Pf}
We may assume the $\CF L$ system is axiomatized by \eqref{eq:90}, \eqref{eq:91}, and axiom schemata. We apply the
$-^\RI$ translation to each line in the proof: modus ponens translates to modus ponens, the translation of
\eqref{eq:91} is trivial, and since $-^\RI$ commutes with substitution, instances of a fixed axiom schema valid in~$L$
translate to instances of a fixed axiom schema, which is valid in~$\CPC$ by~\eqref{eq:77}, and as such has linear-size
$\CF\CPC$ proofs.
\end{Pf}

So far we discussed specific proof systems for a given logic. In general, a (Cook--Reckhow) \emph{proof system} for a
logic~$L$ is a polynomial-time function~$P$ whose image is~$L$. (Here, each string~$w$ is considered
a $P$-proof of the $L$-tautology $P(w)$.) For classical logic, $\np\ne\conp$ implies superpolynomial lower bounds on
\emph{all} proof systems because of the $\conp$-completeness of the set of tautologies.

For the modal logics we are interested in, we will obtain similar automatic lower bounds from $\psp\ne\np$, because
they are $\psp$-hard. Ladner~\cite{ladner} proved that $\lgc K$, $\lgc T$, and~$\lgc{S4}$ are $\psp$-complete, and that
all logics $\lgc{K}\sset L\sset\lgc{S4}$ are $\psp$-hard. It is in fact not difficult to extend Ladner's proof to show
the $\psp$-hardness of all normal modal logics with the \emph{disjunction property} (see Section~\ref{sec:disj-prop} for
precise definition), but the author is not aware of this argument being published anywhere. (Cf.
Lemmas \ref{lem:qbf-dec-sf} and~\ref{lem:aphi-phi}. The $\psp$-hardness of
\emph{superintuitionistic} logics with the DP was proved in Chagrov~\cite{chagl}.) The following stronger result
was shown in Je\v r\'abek~\cite{ej:modparamii}:
\begin{Thm}\label{thm:psp-hard}
All logics $L\Sset\kiv$ with the disjunction property are $\psp$-hard. More generally, if for every finite binary
tree~$T$, there exists a weak subreduction from an $L$-frame to~$T$, then $L$ is $\psp$-hard.
\noproof\end{Thm}
\begin{Cor}\label{cor:psp-lb}
If $L$ is a logic as in Theorem~\ref{thm:psp-hard}, then no proof system for~$L$ is polynomially bounded
unless $\psp=\np=\conp$.
\noproof\end{Cor}

The only conditional superpolynomial lower bounds on $\SF L$ we know of follow from Corollary~\ref{cor:psp-lb} (assuming
$\psp\ne\np$) and from an $\SF{}$ version of Lemma~\ref{lem:one-point} (assuming lower bounds on $\EF\CPC$).

\subsection{Computational complexity}\label{sec:complexity-classes}

We assume the reader is familiar with basic notions from complexity theory, in particular the complexity classes
$\ptime$, $\np$, $\conp$, and~$\psp$, and the notions of polynomial-time reductions, completeness, and hardness.

Recall that a \emph{quantified Boolean formula} (QBF) is a propositional formula that, in addition to the usual
Boolean connectives, also allows quantifiers $\exists p$ and $\forall p$ ranging over the set of truth values~$\two$. We
will generally assume that QBFs are given in prenex normal form, i.e., they consist of a quantifier prefix followed by
a quantifier-free formula. A QBF~$\Phi$ in prenex normal form is in \emph{negation normal form} if its quantifier-free
matrix~$\fii$ is in negation normal form, and it is \emph{monotone in $\vec p$} if the $\vec p$ variables are not bound
in~$\Phi$, and $\fii$ is monotone in~$\vec p$.

The validity problem for QBF is a $\psp$-complete language. More uniformly, for any $\psp$-language $L\sset\two^*$,
there exists a sequence of QBFs $\{\Phi_n(p_0,\dots,p_{n-1}):n\in\omega\}$ constructible in time $n^{O(1)}$ such that
\[w\in L\iff\Phi_n(w_0,\dots,w_{n-1})\]
for all $w\in\two^n$. If $L\in\np$ ($L\in\conp$), the $\Phi_n$ can be taken existential (universal, respectively).

The computational problems studied in this paper are mostly not YES--NO decision problems, but \emph{search problems}.
Here, the search problem~$S_R$ associated with a relation $R(x,y)$ is the following computational task: given $x$, find
a~$y$ such that $R(x,y)$, if one exists. The class of search problems solvable in polynomial time is denoted~$\fp$. A
search problem~$S_R$ is \emph{total%
\footnote{In practice, we will usually deal with search problems whose input is constrained by syntactic prerequisites,
such as ``given a proof of~$\fii$, \dots''. We can consider them to be total by stipulating that, say, $0$ is a valid
output if the input does not meet the requirements; this does not change the computational complexity of the problem,
as the input condition is checkable in polynomial time.}%
} if $\forall x\,\exists y\,R(x,y)$.

A search problem $S_{R_1}$ is (many-one) \emph{reducible} to $S_{R_0}$, written as $S_{R_1}\le S_{R_0}$, if there are
poly-time functions $f$ and~$g$ such that
\[R_0\bigl(f(x),y\bigr)\implies R_1\bigl(x,g(x,y)\bigr)\]
for all $x$ and~$y$ (i.e., $f$ translates instances of~$S_{R_1}$ to instances of~$S_{R_0}$, and $g$ translates
solutions back). We write $S_{R_0}\equiv S_{R_1}$ if $S_{R_0}\le S_{R_1}\le S_{R_0}$.

This standard notion of search problem reduction is suitable for ``open-ended'' search problems with many solutions,
such as when looking for proofs of some formula. However, we will also encounter search problems with a fixed finite
set of possible outcomes that may be better thought of as many-valued decision problems (possibly with non-unique
answers). In such cases, it is not appropriate to translate solutions. (Notice that many-one reductions between
languages likewise do not allow swapping a language for its complement.)

Thus, we define $S_{R_1}$ to be \emph{strictly reducible} to $S_{R_0}$, written as $S_{R_1}\le_sS_{R_0}$, if there
exists a reduction of $S_{R_1}$ to~$S_{R_0}$ with $g(x,y)=y$. Again, we put $S_{R_0}\equiv_sS_{R_1}$ iff
$S_{R_0}\le_sS_{R_1}\le_sS_{R_0}$. An even stricter notion of reduction is when $f$ is identity as well, i.e.,
$R_0\sset R_1$: then we say $S_{R_1}$ is \emph{subsumed} by~$S_{R_0}$.

We will also refer to \emph{nonuniform poly-time} reductions, where the reduction functions are computable in
polynomial time using an extra polynomial-size \emph{advice string} that only depends on the length of the input.

We define $S_R$ to be a \emph{$\conp$~search problem%
\footnote{Confusingly, $\np$~search problems are those where
$R\in\ptime$. To be consistent with this terminology, we should perhaps call $\conp$~search problems
\emph{$\Sigma^\ptime_2$~search problems}. We do not, because we consider the naming of $\np$~search problems somewhat
of a misnomer in the first place, and moreover, the idea behind this nomenclature (that $\Sigma^\ptime_2$~search
problems seek witnesses for $\Sigma^\ptime_2$ predicates) does not apply to our problems, which have a
bounded range, hence the corresponding decision problems are in~$\cxt{BH}$ rather than full $\Sigma^\ptime_2$.
(Calling them \emph{$\cxt{BH}$~search problems} would be probably even more confusing.)}%
} if $R\in\conp$.

Two-valued search problems are closely related to promise problems, i.e., disjoint pairs. In particular, a
\emph{disjoint $\np$~pair} is $\p{A_0,A_1}$, where $A_0,A_1\in\np$ and $A_0\cap A_1=\nul$. This represents the
following computational task: given $x\in A_0\cup A_1$, output $i<2$ such that $x\in A_i$ (if $x\notin A_0\cup A_1$,
any output is valid). A disjoint $\np$~pair $A=\p{A_0,A_1}$ reduces to $B=\p{B_0,B_1}$, written $A\le B$, if there
exists a poly-time function~$f$ such that
\[x\in A_i\implies f(x)\in B_i,\qquad i=0,1.\]
Now, a disjoint $\np$~pair $\p{A_0,A_1}$ represents the same task as the total $\two$-valued $\conp$~search problem
$S_R$, where $R(x,i)\iff x\notin A_{1-i}$. On the other hand, if $S_R$ is a total $\two$-valued $\conp$~search
problem, it represents the same task as the disjoint $\np$~pair $\p{A_0,A_1}$, where $A_i=\{x:\neg R(x,1-i)\}$.
Moreover, if $S_R$ and~$S_{R'}$ are total $\two$-valued $\conp$~search problems, and $A$ and~$A'$ the corresponding
disjoint $\np$~pairs, we have
\[S_R\le_sS_{R'}\iff A\le A',\]
using the same reduction function. For these reasons, we may identify total two-valued $\conp$ search problems with
disjoint $\np$~pairs. (More generally, total two-valued search problems may be identified with promise problems.)

\subsection{Disjunction properties}\label{sec:disj-prop}

A consistent modal logic~$L$ has the \emph{disjunction property} (DP) if for all formulas $\fii_0$ and~$\fii_1$,
$L$~proves $\Box\fii_0\lor\Box\fii_1$ only if it proves $\fii_0$ or~$\fii_1$. (We note that it is conceptually more
appropriate to define DP so that for every \emph{finite set} of formulas $\{\fii_i:i\in I\}$, $L$~proves
$\LOR_{i\in I}\Box\fii_i$ only if it proves $\fii_i$ for some $i\in I$. However, for transitive logics, this more
general definition is equivalent to its special cases with $I=\nul$, which amounts to the consistency of~$L$, and
$\lh I=2$, which is how we introduced DP above. We prefer the definition with $\lh I=2$ as it simplifies the
presentation of DP as a computational problem, see below.)

DP is an example of a multi-conclusion admissible rule. In general, a \emph{consecution} is a pair of finite sets of
formulas, written as $\Gamma\ru\Delta$, and a \emph{multi-conclusion rule%
\footnote{In structural theory of propositional logics, the term ``admissible rule'' is usually reserved for
\emph{schematic} rules, i.e., rules that consist of all substitutions instances of a single consecution, similarly to
Frege rules (see e.g.\ Rybakov~\cite{ryb:bk}); however, it will be more convenient for our purposes to adopt a more
relaxed definition.}%
} is a set~$R$ of consecutions (called the \emph{instances} of~$R$). A rule~$R$ is \emph{$L$-admissible} if for all
instances $\Gamma\ru\Delta$ of~$R$, if $\vdash_L\fii$ for all $\fii\in\Gamma$, then $\vdash_L\psi$ for some
$\psi\in\Delta$. We will write rules in a schematic form (analogous to axiom schemata) whenever possible. Thus, $L$ has
DP iff the rule $\Box\fii_0\lor\Box\fii_1\ru\fii_0,\fii_1$ is admissible, and the finite-set formulation of DP amounts
to the admissibility of the rules
\[\tag{$\DP_n$}\Box\fii_0\lor\dots\lor\Box\fii_{n-1}\ru\fii_0,\dots,\fii_{n-1}\]
for $n\in\omega$.

Semantically, the disjunction property corresponds to the following closure property on $L$-frames (see
\cite[Thm.~15.1]{cha-zax}): given two (or finitely many) rooted $L$-frames $F_0$ and~$F_1$, there exists a rooted
$L$-frame $F$ that includes disjoint isomorphic copies of $F_0$ and~$F_1$ as generated subframes. In particular, if for
each $i=0,1$, $W_i$ is a model based on $F_i$ that refutes $\fii_i$, then $\Box\fii_0\lor\Box\fii_1$ is refuted at the
root of $F$ under any valuation that extends that of $W_0$ and~$W_1$.

The simplest way how to construct a rooted frame that includes given rooted frames $\{F_i:i<n\}$ as generated subframes
is to take their disjoint sum $\sum_{i<n}F_i$, and attach to it a new root: we denote the resulting frame
$\bigl(\sum_{i<n}F_i\bigr)^\I$ if the new root is irreflexive, and $\bigl(\sum_{i<n}F_i\bigr)^\R$ if it is reflexive.
Many common transitive modal logics with DP are in fact closed under this frame construction; if $\RI\in\{\I,\R\}$,
we say that a logic~$L$ is \emph{$\RI$-extensible} if for every $n\in\omega$ and rooted $L$-frames $\{F_i:i<n\}$, the
frame $\bigl(\sum_{i<n}F_i\bigr)^\RI$ is an $L$-frame. (We also say that $L$ is \emph{extensible} if it is
$\I$-extensible unless $L\Sset\lgc{S4}$, and $\R$-extensible unless $L\Sset\lgc{GL}$.)

It turns out that $\RI$-extensible logics do not have just DP, but
they admit more general \emph{extension rules%
\footnote{By an unfortunate clash of terminology, \emph{extension rule} is also a standard name in proof complexity for
the ``rule'' that warrants postulation of extension axioms in $\EF{}$ proofs. We refrain from this usage to avoid
confusion.}%
}
\[\tag{$\Ext^\RI_{n,m}$}\ET_{j<m}B^\RI(\chi_j)\to\Box\fii_0\lor\dots\lor\Box\fii_{n-1}
   \Ru\ET_{j<m}\boxdot\chi_j\to\fii_0,\dots,\ET_{j<m}\boxdot\chi_j\to\fii_{n-1}\]
for $n,m\in\omega$, where
\begin{align*}
B^\I(\fii)&=\Box\fii,\\
B^\R(\fii)&=(\fii\eq\Box\fii).
\end{align*}
We also put $\Ext^\RI=\bigcup\{\Ext^\RI_{n,m}:n,m\in\omega\}$ and $\Ext^\RI_n=\bigcup\{\Ext^\RI_{n,m}:m\in\omega\}$.

For example, the logics $\lgc{K4}$, $\lgc{S4}$, $\lgc{GL}$, $\lgc{K4Grz}$, $\lgc{K4.1}$, $\lgc{K4BC}_k$,
$\lgc{S4.1.4}$, and their arbitrary combinations, are extensible. The logics $\lgc{D4}$, $\lgc{D4.1}$, $\lgc{D4Grz}$,
and $\lgc{D4BC}_k$ are $\R$-extensible, but not $\I$-extensible (though they only fail the condition for $n=0$, hence
they admit $\Ext^\I_n$ for all $n>0$, and most results below on $\I$-extensible logics can be easily adapted to them).

The following characterization was essentially proved in \cite{ej:modfrege}:
\begin{Thm}\label{thm:ext-char}
Let $L\Sset\kiv$, and $\RI\in\{\I,\R\}$. The following are equivalent:
\begin{enumerate}
\item\label{item:1} $L$ is $\RI$-extensible.
\item\label{item:2} The rules $\Ext^\RI$ are $L$-admissible.
\item\label{item:3} $L$ can be axiomatized over $\kiv$ by \brak{substitution instances of} axioms each of which has the form
\addtocounter{equation}{1}
\ifnum\value{equation}=4 \else\errmessage{Fixed the wrong equation}\fi
\begin{equation}\label{eq:1}
\Box\beta\land\Box(\Box\alpha\to\alpha)\to\Box\alpha
\end{equation}
if $\RI=\I$, and one of the forms
\begin{equation}\label{eq:2}
\beta\land\Box\alpha\to\alpha
\end{equation}
or
\begin{equation}\label{eq:3}
\Box\gamma\land\Box(\Box\alpha\to\beta)\land\Box(\Box\beta\to\alpha)\land\Box(\alpha\lor\beta)\to\Box\alpha
\end{equation}
if $\RI=\R$.
\end{enumerate}
\end{Thm}
\begin{Pf}
The equivalence of \ref{item:1} and~\ref{item:2} is from \cite[Thm.~3.5]{ej:modfrege}. \ref{item:3}\txto\ref{item:1}:
It is straightforward to check that a valuation in $\bigl(\sum_{i<n}F_i\bigr)^\RI$ that makes an axiom of such form true
in each~$F_i$ also makes it true in the root.

\ref{item:2}\txto\ref{item:3}: First, assume $\RI=\I$. Even though \cite[Thm.~3.11]{ej:modfrege} is stated only for
extensible logics, the argument (using Claim~1) applies directly to $\I$-extensible logics, showing they are
axiomatizable over $\kiv$ by Zakharyaschev's canonical formulas $\alpha(F,D,\bot)$ (see \cite[\S9.4]{cha-zax} and
\cite[3.6--3.10]{ej:modfrege}) where the root of $F$ is reflexive. Considering that $\boxdot$ commutes with~$\land$,
such a canonical formula can be brought to the syntactic form
\begin{equation}\label{eq:5}
\boxdot\beta\land\boxdot(\Box\alpha\to\alpha)\to\alpha
\end{equation}
for some formulas $\alpha$ and~$\beta$ (in fact, with $\alpha$ being just a variable). Now, for a given $\alpha$ and~$\beta$,
\eqref{eq:5} is equiderivable with~\eqref{eq:1} over~$\kiv$: on the one hand, we can derive \eqref{eq:1}
from~\eqref{eq:5} by \eqref{eq:91} and distributing the boxes; on the other hand, $\eqref{eq:1}\to\eqref{eq:5}$ is a
classical tautology.

If $\RI=\R$, then \cite[Thm.~3.11]{ej:modfrege} shows that $L$ is axiomatizable by canonical formulas
$\alpha(F,D,\bot)$ where the root cluster of~$F$ is either proper or irreflexive. In the former case, the canonical
formula has the form
\[\boxdot\gamma\land\boxdot(\Box\alpha\to\beta)\land\boxdot(\Box\beta\to\alpha)\land\boxdot(\alpha\lor\beta)\to\alpha,\]
which is equiderivable with~\eqref{eq:3} similarly to the argument for $\RI=\I$. In the latter case, the canonical
formula has the form
\[\beta\land\boxdot(\alpha\lor\Box\alpha)\to\alpha,\]
which is equivalent to~\eqref{eq:2}.
\end{Pf}

In contrast to DP, the extension rules are not equivalent to their restrictions with bounded~$n$. For a
fixed~$n$, the $L$-admissibility of $\Ext^\I_n$ or $\Ext^\R_n$ is equivalent to the closure of the
class of rooted $L$-frames under taking $\bigl(\sum_{i<n}F_i\bigr)^\I$ or $\bigl(\sum_{i<n}F_i\bigr)^\R$
(respectively), thus for example, $\lgc{K4BB}_k$ admits $\Ext^\I_n$ and $\Ext^\R_n$ for $n\le k$, but not for any
larger~$n$.

On the other hand, since $\land$ commutes with $\Box$ and~$\boxdot$, $\Ext^\I_n$ is (feasibly) equivalent
to~$\Ext^\I_{n,1}$. The reflexive case is more involved, but it was shown in \cite{ej:indep} that $\Ext^\R_n$ is
equivalent to $\Ext^\R_{n,2}$, and in fact, to its special case
\[\boxdot(\chi\eq\Box\chi)\to\Box\fii_0\lor\dots\lor\Box\fii_{n-1}\ru\boxdot\chi\to\fii_0,\dots,\boxdot\chi\to\fii_{n-1}.\]
However, the reduction as given in \cite[L.~3.3]{ej:indep} involves formulas of size
doubly exponential in~$m$, hence we prefer to state the rules in the more general form above for computational purposes.

The disjunction property gives rise to several computational problems, in particular:
\begin{itemize}
\item Given a proof of $\Box\fii\lor\Box\psi$, decide if $\fii$ or $\psi$ is provable.
\item Given a proof of $\Box\fii\lor\Box\psi$, find a proof of $\fii$ or of~$\psi$.
\end{itemize}
More generally, let $P$ be a proof system for a logic~$L$, and $R$ a (polynomial-time recognizable) multi-conclusion
$L$-admissible rule. The \emph{$R$-decision problem for~$P$}, denoted $\decp(R,P)$, is the total search problem
\begin{itemize}
\item given an instance $\{\fii_i:i<n\}\ru\{\psi_j:j<m\}$ of~$R$, and for each $i<n$, a $P$-proof of $\fii_i$, find 
a $j<m$ such that $\psi_j$ is $P$-provable.
\end{itemize}
The \emph{$R$-proof-construction problem for~$P$}, $\consp(R,P)$, is the total search problem
\begin{itemize}
\item given an instance $\{\fii_i:i<n\}\ru\{\psi_j:j<m\}$ of~$R$, and for each $i<n$, a $P$-proof of $\fii_i$, find a
$P$-proof of some $\psi_j$.
\end{itemize}
(Formally, we make $\decp(R,P)$ and $\consp(R,P)$ total by allowing the output~$0$ if the input does
not have the stated syntactic form.) We say that $P$ has \emph{feasible~$R$} if $\decp(R,P)\in\fp$, and
\emph{constructive feasible~$R$} if $\consp(R,P)\in\fp$.

The extension rules $\Ext^\RI$ have the remarkable feature that they are constructively feasible for Frege, EF, and CF
systems whenever they are admissible at all. This was proved in \cite[Thm.~4.8]{ej:modfrege}. (The result is stated
as a p-simulation of Frege systems for extensible logics using additional single-conclusion admissible rules as new
rules of inference, but the proof, specifically Claims 2 and~3, applies to multi-conclusion rules as well, and only
needs the logic to be $\RI$-extensible. As is the nature of Frege systems, the original formulation allows for
\emph{repeated} applications of the rules, which is something we will not need here.)

Since this is a central tool in this paper, and we will need to adapt the argument later on anyway, we include a
self-contained proof.

If $R$ is a rule, and $S$ a set of formulas, let \emph{$S$-restricted $R$} be the rule consisting of instances
$\Gamma\ru\Delta$ of~$R$ such that $\Gamma\cup\Delta\sset S$.
\begin{Thm}\label{thm:feas-ext}
Let $\RI\in\{\I,\R\}$, and $L\Sset\kiv$ be a $\RI$-extensible logic. Then $\Fr L$ and $\CF L$ have constructive
feasible $\Ext^\RI$, and therefore constructive feasible DP. 
\end{Thm}
\begin{Pf}
Assume first $\RI=\I$. By Theorem~\ref{thm:ext-char} and Corollary~\ref{cor:frege-sim}, we may assume $L$ is axiomatized by the usual axioms and
rules of~$\kiv$, and substitution instances of axioms
\[\Box\beta_j\to\bigl(\Box(\Box\alpha_j\to\alpha_j)\to\Box\alpha_j\bigr),\qquad j<k,\]
for some~$k$ and formulas $\alpha_0,\beta_0,\dots,\alpha_{k-1},\beta_{k-1}$. Given an $\CF L$ proof 
$\pi=\p{\theta_0,\dots,\theta_z}$ of
\[\theta_z=\ET_{j<m}\Box\chi_j\to\LOR_{i<n}\Box\fii_i,\]
let $\Pi$ be the closure of~$\pi\cup\{\chi_j:j<m\}$ under \eqref{eq:90} and $\Sub(\pi)$-restricted \eqref{eq:91}.

Clearly, all circuits in~$\Pi$ are subcircuits of some~$\theta_i$. There are only polynomially many such subcircuits,
and then it is easy to see that $\Pi$ can be computed in polynomial time. Also, $\Pi$ can be arranged into an $\CF L$
derivation from $\chi_j$, $j<m$, as additional axioms. If $\pi$ consists of formulas only, then so does~$\Pi$, i.e., it
is an $\Fr L$ derivation.

Let $v\colon\Form\to\two$ be a Boolean propositional assignment to modal formulas such that $v(p_i)$ is chosen arbitrarily
for each variable~$p_i$, and
\[v(\Box\fii)=1\iff\fii\in\Pi.\]
We claim that
\begin{equation}\label{eq:6}
v(\theta_i)=1
\end{equation}
for all $i\le z$, which we prove by induction on~$i$. If $\theta_i$ is inferred by an axiom or rule of~$\CPC$,
\eqref{eq:6} follows from $v$ being a Boolean assignment. If $\theta_i$ is an instance of \eqref{eq:8}
or~\eqref{eq:4}, then \eqref{eq:6} follows from the closure of $\Pi$ under \eqref{eq:90} or \eqref{eq:91} (respectively).

Assume that $\theta_i$ is
\begin{equation}\label{eq:50}
\Box\beta'_j\to\bigl(\Box(\Box\alpha'_j\to\alpha'_j)\to\Box\alpha'_j\bigr),
\end{equation}
where $j<k$, and $\alpha'_j=\sigma(\alpha_j)$, $\beta'_j=\sigma(\beta_j)$ for some substitution~$\sigma$. If
$v(\Box\beta'_j)=1$ and $v\bigl(\Box(\Box\alpha'_j\to\alpha'_j)\bigr)=1$, then $\beta'_j$ and
$\Box\alpha'_j\to\alpha'_j$ are in~$\Pi$. By closure under~\eqref{eq:91}, $\Pi$ also contains $\Box\beta'_j$ and
$\Box(\Box\alpha'_j\to\alpha'_j)$, thus in view of $\theta_i\in\Pi$, closure under~\eqref{eq:90} gives
$\Box\alpha'_j\in\Pi$, hence (using $\Box\alpha'_j\to\alpha'_j\in\Pi$) also $\alpha'_j\in\Pi$. Thus,
$v(\Box\alpha'_j)=1$.

Taking $i=z$ in~\eqref{eq:6}, $v(\Box\chi_j)=1$ for each~$j$ implies $v\bigl(\LOR_{i<n}\Box\fii_i\bigr)=1$, i.e., there
exists $i<n$ such that $\fii_i\in\Pi$. Thus, $\Pi$ is an $\CF L$ derivation of $\fii_i$ from~$\{\chi_j:j<m\}$, and
we can turn it into an $\CF L$ proof of $\ET_{j<m}\boxdot\chi_j\to\fii_i$ by Lemma~\ref{lem:feas-ded}.

Now, assume $\RI=\R$. By Theorem~\ref{thm:ext-char}, we may assume $L$ is axiomatized over $\kiv$ by substitution instances
of axioms
\begin{align}
\label{eq:9}&\beta_j\land\Box\alpha_j\to\alpha_j,&&j<k,\\
\label{eq:10}
&\Box\gamma_j\to\bigl(\Box(\Box\alpha_j\to\beta_j)\to\bigl(\Box(\Box\beta_j\to\alpha_j)\to\bigl(\Box(\alpha_j\lor\beta_j)\to\Box\alpha_j\bigr)\bigr)\bigr),&&j<l.
\end{align}
Given an $\CF L$ proof $\pi=\p{\theta_0,\dots,\theta_z}$ of
\[\theta_z=\ET_{j<m}(\chi_j\eq\Box\chi_j)\to\LOR_{i<n}\Box\fii_i,\]
define $\Pi$ as above. Again, $\Pi$ is computable in polynomial time, and it is a valid $\CF L$ derivation from
axioms $\chi_j$, $j<m$. We define a Boolean assignment~$v$ such that
\[v(\Box\fii)=1\iff\fii\in\Pi\text{ and }v(\fii)=1.\]
Again, we prove~\eqref{eq:6} by induction on $i\le s$. Axioms and rules of~$\kiv$ are handled as before, and
\eqref{eq:6} holds trivially for instances
\begin{equation}\label{eq:51}
\beta'_j\land\Box\alpha'_j\to\alpha'_j
\end{equation}
of \eqref{eq:9}, as $v(\Box\alpha'_j)=1$ implies
$v(\alpha'_j)=1$ by definition. Assume that $\theta_i$ is
\begin{equation}\label{eq:52}
\Box\gamma'_j\to\bigl(\Box(\Box\alpha'_j\to\beta'_j)\to\bigl(\Box(\Box\beta'_j\to\alpha'_j)\to\bigl(\Box(\alpha'_j\lor\beta'_j)\to\Box\alpha'_j\bigr)\bigr)\bigr),
\end{equation}
where $j<l$, $\alpha'_j=\sigma(\alpha_j)$, $\beta'_j=\sigma(\beta_j)$, and $\gamma'_j=\sigma(\gamma_j)$ for some
substitution~$\sigma$. If $v$ satisfies the four boxed antecedents of~$\theta_i$, the corresponding unboxed circuits are
in~$\Pi$, hence their boxed counterparts as well by closure under~\eqref{eq:91}, hence $\Box\alpha'_j\in\Pi$ by closure
under~\eqref{eq:90}. In view of $\Box\alpha'_j\to\beta'_j\in\Pi$, this gives $\beta'_j\in\Pi$, hence
$\Box\beta'_j\in\Pi$ by~\eqref{eq:91}, hence $\alpha'_j\in\Pi$ using $\Box\beta'_j\to\alpha'_j\in\Pi$. Moreover,
$v(\alpha'_j\lor\beta'_j)=1$. If $v(\alpha'_j)=1$, then $v(\Box\alpha'_j)=1$ and we are done. Otherwise,
$v(\beta'_j)=1$, thus (using $\beta'_j\in\Pi$) $v(\Box\beta'_j)=1$. Since also $v(\Box\beta'_j\to\alpha'_j)=1$, we
obtain $v(\alpha'_j)=1$ and $v(\Box\alpha'_j)=1$ again.

Since $\chi_j\in\Pi$, we have $v(\chi_j\eq\Box\chi_j)=1$ for each $j<m$. Thus, $v(\theta_z)=1$ implies
$v\bigl(\LOR_{i<n}\Box\fii_i)=1$, that is, $\Pi$ is an $\CF L$ derivation of some $\fii_i$ from $\{\chi_j:j<m\}$, and
we can turn it into an $\CF L$ proof of $\ET_{j<m}\boxdot\chi_j\to\fii_i$.
\end{Pf}

We stress that this ``automatic feasibility'' of $\Ext^\RI$ essentially relies on the presence of $\Ext^\RI_n$ for all~$n$.
Indeed, the main part of this paper will be a study of the complexity of $\decp(\Ext^\RI_k,\CF L)$ for logics $L$
involving the $\lgc{BB}_k$ axiom.

\section{Summary of main results}\label{sec:summary-results}

This is a long paper proving a sequence of theorems some of which gradually improve the previous ones, and it is easy
to get lost. For this reason, we provide an overview of the main results, grouping related theorems together, and
omitting some of the more complicated details. The results follow two main threads: first, estimates on the complexity
of the search problems associated with $\DP$ and extension rules for basic logics of bounded branching, and second,
conditional superpolynomial speed-ups of $\SF L$ over $\EF L$ under complexity assumptions.

As for the first thread, the following statement summarizes Theorem~\ref{thm:dec-ext}, part of Theorem~\ref{thm:intern}, and
Theorem~\ref{thm:dec-vis}:
\begin{Thm}
Let $\RI\in\{\I,\R\}$, $L_0$ be a $\RI$-extensible logic, $k\ge t\ge2$, and $L=L_0\oplus\lgc{BB}_k$. Then
$\decp(\Ext^\RI_t,\CF L)$, and therefore $\decp(\DP,\CF L)$, is subsumed by a total $\conp$~search problem.
More precisely, $\decp(\Ext^\RI_t,\CF L)\equiv_s\decp(\rrule_{k,t},\CF\CPC)$ and $\consp(\Ext^\RI_t,\CF
L)\equiv\consp(\rrule_{k,t},\CF\CPC)$. Likewise, $\decp(\VR_t,\CF{\lgc T_k})\equiv_s\decp(\rrule_{k,t},\CF\CPC)$.
\end{Thm}

Here, $\rrule_{k,t}$ is a certain propositional rule introduced in Definition~\ref{def:fi-class}, whose decision problem
reduces to the \emph{interpolation problem} (Lemma~\ref{lem:rkt-itp}). The superintuitionistic Gabbay--de Jongh logics
$\lgc T_k$ and the Visser rules $\VR_t$ are defined in Section~\ref{sec:super-logics}.

Another result in this thread is a form of Hrube\v s-style monotone interpolation in Theorem~\ref{thm:hru-style}, whose
statement is rather technical.

As for the second thread, the following statement summarizes Theorem~\ref{thm:ef-sf-sep},
Corollary~\ref{cor:ef-sf-sep-itp}, and Theorem~\ref{thm:lb-hru-style}
as generalized in Theorem~\ref{thm:sf-ef-posit} (or rather, Example~\ref{exm:dense}), and
Theorems \ref{thm:ef-sf-sep-int} and~\ref{thm:lb-hru-int}.
\begin{Thm}
If $\kiv\sset L\sset\lgc{S4.2GrzBB_2}$, $\kiv\sset L\sset\lgc{GL.2BB_2}$, or $\IPC\sset L\sset\lgc T_2+\lgc{KC}$, then
$\SF L$ has superpolynomial speed-up over $\EF L$ unless the following happen:
\begin{itemize}
\item $\psp=\np=\conp$.
\item The disjoint $\np$~pair version of\/ $\decp(\rrule_{2,2},\CF\CPC)$, and consequently the interpolation $\np$~pair
for $\EF\CPC$, are complete disjoint $\psp$ pairs under nonuniform poly-time reductions.
\item For every monotone $\psp$ language~$P$, there exists a sequence of polynomial-size monotone Boolean circuits
$C_n^\forall$, $C_n^\exists$ in variables $\{p_i:i<n\}$ and $\{s_{l,r}:l<m_n,r<3\}$ that satisfy certain
conditions spelled out in Theorem~\ref{thm:lb-hru-style} \brak{for modal $L$} or Theorem~\ref{thm:lb-hru-int} \brak{for
superintuitionistic $L$}.
\end{itemize}
\end{Thm}

\section{Disjunction properties for logics of bounded branching}\label{sec:disj-prop-bb}

In this section, we will start investigating the complexity of the decision problems for DP and extension rules for
basic logics of bounded branching; more precisely, our results will apply to logics of the form $L=L_0\oplus\lgc{BB}_k$
where $L_0$ is a $\I$-extensible or $\R$-extensible logic. We try to apply the same method as in the proof of
Theorem~\ref{thm:feas-ext}, using Boolean assignments constructed from polynomial-size closures of the given proof under
\eqref{eq:90} and some other rules. In order to handle instances of the $\lgc{BB}_k$ axiom, we need to introduce extra
``rules'' that are not really sound, hence we will not get a valid proof in the end; nevertheless, the combinatorics
of these rules leads to a reduction of the decision problem for $\Ext^\RI_t$ to a certain total $\conp$~search problem
(albeit a rather unnatural one). Even though this does not give a polynomial-time algorithm, it still considerably
lowers the trivial $\psp$ upper bound on the complexity of the problem. As a consequence, we will obtain a
superpolynomial speed-up of $\SF L$ over $\EF L$ conditional on $\psp\ne\np$.
\begin{Thm}\label{thm:dec-ext}
Let $\RI\in\{\I,\R\}$, $L_0$ be a $\RI$-extensible logic, $k\ge t\ge2$, and $L=L_0\oplus\lgc{BB}_k$. Then
$\decp(\Ext^\RI_t,\CF L)$, and therefore $\decp(\DP,\CF L)$, is subsumed by a total $\conp$~search problem.
\end{Thm}
\begin{Pf}
Let $\pi=\p{\theta_0,\dots,\theta_z}$ be a given $\CF L$~proof of
\begin{equation}\label{eq:11}
\ET_{v<s}B^\RI(\chi_v)\to\LOR_{u<t}\Box\fii_u,
\end{equation}
we need to find a $u<t$ such that
\begin{equation}\label{eq:12}
\vdash_L\ET_{v<s}\boxdot\chi_v\to\fii_u
\end{equation}
using a total $\conp$~search problem.

We assume that $L_0$ is axiomatized as in the proof of Theorem~\ref{thm:feas-ext}. Let $\{A_l:l<m\}$ be the list of
instances of the $\lgc{BB}_k$ axiom invoked in~$\pi$, where
\begin{equation}\label{eq:13}
A_l=\Box\Bigl[\LOR_{i\le k}\Box\Bigl(\boxdot\psi_{l,i}\to\LOR_{j\ne i}\boxdot\psi_{l,j}\Bigr)
    \to\LOR_{i\le k}\boxdot\psi_{l,i}\Bigr]
\to\LOR_{i\le k}\Box\LOR_{j\ne i}\boxdot\psi_{l,j},\qquad l<m.
\end{equation}
Let $\Xi_\pi$ be a set of auxiliary circuits consisting of
\begin{align}
\label{eq:19}
&\LOR_{i\le k}\Box\LOR_{j\ne i}\boxdot\psi_{l,j}
  \to\LOR_{i\le k}\Box\Bigl(\boxdot\psi_{l,i}\to\LOR_{j\ne i}\boxdot\psi_{l,j}\Bigr),
&&l<m,\\
\label{eq:31}
&\LOR_{j\ne i}\boxdot\psi_{l,j}\to\Bigl(\boxdot\psi_{l,i}\to\LOR_{j\ne i}\boxdot\psi_{l,j}\Bigr),
&&l<m,\:i\le k,\\
\label{eq:7}
&\psi_{l,i'}\to\Box\psi_{l,i'}\to\LOR_{j\ne i}\boxdot\psi_{l,j},
&&l<m,\:i,i'\le k,\:i\ne i'.
\end{align}
Clearly, $\Xi_\pi$ is polynomial-time constructible, and it consists of $\lgc
K$-tautologies.

Let us write $[k+1]=\{0,\dots,k\}$. For any $\sigma\in[k+1]^m$, let $\Pi_\sigma$ be the closure of
\begin{equation}\label{eq:22}
\pi\cup\Xi_\pi\cup\{\chi_v:v<s\}
\end{equation}
under~\eqref{eq:90}, $\Sub(\pi)$-restricted~\eqref{eq:91}, and under the rules
\begin{equation}\label{eq:14}
\LOR_{i\le k}\boxdot\psi_{l,i}\Ru\LOR_{i\ne r}\boxdot\psi_{l,i},\qquad l<m, r=\sigma_l.
\end{equation}
(We stress that we take \eqref{eq:14} only literally, we do not consider its substitution
instances.) Likewise, let $\Pi^\sigma$ denote the closure of \eqref{eq:22}
under~\eqref{eq:90}, $\Sub(\pi)$-restricted~\eqref{eq:91}, and under the rules \eqref{eq:14} for all $l<m$
and~$r\ne\sigma_l$. The sets $\Pi_\sigma$ and $\Pi^\sigma$ are computable in polynomial time given $\pi$ and~$\sigma$.

We consider the following $\conp$-search problem $D(\pi)$: \emph{given an $\CF L$~proof $\pi$ of~\eqref{eq:11}, find
$u<t$ such that $\forall\tau\in[k+1]^m\,\fii_u\in\Pi^\tau$} (with a suitable convention if the input does not have the
right form). We are going to show that $D(\pi)$ is total, and that it subsumes $\decp(\DP,\CF L)$.

As in Theorem~\ref{thm:ext-char}, given $\sigma\in[k+1]^m$, we define a Boolean assignment $v_\sigma$ to modal formulas such
that
\[v_\sigma(\Box\fii)=1\iff\begin{cases}
  \fii\in\Pi_\sigma,&\RI=\I,\\
  \fii\in\Pi_\sigma\et v_\sigma(\fii)=1,&\RI=\R.
\end{cases}\]
\pagebreak[2]
\begin{Cl}\label{cl:val}
For all $g\le z$, $v_\sigma(\theta_g)=1$.
\end{Cl}
\begin{Pf*}
By induction on~$g$. Since $\Pi_\sigma$ is closed under~\eqref{eq:90} and $\Sub(\pi)$-restricted~\eqref{eq:91}, the
proof of Theorem~\ref{thm:ext-char} shows that the claim holds if $\theta_g$ was derived by an axiom or rule of~$L_0$. Thus,
we only need to prove $v_\sigma(A_l)=1$ for all~$l<m$. Assume that
\begin{equation}\label{eq:16}
v_\sigma\biggl(\Box\Bigl[\LOR_{i\le k}\Box\Bigl(\boxdot\psi_{l,i}\to\LOR_{j\ne i}\boxdot\psi_{l,j}\Bigr)
    \to\LOR_{i\le k}\boxdot\psi_{l,i}\Bigr]\biggr)=1.
\end{equation}
Then the following circuits are in~$\Pi_\sigma$:
\begin{align}
\label{eq:17}
&\LOR_{i\le k}\Box\Bigl(\boxdot\psi_{l,i}\to\LOR_{j\ne i}\boxdot\psi_{l,j}\Bigr)\to\LOR_{i\le k}\boxdot\psi_{l,i}
&&\text{definition of~$v_\sigma$,}\\
\label{eq:20}
&\Box\Bigl[\LOR_{i\le k}\Box\Bigl(\boxdot\psi_{l,i}\to\LOR_{j\ne i}\boxdot\psi_{l,j}\Bigr)\to\LOR_{i\le k}\boxdot\psi_{l,i}\bigr]
&&\text{\eqref{eq:91},}\\
\label{eq:21}
&\LOR_{i\le k}\Box\LOR_{j\ne i}\boxdot\psi_{l,j}
&&\text{\eqref{eq:90} with~\eqref{eq:13},}\\
\label{eq:28}
&\LOR_{i\le k}\Box\Bigl(\boxdot\psi_{l,i}\to\LOR_{j\ne i}\boxdot\psi_{l,j}\Bigr)
&&\text{\eqref{eq:90} with~\eqref{eq:19},}\\
\label{eq:29}
&\LOR_{i\le k}\boxdot\psi_{l,i}
&&\text{\eqref{eq:90} with~\eqref{eq:17},}\\
\label{eq:30}
&\LOR_{i\ne\sigma_l}\boxdot\psi_{l,i}
&&\text{by \eqref{eq:14}.}
\end{align}
If $\RI=\I$, this means
\[v_\sigma\Bigl(\Box\LOR_{i\ne\sigma_l}\boxdot\psi_{l,i}\Bigr)=1\]
and we are done. If $\RI=\R$, we need more work. We have
\begin{equation}\label{eq:32}
v_\sigma\biggl(\LOR_{i\le k}\Box\Bigl(\boxdot\psi_{l,i}\to\LOR_{j\ne i}\boxdot\psi_{l,j}\Bigr)
    \to\LOR_{i\le k}\boxdot\psi_{l,i}\biggr)=1
\end{equation}
from~\eqref{eq:16}. Notice that
\begin{equation}\label{eq:33}
\boxdot\psi_{l,\sigma_l}\to\LOR_{i\ne\sigma_l}\boxdot\psi_{l,i}\in\Pi_\sigma
\end{equation}
by \eqref{eq:30} and~\eqref{eq:31}. Thus, if
\begin{equation}\label{eq:37}
v_\sigma\Bigl(\boxdot\psi_{l,\sigma_l}\to\LOR_{i\ne\sigma_l}\boxdot\psi_{l,i}\Bigr)=1,
\end{equation}
then $v_\sigma\bigl(\Box\bigl(\boxdot\psi_{l,\sigma_l}\to\LOR_{i\ne\sigma_l}\boxdot\psi_{l,i}\bigr)\bigr)=1$, hence
$v_\sigma\bigl(\LOR_i\boxdot\psi_{l,i}\bigr)=1$ by~\eqref{eq:32}, and
\[v_\sigma\Bigl(\LOR_{i\ne\sigma_l}\boxdot\psi_{l,i}\Bigr)=v_\sigma\Bigl(\Box\LOR_{i\ne\sigma_l}\boxdot\psi_{l,i}\Bigr)=1\]
using \eqref{eq:37} and~\eqref{eq:30}.

On the other hand, if $v_\sigma\bigl(\boxdot\psi_{l,\sigma_l}\to\LOR_{i\ne\sigma_l}\boxdot\psi_{l,i}\bigr)=0$, then
$v_\sigma(\boxdot\psi_{l,\sigma_l})=1$. This implies $\psi_{l,\sigma_l}\in\Pi_\sigma$, hence
$\Box\psi_{l,\sigma_l}\in\Pi_\sigma$ by closure under~\eqref{eq:91}. Using~\eqref{eq:7}, we get $\LOR_{j\ne
i}\boxdot\psi_{l,j}\in\Pi_\sigma$ and
\[v_\sigma\Bigl(\Box\LOR_{j\ne i}\boxdot\psi_{l,j}\Bigr)=1\]
for any fixed $i\ne\sigma_l$.
\end{Pf*}

Since $\chi_v\in\Pi_\sigma$, we have $v_\sigma\bigl(B^\RI(\chi_v)\bigr)=1$ for all~$v<s$. Thus, Claim~\ref{cl:val} for
$\theta_z=\eqref{eq:11}$ implies $v_\sigma(\Box\fii_u)=1$ for some~$u<t$, that is,
\begin{equation}\label{eq:34}
\forall\sigma\in[k+1]^m\:\exists u<t\:\fii_u\in\Pi_\sigma.
\end{equation}
If $\sigma,\tau\in[k+1]^m$, let us write $\sigma\#\tau$ if $\sigma_l\ne\tau_l$ for all $l<m$. We claim that
\begin{equation}\label{eq:35}
\exists u<t\:\forall\tau\in[k+1]^m\:\exists\sigma\in[k+1]^m\:(\sigma\#\tau\et\fii_u\in\Pi_\sigma).
\end{equation}
If not, let us fix for each $u<t$ a counterexample~$\tau^u$. Since $t<k+1$, there exists $\sigma$ such that
$\sigma\#\tau^0,\dots,\tau^{t-1}$, say, $\sigma_l=\min\bigl([k+1]\bez\{\tau^u_l:u<t\}\bigr)$ for each $l<m$. But then
$\fii_0,\dots,\fii_{t-1}\notin\Pi_\sigma$, contradicting~\eqref{eq:34}.

Clearly, $\Pi^\tau\Sset\Pi_\sigma$ for any $\sigma\#\tau$, thus \eqref{eq:35}~implies
\[\exists u<t\:\forall\tau\in[k+1]^m\:\fii_u\in\Pi^\tau,\]
i.e., $D(\pi)$ is total. It remains to verify that a solution to $D(\pi)$ gives a valid solution to
$\decp(\Ext^\RI_t,\CF L)$, i.e.,
\begin{equation}\label{eq:15}
\forall\tau\in[k+1]^m\:\fii_u\in\Pi^\tau\implies{}\vdash_L\ET_{v<s}\boxdot\chi_v\to\fii_u.
\end{equation}
Apart from $\{\chi_v:v<s\}$, the elements of~$\Pi^\tau$ are $L$-tautologies, or they are derived by rules of~$L$ (modus
ponens, necessitation), or by \eqref{eq:14} for $r\ne\tau_l$. Thus, we see by induction on the length of the derivation
that
\[\fii\in\Pi^\tau\implies
   {}\vdash_L\ET_{v<s}\boxdot\chi_v\land\ET_{l<m}\Bigl(\LOR_{i\le k}\boxdot\psi_{l,i}\to\boxdot\psi_{l,\tau_l}\Bigr)
          \to\boxdot\fii.\]
In particular, if $\fii_u\in\Pi^\tau$ for all $\tau\in[k+1]^m$, then
\[\vdash_L\ET_{v<s}\boxdot\chi_v\land\LOR_{\tau\in[k+1]^m}\ET_{l<m}\Bigl(\LOR_{i\le k}\boxdot\psi_{l,i}\to\boxdot\psi_{l,\tau_l}\Bigr)\to\fii_u.\]
However,
\[\LOR_{\tau\in[k+1]^m}\ET_{l<m}\Bigl(\LOR_{i\le k}\boxdot\psi_{l,i}\to\boxdot\psi_{l,\tau_l}\Bigr)\]
is a classical tautology, as it follows from
\[\ET_{l<m}\LOR_{j\le k}\Bigl(\LOR_{i\le k}\boxdot\psi_{l,i}\to\boxdot\psi_{l,j}\Bigr)\]
by distributivity. Thus, we obtain~\eqref{eq:15}.
\end{Pf}

Our main application of the bounds on the complexity of DP are lower bounds, or more precisely separations between $\EF
L$ and~$\SF L$ systems. We will make use of the following translation of quantified Boolean formulas to modal circuits.
\begin{Def}\label{def:qbf-mod}
Given a quantified Boolean formula $\Phi(\vec p)$ in prenex normal form with bound propositional variables~$\vec
q$, we construct a modal circuit $A_\Phi(\vec p,\vec q)$ as follows:
\begin{align*}
A_\Phi&=\Phi\qquad\text{if $\Phi$ is quantifier-free,}\\
A_{\forall q\,\Phi}&=\boxdot q\lor\boxdot\neg q\to A_\Phi,\\
A_{\exists q\,\Phi}&=\Box(\boxdot q\to A_\Phi)\lor\Box(\boxdot\neg q\to A_\Phi).
\end{align*}
(In order to make a polynomial-size circuit, both disjuncts in the definition of~$A_{\exists q\,\Phi}$ use the same
copy of~$A_\Phi$.) Let $\ob\Phi$ denote the prenex normal form of~$\neg\Phi$ obtained by dualizing all quantifiers and
negating the quantifier-free matrix of~$\Phi$.
\end{Def}
\begin{Lem}\label{lem:bool-dec}
Given a Boolean circuit $\fii(p_0,\dots,p_{n-1})$, there are poly-time constructible $\CF{\lgc K}$ proofs of
\begin{equation}\label{eq:26}
\ET_{i<n}(\Box p_i\lor\Box\neg p_i)\to\Box\fii\lor\Box\neg\fii.
\end{equation}
\end{Lem}
\begin{Pf}
By induction on the size of~$\fii$, using instances of the tautologies
\begin{align*}
\Box\fii\lor\Box\neg\fii&\to\Box\neg\fii\lor\Box\neg\neg\fii,\\
(\Box\fii\lor\Box\neg\fii)\land(\Box\psi\lor\Box\neg\psi)&\to\Box(\fii\circ\psi)\lor\Box\neg(\fii\circ\psi)
\end{align*}
for $\circ\in\{\land,\lor,\to\}$, which have linear-size proofs by Observation~\ref{obs:frege-schema}.
\end{Pf}
\begin{Lem}\label{lem:qbf-dec-sf}
Given a QBF $\Phi(p_0,\dots,p_{n-1})$, there are poly-time constructible $\SCF\kiv$ proofs of
\begin{equation}\label{eq:24}
\ET_{i<n}(\Box p_i\lor\Box\neg p_i)\to\Box A_\Phi\lor\Box A_{\ob\Phi}.
\end{equation}
\end{Lem}
\begin{Pf}
By induction on the number of quantifiers. The base case is Lemma~\ref{lem:bool-dec}. For the induction step, we may
assume $\Phi=\exists q\,\Phi_0(q,\vec p)$ by swapping the roles of $\Phi$ and~$\ob\Phi$ if necessary. By the induction
hypothesis, we have a proof of
\[\ET_{i<n}(\Box p_i\lor\Box\neg p_i)\land(\Box q\lor\Box\neg q)\to\Box A_{\Phi_0}(q)\lor\Box A_{\ob\Phi_0}(q)\]
(not showing other variables). Using the substitution rule twice, we obtain
\begin{align*}
\ET_{i<n}(\Box p_i\lor\Box\neg p_i)
&\to\bigl(\Box A_{\Phi_0}(\top)\lor\Box A_{\ob\Phi_0}(\top)\bigr)
        \land\bigl(\Box A_{\Phi_0}(\bot)\lor\Box A_{\ob\Phi_0}(\bot)\bigr)\\
&\to\bigl(\Box A_{\Phi_0}(\top)\lor\Box A_{\Phi_0}(\bot)\bigr)
    \lor\Box\bigl(A_{\ob\Phi_0}(\top)\land A_{\ob\Phi_0}(\bot)\bigr)\\
&\to\bigl(\Box(\boxdot q\to A_{\Phi_0})\lor\Box(\boxdot\neg q\to A_{\Phi_0})\bigr)
    \lor\Box(\boxdot q\lor\boxdot\neg q\to A_{\ob\Phi_0})\\
&\to\Box A_\Phi\lor\Box A_{\ob\Phi}
\end{align*}
with the help of Lemma~\ref{lem:subset-eq}.
\end{Pf}
\begin{Lem}\label{lem:aphi-phi}
Let $\Phi$ be a QBF in free variables~$\vec p$, let $\vec a$ be a Boolean assignment to~$\vec p$, and $\vec p/\vec a$
be the corresponding substitution. If $L$ is a logic with DP, and
\[\vdash_LA_\Phi(\vec p/\vec a),\]
then $\Phi(\vec a)$ is true.
\end{Lem}
\begin{Pf}
By induction on the number of quantifiers in~$\Phi$. If $\Phi$ is quantifier-free, then $A_\Phi(\vec p/\vec a)$ is just
$\Phi(\vec a)$. If $\Phi=\exists q\,\Phi_0(\vec p,q)$, and
\[\vdash_L\Box\bigl(\boxdot q\to A_{\Phi_0}(\vec p/\vec a)\bigr)
    \lor\Box\bigl(\boxdot\neg q\to A_{\Phi_0}(\vec p/\vec a)\bigr),\]
then by DP,
\[\vdash_L\boxdot q\to A_{\Phi_0}(\vec p/\vec a)\quad\text{or}\quad\vdash_L\boxdot\neg q\to A_{\Phi_0}(\vec p/\vec a),\]
hence there exists $b\in\{\bot,\top\}$ such that
\[\vdash_LA_{\Phi_0}(\vec p/\vec a,q/b).\]
By the induction hypothesis, $\Phi_0(\vec a,b)$ is true, hence so is $\Phi(\vec a)$.

If $\Phi=\forall q\,\Phi_0(\vec p,q)$, then $\vdash_L\boxdot q\lor\boxdot\neg q\to A_{\Phi_0}(\vec p/\vec a)$ implies
\[\vdash_LA_{\Phi_0}(\vec p/\vec a,q/\bot)\land A_{\Phi_0}(\vec p/\vec a,q/\top),\]
hence $\Phi_0(\vec a,\bot)$ and $\Phi_0(\vec a,\top)$ are true, hence so is $\Phi(\vec a)$.
\end{Pf}

We come to our basic separation between $\EF{}$ and~$\SF{}$. We use the same tautologies for all logics in question,
and while we apply Theorem~\ref{thm:dec-ext} to get the $\EF{}$ lower bounds, the $\SF{}$ upper bounds hold already for
the base logic~$\kiv$. This implies a separation for all \emph{sublogics} of logics satisfying the
assumptions of Theorem~\ref{thm:dec-ext}, which allows us to formulate the result without explicit reference to
$\RI$-extensible logics~$L_0$: the largest $\I$-extensible logic is $\lgc{GL}$ (being complete w.r.t.\ finite
irreflexive trees), and likewise, the largest $\R$-extensible logic is $\lgc{S4Grz}$. For the same reason, we only need
to refer to the strongest among the $\lgc{BB}_k$ axioms, viz.~$\lgc{BB_2}$.

\begin{Thm}\label{thm:ef-sf-sep}
If $\kiv\sset L\sset\lgc{S4GrzBB_2}$ or $\kiv\sset L\sset\lgc{GLBB_2}$, then $\SF L$ has superpolynomial
speed-up over $\EF L$ unless $\psp=\np=\conp$.

More precisely, if\/ $\psp\ne\np$, there exists a sequence of formulas that have polynomial-time
constructible $\SF\kiv$ proofs, but require proofs of superpolynomial size in $\EF{\lgc{S4GrzBB_2}}$ or
$\EF{\lgc{GLBB_2}}$.
\end{Thm}
\begin{Pf}
We may work with $\CF{}$ and~$\SCF{}$ in place of $\EF{}$ and~$\SF{}$ (respectively), and then it is enough to
construct a sequence of circuits rather than formulas by Lemmas \ref{lem:cf-ef} and~\ref{lem:scf-sf}.

Given a QBF $\Phi$ without free variables, the circuits
\begin{equation}\label{eq:18}
\Box A_\Phi\lor\Box A_{\ob\Phi}
\end{equation}
have polynomial-time constructible $\SCF\kiv$ proofs by Lemma~\ref{lem:qbf-dec-sf}. Assume for
not-quite-a-contradiction that they have $\CF L$ proofs of size $\lh\Phi^c$ for some constant~$c$, where w.l.o.g.\
$L=\lgc{S4GrzBB_2}$ or $L=\lgc{GLBB_2}$ using Corollary~\ref{cor:frege-sim}. By Theorem~\ref{thm:dec-ext}, there are $\conp$ predicates $D_0$
and~$D_1$ such that
\begin{align*}
\text{$\pi$ is an $\CF L$ proof of $\Box A_\Phi\lor\Box A_{\ob\Phi}$}&\implies D_0(\Phi,\pi)\lor D_1(\Phi,\pi),\\
D_1(\Phi,\pi)&\implies{}\vdash_LA_\Phi,\\
D_0(\Phi,\pi)&\implies{}\vdash_LA_{\ob\Phi}.
\end{align*}
Since
\begin{align*}
\vdash_LA_\Phi&\implies\Phi\text{ is true,}\\
\vdash_LA_{\ob\Phi}&\implies\Phi\text{ is false}
\end{align*}
by Lemma~\ref{lem:aphi-phi}, we obtain
\begin{align*}
\Phi\text{ is true}
&\iff\forall\pi\:\bigl(\lh\pi\le\lh\Phi^c\to D_1(\Phi,\pi)\bigr)\\
&\iff\exists\pi\:\bigl(\lh\pi\le\lh\Phi^c\et\neg D_0(\Phi,\pi)\bigr),
\end{align*}
which gives an $\np$ and $\conp$ expression for a $\psp$-complete language.
\end{Pf}

\begin{Rem}\label{rem:ef-sf-sep}
We can improve the speed-up to exponential ($2^{n^{\epsilon}}$) under the stronger hypothesis $\psp\nsset\cxt{NSUBEXP}$.

With some care, we could make sure the formulas had poly-time proofs even in $\SF{\lgc K}$. (Basically, in
Definition~\ref{def:qbf-mod}, we need to replace $\boxdot$ with $\Box^d$ (i.e., $\Box\dots\Box$ with $d$~boxes) where $d$ is the number of quantifiers in~$\Phi$, and
add an extra $\Box$ in front of the definition of~$A_{\forall q\,\Phi}$. We also replace $\Box$ with $\Box^{d+1}$ in the
premise of~\eqref{eq:24}.)
\end{Rem}

\section{The argument internalized}\label{sec:intern}

Theorem~\ref{thm:dec-ext} does not satisfactorily determine the complexity of $\decp(\Ext^\RI_t,\CF L)$: the upper bound it
gives (total $\conp$~search problem) does not come with a matching lower bound, and in fact, the true complexity of the
problem is most probably strictly weaker. The reason for this is that there likely exist no \emph{complete} total
$\conp$~search problems (see Pudl\'ak~\cite{pud:inco} for a detailed discussion of conjectures related to the
nonexistence of complete disjoint $\np$~pairs---recall that disjoint $\np$~pairs can be identified with two-valued
total $\conp$~search problems).

Thus, unlike classes such as $\np$, the class of total $\conp$~search problems forms an (upwards directed) preorder of
problems of ever growing complexity with no maximum, and any particular total $\conp$~search problem has complexity
strictly smaller than the whole class. For this reason, it is desirable to gauge the complexity of
$\decp(\Ext^\RI_t,\CF L)$ more precisely by reducing it to \emph{specific} natural and/or previously studied total
$\conp$ search problems (more informative than the opaque ad hoc problem $D(\pi)$ from the proof of
Theorem~\ref{thm:dec-ext}), and ideally, to prove it equivalent to such a problem.

In this section, we are going to reduce $\decp(\Ext^\RI_t,\CF L)$ to the well known \emph{feasible interpolation}
problem for the \emph{classical} extended Frege system, and in fact, we will show that it is equivalent to its special
case $\decp(\rrule_{k,t},\CF\CPC)$, where $\rrule_{k,t}$ is a certain rule introduced in Definition~\ref{def:fi-class}.
Moreover, the equivalence lifts to the corresponding proof-construction problems. As a consequence, we can improve
Theorem~\ref{thm:ef-sf-sep}: if, for the logics in question, $\SF L$ has no speed-up over $\EF L$, then $\psp$ collapses not
just to $\np$, but to the interpolation disjoint $\np$~pair for $\EF\CPC$ (and even to the corresponding problem
involving $\rrule_{2,2}$), albeit with nonuniform advice.

The argument is based on \emph{internalizing} parts of the proof of Theorem~\ref{thm:dec-ext}: we express some of the
polynomial-time constructions employed in the proof by explicit Boolean circuits, and we derive some of their
properties used in the argument by short $\CF\CPC$ or $\CF L$ proofs. As a bonus, we will obtain additional information
on feasibility of some weaker forms of the $\Ext^\RI_t$ rules (see the statement of Theorem~\ref{thm:intern} for details).

From now on, let us fix $k\ge t\ge2$, $\RI\in\{\I,\R\}$, a $\RI$-extensible logic~$L_0$, and $L=L_0\oplus\lgc{BB}_k$.
Moreover, assume we are given an $\CF L$ proof $\pi=\p{\theta_0,\dots,\theta_z}$ of
\begin{equation}\label{eq:39}
\ET_{v<s}B^\RI(\chi_v)\to\LOR_{u<t}\Box\fii_u,
\end{equation}
and let $\{A_l:l<m\}$ and $\Xi_\pi$ be as in the proof of Theorem~\ref{thm:dec-ext}. Put $S=\Sub(\pi\cup\Xi_\pi)$ and
$N=\lh S$.

We start by describing the sets $\Pi_\sigma$ and~$\Pi^\tau$ from the proof of Theorem~\ref{thm:dec-ext} with (Boolean)
circuits. More generally, if $a$ is any assignment to the propositional variables $\{s_{l,r}:l<m,r\le k\}$ (which we
assume to be distinct from all variables used in~$\pi$), let $\Pi_a\sset S$ be the closure of
$\pi\cup\Xi_\pi\cup\{\chi_v:v<s\}$ under~\eqref{eq:90}, $S$-restricted~\eqref{eq:91}, and the rules \eqref{eq:14} for
$l<m$ and $r\le k$ such that $a(s_{l,r})=1$. We may stratify it by putting $\Pi_{a,0}=\pi\cup\Xi_\pi\cup\{\vec\chi\}$,
and inductively defining $\Pi_{a,h+1}$ as $\Pi_{a,h}$ plus conclusions of all the above-mentioned rules whose premises
are in~$\Pi_{a,h}$. We have $\Pi_{a,N}=\Pi_a$.

In order to describe $\Pi_{a,h}$, we construct monotone Boolean circuits $C_{\fii,h}(\vec s)$ for $\fii\in S$ and
$h\le N+1$ as follows:
\begin{align*}
C_{\fii,0}&=\begin{cases}
    \top,&\fii\in\pi\cup\Xi_\pi\cup\{\chi_v:v<s\},\\
    \bot,&\text{otherwise,}
  \end{cases}\\
C_{\fii,h+1}&=C_{\fii,h}
  \lor\underbrace{\LOR\nolimits_{\!\!\psi}(C_{\psi,h}\land C_{\psi\to\fii,h})}_{\text{for $\psi$ s.t.\ $\psi\to\fii\in S$}}
  \underbrace{\vphantom{\LOR\nolimits_\psi}\lor C_{\psi,h}}_{\text{if $\fii=\Box\psi$}}
  \lor\underbrace{\LOR\nolimits_{\!\!\psi}(C_{\psi,h}\land s_{l,r})}
       _{\substack{\text{for $\psi=\LOR_i\boxdot\psi_{l,i}$}\\\text{s.t.\ $\fii=\LOR_{i\ne r}\boxdot\psi_{l,i}$}}}
.
\end{align*}
Finally, we define $C_\fii=C_{\fii,N}$. It should be clear from the definition that
\begin{align*}
C_{\fii,h}(a)=1&\iff\fii\in\Pi_{a,h},\\
C_\fii(a)=1&\iff\fii\in\Pi_a.
\end{align*}
We need to internally verify two basic properties of $\{\fii:C_\fii=1\}$: that it is closed under the
above-mentioned rules, and that its elements are provable from appropriate hypotheses. These are formalized by the
next two lemmas.
\begin{Lem}\label{lem:closed}
The following have poly-time constructible $\CF\CPC$ proofs.
\begin{align}
\label{eq:40}&C_{\fii,h}\to C_{\fii,h'},&&h<h'\le N+1,\:\fii\in S,\\
\label{eq:41}&\ET_{\fii\in S}(C_{\fii,h+1}\to C_{\fii,h})\to\ET_{\fii\in S}(C_{\fii,h+2}\to C_{\fii,h+1}),&&h<N,\\
\label{eq:42}&C_{\fii,N+1}\to C_{\fii,N},&&\fii\in S,\\
\label{eq:49}&C_\fii,&&\fii\in\pi\cup\Xi_\pi\cup\{\chi_v:v<s\},\\
\label{eq:43}&C_\fii\land C_{\fii\to\psi}\to C_\psi,&&\fii\to\psi\in S,\\
\label{eq:44}&C_\fii\to C_{\Box\fii},&&\Box\fii\in S,\\
\label{eq:45}&s_{l,r}\land C_{\LOR_i\boxdot\psi_{l,i}}\to C_{\LOR_{i\ne r}\boxdot\psi_{l,i}},&&l<m,\:r\le k.
\end{align}
\end{Lem}
\begin{Pf}
\eqref{eq:40} follows by chaining the implications $C_{\fii,h}\to C_{\fii,h+1}$, which are immediate consequences of
the definition.

\eqref{eq:41}: For any $\fii'\in S$, we can prove
\[\ET_{\fii\in S}(C_{\fii,h+1}\to C_{\fii,h})
   \to\Bigl(\LOR\nolimits_{\!\!\psi}(C_{\psi,h+1}\land C_{\psi\to\fii',h+1})
            \to\LOR\nolimits_{\!\!\psi}(C_{\psi,h}\land C_{\psi\to\fii',h})\Bigr),\]
and similarly for the other disjuncts in the definition of~$C_{\fii',h+2}$, hence
\[\ET_{\fii\in S}(C_{\fii,h+1}\to C_{\fii,h})\to(C_{\fii',h+2}\to C_{\fii',h+1}).\]
Combining these for all $\fii'\in S$ gives~\eqref{eq:41}.

\eqref{eq:42}: In view of~\eqref{eq:41}, it suffices to prove
\begin{equation}\label{eq:46}
\LOR_{h\le N}\ET_{\fii\in S}(C_{\fii,h+1}\to C_{\fii,h}).
\end{equation}
Let $\alpha_{h,\fii}=C_{\fii,h+1}\land\neg C_{\fii,h}$. Using~\eqref{eq:40}, we can construct a proof of
\[\ET_{\substack{\fii\in S\\h<h'\le N}}\neg(\alpha_{h,\fii}\land\alpha_{h',\fii}),\]
while obviously
\[\neg\LOR_{h\le N}\ET_{\fii\in S}(C_{\fii,h+1}\to C_{\fii,h})\to\ET_{h\le N}\LOR_{\fii\in S}\alpha_{h,\fii}.\]
Thus, \eqref{eq:46} follows from an instance of $\php^{N+1}_N$, which has short $\CF\CPC$ proofs~\cite{cookrek}.

\eqref{eq:49} follows from~\eqref{eq:40}, as $C_{\fii,0}=\top$ by definition.

\eqref{eq:43}: We derive
\begin{align*}
C_{\fii,N}\land C_{\fii\to\psi,N}&\to C_{\psi,N+1}
&&\text{definition of $C_{\psi,N+1}$,}\\
&\to C_{\psi,N}
&&\text{by~\eqref{eq:42}.}
\end{align*}
The proofs of \eqref{eq:44} and~\eqref{eq:45} are analogous.
\end{Pf}
\begin{Lem}\label{lem:back}
For any $\fii\in S$ and $h\le N$, there are poly-time constructible $\CF L$ proofs of
\begin{equation}\label{eq:54}
\ET_{\substack{l<m\\r\le k}}\Bigl(s_{l,r}\land\boxdot\psi_{l,r}\to\LOR_{i\ne r}\boxdot\psi_{l,i}\Bigr)
           \land C_{\fii,h}(\vec s)\land\ET_{v<s}\boxdot\chi_v
  \to\boxdot\fii.
\end{equation}
\end{Lem}
\begin{Pf}
By induction on~$h$. For $h=0$, the cases $\fii=\chi_v$ are trivial, $\pi$ itself gives a proof of $\fii$ (whence
$\boxdot\fii$) for all $\fii\in\pi$, and it is straightforward to construct short $\CF{\lgc K}$ proofs of
$\fii\in\Xi_\pi$.

For $h+1$, we unwind the definition of $C_{\fii,h+1}$, and use short subproofs of
\begin{align*}
\boxdot\psi\land\boxdot(\psi\to\fii)&\to\boxdot\fii,\\
\boxdot\psi&\to\boxdot\Box\psi,\\
s_{l,r}\land\Bigl(s_{l,r}\land\boxdot\psi_{l,r}\to\LOR_{i\ne r}\boxdot\psi_{l,i}\Bigr)
  \land\boxdot\LOR_{i\le k}\boxdot\psi_{l,i}
&\to\boxdot\LOR_{i\ne r}\boxdot\psi_{l,i},
\end{align*}
where the last one employs $\LOR_{i\ne r}\boxdot\psi_{l,i}\to\Box\LOR_{i\ne r}\boxdot\psi_{l,i}$.
\end{Pf}

We remark that the same proof shows that if $\alpha(p)$ is a formula such that $L$ proves
$\alpha(\top)$, $\alpha(p)\to\alpha(\Box p)$, and $\alpha(p)\land\alpha(p\to q)\to\alpha(q)$, then there are poly-time
constructible $\CF L$ proofs of
\[\ET_{\substack{l<m\\r\le k}}\Bigl[s_{l,r}\land\alpha\Bigl(\LOR_{i\le k}\boxdot\psi_{l,i}\Bigr)
                    \to\alpha\Bigl(\LOR_{i\ne r}\boxdot\psi_{l,i}\Bigr)\Bigr]
  \land C_\fii(\vec s)\land\ET_{v<s}\alpha(\chi_v)
\to\alpha(\fii).\]
However, we do not have a use for this more general statement.

The heart of the argument is to show that $C_{\fii_u}$ holds for some~$u<t$ (under suitable conditions). To this
end, we define Boolean circuits $V_\fii(\vec s)$ for $\fii\in S$, representing the Boolean assignments~$v_\sigma$
from the proof of Theorem~\ref{thm:dec-ext}: we let $V_\fii$ be arbitrary (say, $\top$) if $\fii$ is a variable, and we put
\begin{align*}
V_{c(\fii_0,\dots,\fii_{d-1})}&=c(V_{\fii_0},\dots,V_{\fii_{d-1}}),&&c\in\{\land,\lor,\to,\neg,\top,\bot\},\\
V_{\Box\fii}&=\begin{cases}
    C_\fii,&\RI=\I,\\
    C_\fii\land V_\fii,&\RI=\R.
\end{cases}
\end{align*}
\begin{Lem}\label{lem:sound}
There are poly-time constructible $\CF\CPC$ proofs of
\begin{align}
\label{eq:47}&\ET_{l<m}\LOR_{r\le k}s_{l,r}\to V_{\theta_g},&&g\le z,\\
\label{eq:48}&\ET_{l<m}\LOR_{r\le k}s_{l,r}\to\LOR_{u<t}C_{\fii_u}.
\end{align}
\end{Lem}
\begin{Pf}
\eqref{eq:47}: By induction on~$g$, using the structure of~$\pi$. If $\theta_g$ is derived by~\eqref{eq:90} from
$\theta_h=\theta_i\to\theta_g$ and~$\theta_i$, we have
\[V_{\theta_h}\land V_{\theta_i}\to V_{\theta_g}\]
from the definition of~$V_{\theta_h}$. Likewise, if $\theta_g$ is an instance of an axiom of $\CPC$, then
$V_{\theta_g}$ unwinds to an instance of the same axiom. If $\theta_g=\Box\theta_h$ is derived by~\eqref{eq:91}, we have
\[C_{\theta_h}\land V_{\theta_h}\to V_{\theta_g}\]
by the definition of $V_{\theta_g}$, while $C_{\theta_h}$ is provable by~\eqref{eq:49}. If $\theta_g$ is an instance
of~\eqref{eq:8}, then depending on~$\RI$, $V_{\theta_g}$ is one of
\begin{align*}
C_{\fii\to\psi}&\to(C_\fii\to C_\psi),\\
C_{\fii\to\psi}\land(V_\fii\to V_\psi)&\to(C_\fii\land V_\fii\to C_\psi\land V_\psi),
\end{align*}
which have short proofs using~\eqref{eq:43}. If $\theta_g$ is an instance of~\eqref{eq:4}, $V_{\theta_g}$ is one of
\begin{align*}
C_\fii&\to C_{\Box\fii},\\
C_\fii\land V_\fii&\to C_{\Box\fii}\land C_\fii\land V_\fii,
\end{align*}
which follow from~\eqref{eq:44}. This completes the axioms and rules of~$\kiv$.

If $\RI=\I$ and $\theta_g$ is~\eqref{eq:50}, $V_{\theta_g}$ is
\[C_{\beta'_j}\to(C_{\Box\alpha'_j\to\alpha'_j}\to C_{\alpha'_j}).\]
We can prove
\begin{align*}
C_{\beta'_j}\land C_{\Box\alpha'_j\to\alpha'_j}
&\to C_{\Box\beta'_j}\land C_{\Box(\Box\alpha'_j\to\alpha'_j)}&&\text{by \eqref{eq:44},}\\
&\to C_{\Box\alpha'_j}&&\text{by \eqref{eq:49} for $\theta_g$, and \eqref{eq:43},}\\
&\to C_{\alpha'_j}&&\text{by \eqref{eq:43}.}
\end{align*}

If $\RI=\R$ and $\theta_g$ is~\eqref{eq:51}, $V_{\theta_g}$ is the tautology
\[V_{\beta'_j}\land C_{\alpha'_j}\land V_{\alpha'_j}\to V_{\alpha'_j}.\]
If $\theta_g$ is \eqref{eq:52}, then $V_{\theta_g}$ can be proved by formalizing the relevant part of the proof of
Theorem~\ref{thm:feas-ext}, which we leave to the reader.

The remaining case is $\theta_g=A_l$ for some $l<m$. Let us abbreviate
\begin{align*}
\delta_l&=\LOR_{i\le k}\boxdot\psi_{l,i},\\
\delta_{l,i}&=\LOR_{j\ne i}\boxdot\psi_{l,j},\\
\beta_l&=\LOR_{i\le k}\Box(\boxdot\psi_{l,i}\to\delta_{l,i}),
\end{align*}
so that
\[A_l=\Box(\beta_l\to\delta_l)\to\LOR_{i\le k}\Box\delta_{l,i}.\]
For any $r\le k$, we prove
\begin{align*}
V_{\Box(\beta_l\to\delta_l)}
&\to C_{\beta_l\to\delta_l}
&&\text{by definition,}\\
&\to C_{\Box(\beta_l\to\delta_l)}
&&\text{by \eqref{eq:44},}\\
&\to C_{\LOR_i\Box\delta_{l,i}}
&&\text{by \eqref{eq:49} for $A_l$, and \eqref{eq:43},}\\
&\to C_{\beta_l}
&&\text{by \eqref{eq:49} for \eqref{eq:19}, and \eqref{eq:43},}\\
&\to C_{\delta_l}
&&\text{by \eqref{eq:43},}\\
&\to(s_{l,r}\to C_{\delta_{l,r}})
&&\text{by \eqref{eq:45}.}
\end{align*}
If $\RI=\I$, this gives
\[\LOR_{r\le k}s_{l,r}\land V_{\Box(\beta_l\to\delta_l)}\to\LOR_{r\le k}V_{\Box\delta_{l,r}},\]
thus~\eqref{eq:47}. If $\RI=\R$, we continue with
\begin{align*}
s_{l,r}\land V_{\Box(\beta_l\to\delta_l)}
&\to C_{\boxdot\psi_{l,r}\to\delta_{l,r}}
&&\text{by \eqref{eq:49} for \eqref{eq:31}, and \eqref{eq:43},}\\
&\to\bigl((V_{\boxdot\psi_{l,r}}\to V_{\delta_{l,r}})\to V_{\beta_l}\bigr)
&&\text{definition of $V_{\Box(\boxdot\psi_{l,i}\to\delta_{l,i})}$,}\\
&\to(V_{\beta_l}\to V_{\delta_l})
&&\text{definition of $V_{\Box(\beta_l\to\delta_l)}$,}\\
&\to V_{\boxdot\psi_{l,r}}\lor V_{\delta_{l,r}}
&&\text{using $V_{\delta_l}\to V_{\boxdot\psi_{l,r}}\lor V_{\delta_{l,r}}$,}\\
&\to V_{\boxdot\psi_{l,r}}\lor V_{\Box\delta_{l,r}}
&&\text{definition of $V_{\Box\delta_{l,r}}$.}
\end{align*}
We also have for any fixed $i\ne r$,
\begin{align*}
V_{\boxdot\psi_{l,r}}
&\to C_{\psi_{l,r}}\land V_{\delta_{l,i}}&&\text{definitions,}\\
&\to C_{\Box\psi_{l,r}}&&\text{by \eqref{eq:44},}\\
&\to C_{\delta_{l,i}}&&\text{by \eqref{eq:49} for \eqref{eq:7}, and \eqref{eq:43},}\\
&\to V_{\Box\delta_{l,i}}&&\text{definition,}
\end{align*}
thus
\[s_{l,r}\land V_{\Box(\beta_l\to\delta_l)}\to\LOR_{i\le k}V_{\Box\delta_{l,i}}\]
for all $r\le k$, which implies~\eqref{eq:47}.

\eqref{eq:48}: By applying \eqref{eq:47} to $\theta_z=\eqref{eq:39}$, we obtain
\[\ET_{l<m}\LOR_{r\le k}s_{l,r}\land\ET_{v<s}V_{B^\RI(\chi_v)}\to\LOR_{u<t}V_{\Box\fii_u}.\]
By definition, $V_{\Box\fii_u}$ implies $C_{\fii_u}$, and $V_{B^\RI(\chi_v)}$ is one of the circuits
$C_{\chi_v}$ or $V_{\chi_v}\eq C_{\chi_v}\land V_{\chi_v}$ which follow from~\eqref{eq:49}. Thus, we
obtain~\eqref{eq:48}.
\end{Pf}

As we already stated, we intend to reduce $\decp(\Ext^\RI_t,\CF L)$ to interpolation problems for $\CF\CPC$. We formulate
feasible interpolation in the following way to fit into our framework of multi-conclusion rules. If $P$ is a classical
proof system, the standard interpolation problem for~$P$ (introduced by Pudl\'ak~\cite{pud:np-pair} as a disjoint
$\np$~pair rather than the corresponding search problem) is $\decp(\itp_2,P)$ in our notation.
\begin{Def}\label{def:fi-class}
For classical logic, the $t$-ary \emph{interpolation} multi-conclusion rule is
\[\tag{$\itp_t$}\LOR_{u<t}\fii_u\Ru\fii_0,\dots,\fii_{t-1},\]
where $\fii_u$, $u<t$, are formulas using pairwise disjoint sets of variables.

For any constants $k\ge t\ge2$, we introduce the rule
\[\tag{$\rrule_{k,t}$}\frac{\displaystyle\ET_{l<n}\LOR_{i\le k}p_{l,i}\to\LOR_{u<t}\fii_u}
   {\displaystyle\ET_{\substack{l<n\\\clap{i<j\le k}}}(p_{l,i}\lor p_{l,j})\to\fii_0,\dots,
    \ET_{\substack{l<n\\\clap{i<j\le k}}}(p_{l,i}\lor p_{l,j})\to\fii_{t-1}},\]
where $\fii_u$ are monotone formulas or circuits in the (pairwise distinct) variables $p_{l,i}$ ($l<n$, $i\le k$).
\end{Def}

It is well known that $\itp_t$ is admissible in~$\CPC$ (if no $\fii_u$ is a tautology, we can combine assignments
refuting each $\fii_u$ to an assignment refuting $\LOR_u\fii_u$, using the disjointness of their sets of variables). It
is also easy to see that for proof systems~$P$ dealing with circuits such as $\CF\CPC$, we may allow $\fii_u$ to be
circuits without changing the complexity of $\decp(\itp_t,P)$, as we can choose disjoint sets of extension variables
for each $\fii_u$ to express them as formulas.
\begin{Lem}\label{lem:rkt-itp}
For any $k\ge t\ge2$, the rules $\rrule_{k,t}$ are admissible in~$\CPC$. Moreover, if $P=\CF\CPC$, then
$\decp(\rrule_{k,t},P)\le_s\decp(\itp_t,P)$ and $\consp(\rrule_{k,t},P)\le\consp(\itp_t,P)$.
\end{Lem}
\begin{Pf}
It is enough to prove the latter. Assume we are given a $P$-proof of
\begin{equation}\label{eq:27}
\ET_{l<n}\LOR_{i\le k}p_{l,i}\to\LOR_{u<t}\fii_u(\vec p)
\end{equation}
where the $\fii_u$ are monotone. Using $t$ copies $\{p_{l,i}^u:u<t\}$ of each original $p_{l,i}$ variable, it suffices to
construct a $P$-proof of
\[\LOR_{u<t}\Bigl(\ET_{l<n}\ET_{i<j\le k}(p^u_{l,i}\lor p^u_{l,j})\to\fii_u(\vec p^u)\Bigr).\]
Since this is clearly implied by $\LOR_{u<t}\neg\ET_l\ET_{i<j}(p^u_{l,i}\lor p^u_{l,j})$, it is enough to prove
\begin{equation}\label{eq:38}
\ET_{u<t}\ET_{l<n}\ET_{i<j\le k}(p^u_{l,i}\lor p^u_{l,j})\to\LOR_{u<t}\fii_u(\vec p^u).
\end{equation}
Now, using $n$ instances of the constant-size tautology
\[\ET_{u<t}\ET_{i<j\le k}(q_i^u\lor q_j^u)\to\LOR_{i\le k}\ET_{u<t}q_i^u\]
(a form of $\php^{k+1}_t$), we can construct a proof of
\[\ET_{l<n}\ET_{u<t}\ET_{i<j\le k}(p^u_{l,i}\lor p^u_{l,j})\to\ET_{l<n}\LOR_{i\le k}\ET_{u<t}p_{l,i}^u,\]
hence also
\begin{align*}
\ET_{l<n}\ET_{u<t}\ET_{i<j\le k}(p^u_{l,i}\lor p^u_{l,j})
&\to\LOR_{u<t}\fii_u\Bigl(\dots,\ET_{v<t}p_{l,i}^v,\dots\Bigr)\\
&\to\LOR_{u<t}\fii_u(\vec p^u)
\end{align*}
using a substitution instance of~\eqref{eq:27} and Lemma~\ref{lem:mon}. This establishes~\eqref{eq:38}.
\end{Pf}
\begin{Rem}\label{rem:canon}
For $P=\CF\CPC$ (or equivalently, $P=\EF\CPC$), the interpolation $\np$~pair is equivalent to the \emph{canonical} pair
$\p{\mathit{SAT}^*,\mathit{REF}(P)}$ of Razborov~\cite{razb:pair} by a folklore argument using the fact that $P$ has
polynomial-time constructible proofs of its own reflection principle.
\end{Rem}

\begin{Lem}\label{lem:rkt-hardness}
Under our running assumptions, $\decp(\rrule_{k,t},\CF\CPC)\le_s\decp(\Ext^\RI_t,\CF L)$ and
$\consp(\rrule_{k,t},\CF\CPC)\le\consp(\Ext^\RI_t,\CF L)$.
\end{Lem}
\begin{Pf}
Assume we are given a $\CF\CPC$ proof of
\begin{equation}\label{eq:25}
\ET_{l<n}\LOR_{i\le k}p_{l,i}\to\LOR_{u<t}\fii_u,
\end{equation}
where $\fii_u$ are monotone circuits. For each $l<n$ and $i\le k$, put
\begin{align*}
\beta_{l,i}&=\boxdot q_{l,i}\to\LOR_{j\ne i}\boxdot q_{l,j},\\
\alpha_l&=\LOR_{i\le k}\Box\beta_{l,i}\to\LOR_{i\le k}\boxdot q_{l,i}.
\end{align*}
We can construct for each $l<n$ short $\CF L$ proofs of
\begin{align*}
B^\RI(\alpha_l)&\to\Box\alpha_l\lor\neg\alpha_l&&\text{from definition,}\\
&\to\Box\alpha_l\lor\LOR_{i\le k}\Box\beta_{l,i}\\
&\to\LOR_{i\le k}\Box\LOR_{j\ne i}\boxdot q_{l,j}\lor\LOR_{i\le k}\Box\beta_{l,i}
&&\text{by $\lgc{BB}_k$,}\\
&\to\LOR_{i\le k}\Box\beta_{l,i},
\end{align*}
hence of
\begin{align*}
\ET_{l<n}B^\RI(\alpha_l)&\to\ET_{l<n}\LOR_{i\le k}\Box\beta_{l,i}\\
&\to\LOR_{u<t}\fii_u(\dots,\Box\beta_{l,i},\dots)
&&\text{substitution instance of \eqref{eq:25},}\\
&\to\LOR_{u<t}\Box\fii_u(\dots,\beta_{l,i},\dots)
&&\text{Lemma~\ref{lem:mon-box}.}
\end{align*}
This is our reduction to $\decp(\Ext_t^\RI,\CF L)$. We need to show that if $u<t$ is such that $L$ proves
\begin{equation}\label{eq:60}
\ET_{l<n}\boxdot\alpha_l\to\fii_u(\dots,\beta_{l,i},\dots),
\end{equation}
then $\CPC$ proves
\begin{equation}\label{eq:61}
\ET_{\substack{l<n\\\clap{i<j\le k}}}(p_{l,i}\lor p_{l,j})\to\fii_u,
\end{equation}
and that given an $\CF L$ proof of~\eqref{eq:60}, we can construct a $\CF\CPC$ proof of~\eqref{eq:61}.

Using short $\CF L$ proofs of
\begin{align*}
\LOR_{i\le k}\boxdot q_{l,i}&\to\boxdot\alpha_l,\\
\LOR_{i\le k}\boxdot q_{l,i}&\to\Bigl(\beta_{l,i}\to\LOR_{j\ne i}\boxdot q_{l,j}\Bigr),
\end{align*}
and Lemma~\ref{lem:mon}, \eqref{eq:60} yields an $\CF L$ proof of
\[\ET_{l<n}\LOR_{i\le k}\boxdot q_{l,i}\to\fii_u\Bigl(\dots,\LOR_{j\ne i}\boxdot q_{l,j},\dots\Bigr).\]
By Lemma~\ref{lem:one-point}, we can construct a $\CF\CPC$ proof of
\[\ET_{l<n}\LOR_{i\le k}q_{l,i}\to\fii_u\Bigl(\dots,\LOR_{j\ne i}q_{l,j},\dots\Bigr).\]
We now substitute $\ET_{j\ne i}p_{l,j}$ for~$q_{l,i}$ in the proof. Using short proofs of
\begin{align*}
\ET_{i<j\le k}(p_{l,i}\lor p_{l,j})&\to\LOR_{i\le k}\ET_{j\ne i}p_{l,j},\\
\LOR_{j\ne i}\ET_{r\ne j}p_{l,r}&\to p_{l,i},
\end{align*}
and Lemma~\ref{lem:mon}, we obtain a $\CF\CPC$ proof of~\eqref{eq:61}.
\end{Pf}

We can now put everything together.
\begin{Thm}\label{thm:intern}
Let $\RI\in\{\I,\R\}$, $L_0$ be a $\RI$-extensible logic, $k\ge t\ge2$, and $L=L_0\oplus\lgc{BB}_k$.
\begin{enumerate}
\item\label{item:4}
$\decp(\Ext^\RI_t,\CF L)\equiv_s\decp(\rrule_{k,t},\CF\CPC)$, $\consp(\Ext^\RI_t,\CF L)\equiv\consp(\rrule_{k,t},\CF\CPC)$.
\item\label{item:5}
Given an $\CF L$ proof of
\begin{equation}\label{eq:59}
\ET_{v<s}B^\RI(\chi_v)\to\LOR_{u<t}\Box\fii_u
\end{equation}
using variables $\{p_i:i<n\}$, we can construct in polynomial time an $\CF L$ proof of
\begin{equation}\label{eq:57}
\LOR_{u<t}\sigma^u\Bigl(\ET_{v<s}\boxdot\chi_v\to\fii_u\Bigr),
\end{equation}
where we choose pairwise distinct variables $\{p_i^u:u<t,i<n\}$, and define $\sigma^u$ as the substitution such
that $\sigma^u(p_i)=p_i^u$ for each $i<n$.
\item\label{item:6}
$\consp(\Ext^\RI_1,\CF L)\in\fp$.
\end{enumerate}
\end{Thm}
\begin{Pf}
\ref{item:4}: The right-to-left reductions were given in Lemma~\ref{lem:rkt-hardness}. For the left-to-right directions,
assume we are given an $\CF L$ proof of $\eqref{eq:59}=\eqref{eq:39}$. By Lemma~\ref{lem:sound}, we can construct in
polynomial time a $\CF\CPC$ proof of~\eqref{eq:48}. We claim that this gives the desired reduction to
$\decp(\rrule_{k,t},\CF\CPC)$: that is, if $u<t$ is such that
\begin{equation}\label{eq:53}
\ET_{\substack{l<m\\i<j\le k}}(s_{l,i}\lor s_{l,j})\to C_{\fii_u}
\end{equation}
is a classical tautology, then $L$ proves
\begin{equation}\label{eq:55}
\ET_{v<s}\boxdot\chi_v\to\fii_u,
\end{equation}
and moreover, given a $\CF\CPC$ proof of~\eqref{eq:53}, we can construct in polynomial time an $\CF L$ proof
of~\eqref{eq:55}.

To see this, let $\sigma$ be the substitution such that
$\sigma(s_{l,r})=\boxdot\psi_{l,r}\to\LOR_{i\ne r}\boxdot\psi_{l,i}$ for each $l<m$ and~$r\le k$. Applying $\sigma$ to
Lemma~\ref{lem:back}, we can construct in polynomial time an $\CF L$ proof of
\begin{equation}\label{eq:56}
\sigma(C_{\fii_u})\land\ET_{v<s}\boxdot\chi_v\to\boxdot\fii_u.
\end{equation}
We can also easily construct a proof of the tautology
\begin{equation}\label{eq:58}
\ET_{\substack{l<m\\i<j\le k}}\sigma(s_{l,i}\lor s_{l,j}),
\end{equation}
hence by applying $\sigma$ to a proof of~\eqref{eq:53}, we obtain an $\CF L$ proof of~$\sigma(C_{\fii_u})$, which
together with~\eqref{eq:56} yields~\eqref{eq:55}.

\ref{item:5}: Again, we can construct in polynomial time a $\CF\CPC$ proof of~\eqref{eq:48}. By the argument in
Lemma~\ref{lem:rkt-itp}, we can construct a $\CF\CPC$ proof of
\[\LOR_{u<t}\Bigl(\ET_{l<m}\ET_{i<j\le k}(s_{l,i}^u\lor s_{l,j}^u)\to C_{\fii_u}(\vec s^u)\Bigr).\]
Applying the substitution~$\sigma'$ such that $\sigma'(s_{l,r}^u)=\sigma^u(\sigma(s_{l,r}))$ gives
\[\LOR_{u<t}\sigma^u(\sigma(C_{\fii_u})),\]
using short proofs of $\sigma^u\eqref{eq:58}$. Using Lemma~\ref{lem:back} as above, we construct for each $u<t$ an $\CF L$
proof of
\[\sigma^u(\sigma(C_{\fii_u}))\to\sigma^u\Bigl(\ET_{v<s}\boxdot\chi_v\to\boxdot\fii_u\Bigr).\]
This yields~\eqref{eq:57}.

\ref{item:6} follows from~\ref{item:5}, either by noting that the proof above directly works also for $t=1$, or
formally by putting $\fii_1=\fii_0$, applying \ref{item:5} with $t=2$, and substituting $p_i$ back for $p_i^0$ and~$p_i^1$.
\end{Pf}
\begin{Rem}\label{rem:dp-gt-k}
Theorems \ref{thm:dec-ext} and~\ref{thm:intern} put bounds on the complexity of $\decp(\DP_t,\CF L)$ for $t\le k$. The rules $\DP_t$
are in fact $L$-admissible for all~$t$, and we can derive them by iterating $\DP_2$ (or $\DP_k$). Nevertheless, we do
not directly get any nontrivial bounds on the complexity of $\decp(\DP_t,\CF L)$ for $t>k$: in particular, we cannot
simply iterate Theorem~\ref{thm:intern}, as we will not have an $\CF L$ proof at hand for the second iteration.

We could in principle iterate $\consp(\DP_2,\CF L)$, but this would only work in the unlikely case that it is
polynomially bounded. That is, if $\EF\CPC$ has constructive feasible interpolation, then $\consp(\DP_t,\CF L)\in\fp$
for all~$t$; more generally, if $\consp(\rrule_{k,2},\CF\CPC)$ is polynomially bounded, then $\consp(\DP_t,\CF L)$ is
polynomially bounded for each~$t$, and it is poly-time bounded-query Turing reducible to $\consp(\rrule_{k,2},\CF\CPC)$.
\end{Rem}
\begin{Rem}\label{rem:single}
It would be very interesting if we could strengthen \eqref{eq:57} to
\[\LOR_{u<t}\boxdot\Bigl(\ET_{v<s}\boxdot\chi_v\to\fii_u\Bigr)\]
(note that if desired, we could reinsert the $\sigma^u$'s by the form of Theorem~\ref{thm:intern} already proved), or even
better, if we could prove that the following single-conclusion version of the $\Ext^\RI_t$ rule is feasible for
$\CF L$:
\[\tag{$\Ext^{\RI,\lor}_t$}
\Box\omega\lor\Box\Bigl(\ET_{v<s}B^\RI(\chi_v)\to\LOR_{u<t}\Box\fii_u\Bigr)
  \Ru\boxdot\omega\lor\LOR_{u<t}\boxdot\Bigl(\ET_{v<s}\boxdot\chi_v\to\fii_u\Bigr).\]
For one thing, this would imply $\decp(\DP_t,\CF L)\equiv_s\decp(\rrule_{k,t},\CF\CPC)$, but the main significance
of the $\Ext^{\RI,\lor}_t$ rules is that they form a \emph{basis} of schematic single-conclusion admissible rules of~$L$ (see
\cite{ej:modparami}), hence it would follow that all schematic single-conclusion admissible rules of~$L$ are feasible
for $\CF L$. Moreover, if the construction remained polynomial for repeated usage of such rules, we could generalize to
the logics $L=L_0\oplus\lgc{BB}_k$ (the $\EF{}$ version of) the main result of \cite{ej:modfrege}: all extended Frege
systems for~$L$ are equivalent, where we relax the definition of Frege and $\EF{}$ systems such that the consequence
relation defined by the Frege rules extends $\vdash_L$, and generates the same set of tautologies, but may include
non-derivable rules.
\end{Rem}

Back to earth, Theorem~\ref{thm:intern} allows us to improve Theorem~\ref{thm:ef-sf-sep}:
\begin{Cor}\label{cor:ef-sf-sep-itp}
If $\kiv\sset L\sset\lgc{S4GrzBB_2}$ or $\kiv\sset L\sset\lgc{GLBB_2}$, then $\SF L$ has superpolynomial
speed-up over $\EF L$ unless the disjoint-$\np$-pair version of\/ $\decp(\rrule_{2,2},\CF\CPC)$, and consequently the
interpolation $\np$~pair for $\EF\CPC$, are complete disjoint $\psp$ pairs under nonuniform poly-time reductions.
\end{Cor}
\begin{Pf}
It is enough to prove hardness w.r.t.\ complementary $\psp$ pairs, i.e., $\psp$ languages. Any such language
$P\sset\two^*$ can be defined by a poly-time constructible sequence of QBFs $\Phi_n(p_0,\dots,p_{n-1})$. By
Lemma~\ref{lem:qbf-dec-sf}, there are poly-time constructible $\SCF\kiv$ proofs of
\[\ET_{i<n}(\Box p_i\lor\Box\neg p_i)\to\Box A_{\Phi_n}\lor\Box A_{\ob\Phi_n}.\]
Assume that these circuits have polynomial-size $\CF L$ proofs~$\pi_n$, where w.l.o.g.\ $L=\lgc{GLBB_2}$ or
$L=\lgc{S4GrzBB_2}$. Then the following makes a poly-time reduction of $P$ to
$\decp(\DP_2,\CF L)$ with nonuniform advice~$\pi_n$: given $\vec w\in\two^n$, substitute the bits of~$\vec w$ for
the $p_i$~variables in~$\pi_n$, and derive $\Box A_{\Phi_n}(\vec p/\vec w)\lor\Box A_{\ob\Phi_n}(\vec p/\vec w)$; pass
the resulting proof to $\decp(\DP_2,\CF L)$ to find which disjunct is provable, which by Lemma~\ref{lem:aphi-phi} tells us
whether $\vec w\in P$. By Theorem~\ref{thm:intern} and Lemma~\ref{lem:rkt-itp}, $\decp(\DP_2,\CF
L)\le_s\decp(\rrule_{2,2},\CF\CPC)\le_s\decp(\itp_2,\EF\CPC)$.
\end{Pf}
\begin{Rem}\label{rem:ef-sf-sep-itp}
With more care, one can prove the following strengthening of Corollary~\ref{cor:ef-sf-sep-itp} which internalizes circuits
computing the reduction to $\decp(\rrule_{2,2},\CF\CPC)$: if $\EF L$ weakly simulates $\SF L$, then for
every language $P\in\psp$, there exist poly-size circuits $\{C_n^0,C_n^1:n\in\omega\}$ in variables
$\{p_i:i<n\}\cup\{s_{l,r}:l<m_n,r<3\}$ that are monotone in~$\vec s$ such that
\begin{align*}
w\in P&\iff\forall\vec s\:\Bigl(\ET_{l<m_n}\ET_{i<j<3}(s_{l,i}\lor s_{l,j})\to C_n^1(w,\vec s)\Bigr),\\
w\notin P&\iff\forall\vec s\:\Bigl(\ET_{l<m_n}\ET_{i<j<3}(s_{l,i}\lor s_{l,j})\to C_n^0(w,\vec s)\Bigr),
\end{align*}
and there are poly-size $\CF\CPC$ proofs of
\[\ET_{l<m_n}\LOR_{r<3}s_{l,r}\to C_n^0(\vec p,\vec s)\lor C_n^1(\vec p,\vec s).\]
We will prove an even stronger result in the next section.
\end{Rem}

\section{Hrube\v s-style monotone interpolation}\label{sec:hrubes-style}

The original idea of utilizing $\DP$ to prove lower bounds on the proof complexity of nonclassical logics comes from
Buss and Pudl\'ak \cite{buss-pud}: in this setup, feasible $\DP$ serves a role analogous to feasible interpolation for
classical proof systems, and in accordance with that, it implies \emph{conditional} proof-size lower bounds relying on
(unproven) circuit lower bounds. We followed much the same strategy to derive the conditional separations between
$\SF L$ and $\EF L$ from bounds on the complexity of $\decp(\DP,\EF L)$ in Sections \ref{sec:disj-prop-bb}
and~\ref{sec:intern}.

Hrube\v s~\cite{hru:lbmod} discovered another setup where $\DP$ is replaced by a somewhat different admissible rule
(whose feasibility can be proved using similar methods as for $\DP$) which plays a role analogous to \emph{monotone}
feasible interpolation for classical proof systems. This enabled him to prove \emph{unconditional} proof-size lower
bounds, exploiting known exponential lower bounds on monotone circuit size. (The separations between $\EF{}$ and
$\SF{}$ systems for logics of unbounded branching in Je\v r\'abek~\cite{ej:sfef} that make the starting point for this
paper also rely on Hrube\v s's method.)

This suggests that we should try to adapt our arguments from the previous sections to Hrube\v s's setup, with the hope
that it might improve our conditional separations between $\SF L$ and $\EF L$ to weaken the required complexity
assumptions, or even to make them fully unconditional.

We pursue this idea in the present section to see how far it can get us. We can, indeed, easily adapt our method to
Hrube\v s's setup, as we will see shortly in Theorem~\ref{thm:hru-style}. Unfortunately, we do not know how to extract
unconditional lower bounds from the result; while it does furnish an improvement to our conditional lower bounds, the
statement it leads to (Theorem~\ref{thm:lb-hru-style}) is rather complicated, and it is unclear how significant the
improvement really is.

Let $L=L_0\oplus\lgc{BB}_k$ be as in Section~\ref{sec:intern}. We consider $L$-tautologies of the form
\begin{equation}\label{eq:62}
\alpha(\Box\vec p,\vec q)\to\LOR_{u<t}\Box\beta_u(\vec p,\vec r),
\end{equation}
where the indicated lists of variables $\vec p$, $\vec q$, and $\vec r$ are disjoint, $\alpha$ is a Boolean circuit
monotone in the variables $\vec p$, and the $\beta_u$'s are arbitrary modal circuits. (We will actually only use $t=1$
for the modal lower bounds later on.)
\begin{Thm}\label{thm:hru-style}
Given an $\CF L$ proof of~\eqref{eq:62}, we can construct in polynomial time monotone Boolean circuits
$\{C_u(\vec p,\vec s):u<t\}$ using extra variables $\{s_{l,i}:l<m,\,i\le k\}$, a $\CF\CPC$ proof of
\begin{equation}\label{eq:63}
\alpha(\vec p,\vec q)\land\ET_{l<m}\LOR_{r\le k}s_{l,r}\to\LOR_{u<t}C_u(\vec p,\vec s),
\end{equation}
and for each $u<t$, an $\CF L$ proof of
\begin{equation}\label{eq:64}
\ET_{\substack{l<m\\r\le k}}\Bigl(s_{l,r}\land\boxdot\psi_{l,r}\to\LOR_{i\ne r}\boxdot\psi_{l,i}\Bigr)
           \land\ET_i(p_i\to\Box p_i)\land C_u(\vec p,\vec s)
  \to\boxdot\beta_u(\vec p,\vec r)
\end{equation}
for some circuits $\{\psi_{l,i}:l<m,\,i\le k\}$.
\end{Thm}
\begin{Pf}
We fix an $\CF L$ proof $\pi$ of~\eqref{eq:62}, and we modify the argument given in Section~\ref{sec:intern} as follows.
First, the monotone circuits $C_\fii$ and~$C_{\fii,h}$ will use both $\vec s$ and~$\vec p$ variables; we change the definition of
the base case to
\[C_{\fii,0}=\begin{cases}
    \top,&\fii\in\pi\cup\Xi_\pi,\\
    p_i,&\fii=p_i\text{ for some $i$,}\\
    \bot,&\text{otherwise.}
  \end{cases}\]
(Since $L$ is consistent, $p_i\notin\pi$.) We define the circuits $C_u$ from the statement of our theorem
as~$C_{\beta_u}$. Lemma~\ref{lem:closed} holds unchanged, except for an obvious adaptation of~\eqref{eq:49}. It is also
straightforward to prove an analogue of Lemma~\ref{lem:back}, stating that for any $\fii\in S$ and $h\le N$, there are
poly-time constructible $\CF L$ proofs of
\[\ET_{\substack{l<m\\r\le k}}\Bigl(s_{l,r}\land\boxdot\psi_{l,r}\to\LOR_{i\ne r}\boxdot\psi_{l,i}\Bigr)
           \land\ET_i(p_i\to\Box p_i)\land C_{\fii,h}(\vec p,\vec s)
  \to\boxdot\fii.\]
As a special case, this implies~\eqref{eq:64}.

Recall that the definition of~$V_\fii$ was arbitrary in the case of propositional variables. We now fix it more
specifically: we put $V_\fii=\fii$ if $\fii$ is any of the $\vec p$ or~$\vec q$ variables. Since
Lemma~\ref{lem:sound} worked for arbitrary choices of $V_\fii$ for propositional variables, the proof of~\eqref{eq:47}
continues to hold unchanged. Taking $g=z$, we obtain a $\CF\CPC$ proof of
\[\ET_{l<m}\LOR_{r\le k}s_{l,r}\land V_{\alpha(\Box\vec p,\vec q)}\to\LOR_{u<t}V_{\Box\beta_u(\vec p,\vec r)}.\]
Now, by definition, $V_{\Box\beta_u}$ implies $C_{\beta_u}$, i.e., $C_u$, and since $V$ commutes with Boolean
connectives and preserves~$\vec q$, we have
\[V_{\alpha(\Box\vec p,\vec q)}\equiv\alpha(\dots,V_{\Box p_i},\dots,\vec q).\]
Moreover, $V_{\Box p_i}$ is $C_{p_i}$ or $C_{p_i}\land p_i$, and $p_i$ implies $C_{p_i}$ by the definition of
$C_{p_i,0}$, hence there are short proofs of $p_i\to V_{\Box p_i}$. By Lemma~\ref{lem:mon}, we can thus construct short
$\CF\CPC$ proofs of
\[\alpha(\vec p,\vec q)\to\alpha(\dots,V_{\Box p_i},\dots,\vec q).\]
Putting it all together yields~\eqref{eq:63}.
\end{Pf}

We will apply Theorem~\ref{thm:hru-style} with $t=1$. In this case, the circuit~$C_0$ and the stuff around it act as a weird
sort of interpolant between $\alpha(\vec p,\vec q)$ and $\beta_0(\vec p,\vec r)$ that does not depend on the $\vec q$
or~$\vec r$ variables. It is thus easy to see that when trying to use it for lower bounds, the optimal choice
for~$\beta_0$ is the circuit $A_{\exists\vec r\,\alpha(\vec p,\vec r)}(\vec p,\vec r)$. Since we are interested in
separations between $\CF{}$ and~$\SF{}$, let us observe that the resulting tautologies have short $\SF{}$ proofs, at
least for formulas in negation normal form.
\begin{Lem}\label{lem:hru-sf}
Given a monotone Boolean circuit $\alpha(\vec p,\vec p',\vec q,\vec q')$, we can construct in polynomial time a
$\SCF\kiv$ proof of
\begin{equation}\label{eq:65}
\alpha(\Box\vec p,\Box\neg\vec p,\vec q,\neg\vec q)\to\Box A_{\exists\vec r\,\alpha(\vec p,\neg\vec p,\vec r,\neg\vec r)}(\vec p,\vec r).
\end{equation}
\end{Lem}
\begin{Pf}
By induction on $n=\lh{\vec q}$. If $n=0$, \eqref{eq:65} amounts to
$\alpha(\Box\vec p,\Box\neg\vec p)\to\Box\alpha(\vec p,\neg\vec p)$, which is a substitution instance
of~Lemma~\ref{lem:mon-box}. Going from $n$ to~$n+1$, we take the $q$~variable that corresponds to the outermost
existential quantifier, and reconsider it as part of~$\vec p$; then the induction hypothesis gives a proof of
\[\alpha(\Box\vec p,\Box\neg\vec p,\Box q,\Box\neg q,\vec q,\neg\vec q)\to\Box A(\vec p,q,\vec r),\]
where we abbreviate $A=A_{\exists\vec r\,\alpha(\vec p,\neg\vec p,q,\neg q,\vec r,\neg\vec r)}$. Substituting $\top$
and~$\bot$ for~$q$, we obtain proofs of
\begin{align*}
\alpha(\Box\vec p,\Box\neg\vec p,\top,\bot,\vec q,\neg\vec q)&\to\Box A(\vec p,\top,\vec r)\\
&\to\Box\bigl(\boxdot r\to A(\vec p,r,\vec r)\bigr),\\
\alpha(\Box\vec p,\Box\neg\vec p,\bot,\top,\vec q,\neg\vec q)&\to\Box A(\vec p,\bot,\vec r)\\
&\to\Box\bigl(\boxdot\neg r\to A(\vec p,r,\vec r)\bigr)
\end{align*}
using Lemma~\ref{lem:subset-eq}. Since $\alpha$ is Boolean, there is also a short proof of
\[\alpha(\Box\vec p,\Box\neg\vec p,q,\neg q,\vec q,\neg\vec q)
  \to\alpha(\Box\vec p,\Box\neg\vec p,\top,\bot,\vec q,\neg\vec q)
     \lor\alpha(\Box\vec p,\Box\neg\vec p,\bot,\top,\vec q,\neg\vec q),\]
hence we obtain
\begin{align*}
\alpha(\Box\vec p,\Box\neg\vec q,q,\neg q,\vec q,\neg\vec q)
&\to\Box\bigl(\boxdot r\to A(\vec p,r,\vec r)\bigr)\lor\Box\bigl(\boxdot\neg r\to A(\vec p,r,\vec r)\bigr)\\
&\to\Box\bigl[\Box\bigl(\boxdot r\to A(\vec p,r,\vec r)\bigr)\lor\Box\bigl(\boxdot\neg r\to A(\vec p,r,\vec r)\bigr)\bigr],
\end{align*}
where the disjunction inside square brackets is just $A_{\exists r\exists\vec r\,\alpha(\vec p,\neg\vec p,r,\neg r,\vec
r,\neg\vec r)}$.
\end{Pf}

We note that as in Remark~\ref{rem:ef-sf-sep}, slightly modified variants of the tautologies have even short $\SCF{\lgc K}$
proofs.

We come to the final lower bound of this section. The statement of the theorem is somewhat involved as we try to
push the argument as far as possible, but the most important component is the first part stating the existence of
circuits satisfying \eqref{eq:69}--\eqref{eq:72}. In particular, the gap between \eqref{eq:69} and~\eqref{eq:70}
effectively gives a reduction to a certain promise problem (if $w\in P$, then $C^\forall(w,\vec s)$ holds whenever at
least one variable is true in each triple $\{s_{l,0},s_{l,1},s_{l,2}\}$, while if $w\notin P$, $C^\forall(w,\vec s)$
fails under some assignment that makes \emph{two} variables true in each triple), and this does not seem to follow from just
$\psp=\np$.

\begin{Thm}\label{thm:lb-hru-style}
Let $\kiv\sset L\sset\lgc{S4GrzBB_2}$ or $\kiv\sset L\sset\lgc{GLBB_2}$, and assume that $\EF L$ weakly
simulates $\SF L$.

Then for every monotone $\psp$ language~$P$, there exists a sequence of polynomial-size monotone Boolean circuits
$\{C_n^\forall,C_n^\exists:n\in\omega\}$ such that $C_n^\forall$ and~$C_n^\exists$ use variables $\{p_i:i<n\}$ and
$\{s_{l,r}:l<m_n,r<3\}$, and for every $w\in\two^n$, we have
\begin{align}
w\in P
\label{eq:69}&\iff\forall\vec s\:\Bigl(\ET_{l<m_n}\LOR_{r<3}s_{l,r}\to C_n^\forall(w,\vec s)\Bigr)\\
\label{eq:70}&\iff\forall\vec s\:\Bigl(\ET_{l<m_n}\ET_{i<j<3}(s_{l,i}\lor s_{l,j})\to C_n^\forall(w,\vec s)\Bigr)\\
\label{eq:71}&\iff\exists\vec s\:\Bigl(\ET_{l<m_n}\LOR_{r<3}s_{l,r}\land C_n^\exists(w,\neg\vec s)\Bigr)\\
\label{eq:72}&\iff\exists\vec s\:\Bigl(\ET_{l<m_n}\ET_{i<j<3}(s_{l,i}\lor s_{l,j})\land C_n^\exists(w,\neg\vec s)\Bigr).
\end{align}
The circuits
\begin{equation}\label{eq:73}
\ET_{l<m_n}\LOR_{r<3}t_{l,r}\land C_n^\exists(\vec p,\neg\vec t)\land\ET_{l<m_n}\LOR_{r<3}s_{l,r}\to C_n^\forall(\vec p,\vec s)
\end{equation}
have poly-size $\CF\CPC$ proofs. Moreover, if $\{\alpha_n(\vec p,\vec q):n\in\omega\}$ is a sequence of polynomial-size
circuits monotone in~$\vec p$ such that
\begin{equation}\label{eq:66}
w\in P\iff\exists\vec q\:\alpha_n(w,\vec q),
\end{equation}
we can choose $C_n^\forall$ in such a way that there are polynomial-size $\CF\CPC$ proofs of
\begin{equation}\label{eq:67}
\alpha_n(\vec p,\vec q)\land\ET_{l<m_n}\LOR_{r<3}s_{l,r}\to C_n^\forall(\vec p,\vec s),
\end{equation}
and if $\{\beta_n(\vec p,\vec q):n\in\omega\}$ are polynomial-size circuits monotone in~$\vec p$ such that
\begin{equation}\label{eq:74}
w\in P\iff\forall\vec q\:\beta_n(w,\vec q),
\end{equation}
we can choose $C_n^\exists$ such that there are polynomial-size $\CF\CPC$ proofs of
\begin{equation}\label{eq:75}
\ET_{l<m_n}\LOR_{r<3}s_{l,r}\land C_n^\exists(\vec p,\neg\vec s)\to\beta_n(\vec p,\vec q).
\end{equation}

If $P\in\psp$ is not necessarily monotone, the above holds with $C_n^\forall$ and~$C_n^\exists$ monotone in~$\vec s$,
and $\alpha_n$ and~$\beta_n$ arbitrary.
\end{Thm}
\begin{Pf}
Let $P\in\psp$ be monotone. By Theorem~\ref{thm:ef-sf-sep}, $P\in\np$, hence there exists a sequence of poly-size formulas
$\alpha_n(\vec p,\vec q)$ satisfying~\eqref{eq:66}. Since $P$ is monotone, we have
\[w\in P\iff\exists\vec p,\vec q\:\bigl(\vec p\le w\land\alpha_n(\vec p,\vec q)\bigr),\]
hence we can ensure $\alpha_n$ is monotone in~$\vec p$. Let us fix such a sequence~$\alpha_n$, where we also assume
w.l.o.g.\ that $\alpha_n$ is in negation normal form.

By Lemma~\ref{lem:hru-sf} and the assumption, there are poly-size proofs $\CF L$ proofs of
\[\alpha_n(\Box\vec p,\vec q)\to\Box A_{\exists\vec r\,\alpha_n(\vec p,\vec r)}(\vec p,\vec r),\]
where we may assume w.l.o.g.\ that $L=\lgc{S4GrzBB_2}$ or $L=\lgc{GLBB_2}$.
By Theorem~\ref{thm:hru-style}, there exist poly-size monotone circuits $C_n^\forall(\vec p,\vec s)$ such that \eqref{eq:67}
has poly-size $\CF\CPC$ proofs, and
\begin{equation}\label{eq:68}
\ET_{\substack{l<m_n\\r<3}}\Bigl(s_{l,r}\land\boxdot\psi_{l,r}\to\LOR_{i\ne r}\boxdot\psi_{l,i}\Bigr)
           \land\ET_{i<n}(p_i\to\Box p_i)\land C_n^\forall(\vec p,\vec s)
  \to\boxdot A_{\exists\vec r\,\alpha_n(\vec p,\vec r)}(\vec p,\vec r)
\end{equation}
has poly-size $\CF L$ proofs. We claim that this makes
\[\forall\vec s\:\Bigl(\ET_{l<m_n}\ET_{i<j<3}(s_{l,i}\lor s_{l,j})\to C_n^\forall(\vec p,\vec s)\Bigr)
    \to\exists\vec q\:\alpha_n(\vec p,\vec q)\]
a quantified Boolean tautology, which together with~\eqref{eq:67} implies \eqref{eq:69} and~\eqref{eq:70}. Indeed, let
$w\in\two^n$ be such that
\[\forall\vec s\:\Bigl(\ET_{l<m_n}\ET_{i<j<3}(s_{l,i}\lor s_{l,j})\to C_n^\forall(w,\vec s)\Bigr)\]
is true. Substituting the bits of~$w$ as truth constants into~\eqref{eq:68}, we see that
\[\vdash_L
  \ET_{\substack{l<m_n\\r<3}}\Bigl(s_{l,r}\land\boxdot\psi_{l,r}(\vec p/w)\to\LOR_{i\ne r}\boxdot\psi_{l,i}(\vec p/w)\Bigr)
           \land\ET_{l<m_n}\ET_{i<j<3}(s_{l,i}\lor s_{l,j})
  \to\boxdot A_{\exists\vec r\,\alpha_n(\vec p,\vec r)}(w,\vec r).\]
Further substituting $\boxdot\psi_{l,r}(\vec p/w)\to\LOR_{i\ne r}\boxdot\psi_{l,i}(\vec p/w)$ for~$s_{l,r}$, we obtain
\[\vdash_LA_{\exists\vec r\,\alpha_n(\vec p,\vec r)}(w,\vec r),\]
which implies the truth of $\exists\vec q\,\alpha_n(w,\vec q)$ by Lemma~\ref{lem:aphi-phi}.

The dual language $P^\dual=\bigl\{w\in\two^*:(\neg w)\notin P\bigr\}$ is also monotone, hence by the already proved
part, there exist monotone circuits $C_n^{\forall,\dual}$ such that
\begin{align*}
w\in P^\dual
&\iff\forall\vec s\:\Bigl(\ET_{l<m_n}\LOR_{r<3}s_{l,r}\to C_n^{\forall,\dual}(w,\vec s)\Bigr)\\
&\iff\forall\vec s\:\Bigl(\ET_{l<m_n}\ET_{i<j<3}(s_{l,i}\lor s_{l,j})\to C_n^{\forall,\dual}(w,\vec s)\Bigr).
\end{align*}
(The $m_n$ here is a priori different from the one for~$P$, but we can enlarge one of them to make them equal.) Then
\[C_n^\exists(\vec p,\vec s)=\neg C_n^{\forall,\dual}(\neg\vec p,\neg\vec s)\]
is (equivalent to) a monotone circuit, and it satisfies \eqref{eq:71} and~\eqref{eq:72}. Moreover, given \eqref{eq:74},
we can arrange $C_n^\exists$ to satisfy~\eqref{eq:75}; as a special case, we obtain \eqref{eq:73} by taking
\eqref{eq:69} for~\eqref{eq:74}.

In order to prove the last sentence of the theorem, if $P\in\psp$ is not necessarily monotone, it can be still defined
as in~\eqref{eq:66} with $\alpha_n$ poly-size Boolean formulas. Writing $\alpha_n$ in negation normal form, we have
\[w\in P\iff\exists\vec q\:\alpha'_n(w,\neg w,\vec q)\]
for $\alpha'_n(\vec p,\vec p',\vec q)$ monotone in~$\vec p$ and~$\vec p'$. Thus,
\[\p{w,w'}\in P'\iff\exists\vec q\:\alpha'_n(w,w',\vec q)\]
defines a monotone language, hence we can apply the results above to~$P'$, and substitute $\neg\vec p$ back
for~$\vec p'$.
\end{Pf}
\begin{Rem}\label{rem:lb-hru-str}
Since \eqref{eq:73} implies
\[\ET_{l<m_n}\LOR_{r<3}s_{l,r}\to\neg C_n^\exists(\vec p,\neg\vec s)\lor C_n^\forall(\vec p,\vec s),\]
Theorem~\ref{thm:lb-hru-style} further strengthens Corollary~\ref{cor:ef-sf-sep-itp} and Remark~\ref{rem:ef-sf-sep-itp}.
\end{Rem}

\section{Negation-free lower bounds}\label{sec:lower-bounds-separ}

Our results apply to a fairly limited class of logics. This is unavoidable in Theorem~\ref{thm:dec-ext} as the $\Ext^\RI_t$
rules are not admissible in most other extensions of $\lgc{K4BB}_k$ in the first place, but our separations between
$\EF{}$ and~$\SF{}$ may in principle be applicable to a broader class of logics. In this section, we will show how to
generalize them to logics such as $\lgc{S4.2BB_2}$ (which does not even have the disjunction property),
using a reformulation of the tautologies we used for the separations as positive formulas, and a proof-theoretic
analogue of preservation of positive formulas by dense subreductions. A similar approach was used in \cite{ej:sfef} to
generalize separations from logics of depth~$2$ to logics of unbounded branching.

\begin{Def}\label{def:tree}
For any $h\ge0$, let $\bintr_h$ denote the perfect binary tree of height~$h$ (where the tree consisting of a single
node has height~$0$), and let $\bintr_{h,\I}$ ($\bintr_{h,\R}$) denote the irreflexive (reflexive, resp.) Kripke frame
with skeleton~$\bintr_h$. We will number the levels of~$\bintr_h$ bottom-up such that the root is at level~$0$, and
leaves at level~$h$.
\end{Def}
\begin{Lem}\label{lem:refl-irrefl}
Let $L\Sset\kiv$ be a logic such that for every $h\ge0$, there exists a dense subreduction from an $L$-frame to a
Kripke frame with skeleton~$\bintr_h$.

Then there exists $\RI\in\{\I,\R\}$ such that for every $h\ge0$, there exists a dense subreduction from an $L$-frame
to~$\bintr_{h,\RI}$.
\end{Lem}
\begin{Pf}
Since $\bintr_{h',\RI}$ is a generated subframe of~$\bintr_{h,\RI}$ for $h'<h$, it is enough if the conclusion holds for
infinitely many~$h$; thus, by the infinitary pigeonhole principle, it suffices to show that for arbitrarily large~$h$,
there exists a dense subreduction from an $L$-frame to $\bintr_{h,\I}$ or to~$\bintr_{h,\R}$. This in turn follows from
transitivity of dense subreductions and the fact that any Kripke frame~$F$ with skeleton~$\bintr_{(h+1)(g+1)}$
densely subreduces onto $\bintr_{h,\I}$ or~$\bintr_{g,\R}$.

To see this, notice that either $F$ includes~$\bintr_{h,\I}$ as a dense subframe, or for every $x\in F$ of depth $>h$,
there exists a reflexive $y\ge x$ at most $h$~levels above~$x$. In the latter case, we can construct a meet-preserving
embedding $f\colon\bintr_{g,\R}\to F$ by a bottom-up approach: we map the root of~$\bintr_{g,\R}$ to a reflexive point
of~$F$ at level~$\le h$, and if $f(u)=x$ is already defined, $u_0$ and~$u_1$ are the immediate successors of~$u$, and
$x_0$ and~$x_1$ the immediate successors of~$x$, we fix reflexive points $y_0\ge x_0$ and $y_1\ge x_1$ at most $h+1$
levels above~$x$, and we put $f(u_i)=y_i$, $i=0,1$. We extend $f^{-1}$ to a dense subreduction from $F$
to~$\bintr_{g,\R}$ as follows: if $x\in f[\bintr_{g,\R}]\down$, we map $x$ to $\min\{u\in\bintr_{h,\R}:x\le f(u)\}$,
which exists as $f$ is meet-preserving.
\end{Pf}
\begin{Lem}\label{lem:fix-var-free}
Let $\RI\in\{\I,\R\}$, and $L\Sset\kiv$ be a logic such that for every $h\ge0$, there exists a dense subreduction
from an $L$-frame to~$\bintr_{h,\RI}$.

Then for every finite set $\Phi$ of variable-free formulas, there exists $e\colon\Phi\to\two$ such that for every
$h\ge0$, there exists an $L$-frame~$F$ and a dense subreduction $f$ from~$F$ to $\bintr_{h,\RI}$ such that
\begin{equation}\label{eq:78}
F,u\model\ET_{\fii\in\Phi}(\Box\fii)^{e(\fii)}
\end{equation}
for all $u\in\dom(f)$, where we write $\fii^1=\fii$, $\fii^0=\neg\fii$.
\end{Lem}
\begin{Pf}
By induction on~$\lh\Phi$. The base case $\Phi=\nul$ is trivial. Assuming the statement holds for~$\Phi$, we will show
it holds for $\Phi\cup\{\psi\}$; as in Lemma~\ref{lem:refl-irrefl}, it suffices to prove it with reversed order of
quantifiers (for arbitrarily large~$h$, there exists~$e$, etc.).

Let $h\ge0$. By the induction hypothesis, there exist $e\colon\Phi\to\two$, an $L$-frame~$F$, and a dense
subreduction from $F$ to~$T_{2h,\RI}$ satisfying~\eqref{eq:78}. Observe that $\{u\in F:u\model\Box\psi\}$ is an upper
subset of~$F$. Thus, if there exists $v\in\dom(f)$ such that $v\model\Box\psi$ and $f(v)$ is one of the points at
level~$h$ of~$\bintr_{2h,\RI}$, the restriction $g=f\res v\Up$ is a dense subreduction from the $L$-frame $\{v\}\Up$ to
$\{f(v)\}\Up\simeq\bintr_{h,\RI}$ such that, in addition to~\eqref{eq:78}, we have $u\model\Box\psi$ for all
$u\in\dom(g)$. Otherwise, let $T$ be the copy of~$\bintr_{h,\RI}$ consisting of the points of~$\bintr_{2h,\RI}$ at
levels $\le h$; then $g=f\res f^{-1}[T]$ is a dense subreduction from~$F$ to~$\bintr_{h,\RI}$ that
satisfies~\eqref{eq:78} as well as $u\model\neg\Box\psi$ for all $u\in\dom(g)$.
\end{Pf}
\begin{Thm}\label{thm:posit-sim}
Let $\RI\in\{\I,\R\}$, and $L\Sset\kiv$ be a logic such that for every $h\ge0$, there exists a dense subreduction
from an $L$-frame to~$\bintr_{h,\RI}$. Put $\ob L=\lgc{GLBB_2}$ if $\RI=\I$, and $\ob L=\lgc{S4GrzBB_2}$ if
$\RI=\R$. Then $\CF{\ob L}$ weakly simulates $\CF L$ proofs of positive formulas or circuits.
\end{Thm}
\begin{Pf}
If $S$ is a set of circuits and $e\colon S\to\two$, we define a translation $\fii^e$ for circuits $\fii$ such that
$\{\psi:\Box\psi\in\Sub(\fii)\}\sset S$ as follows: $p_i^e=p_i$ for all variables~$p_i$, the translation commutes with Boolean
connectives, and
\[(\Box\fii)^e=\begin{cases}\Box\fii^e,&e(\fii)=1,\\\bot,&e(\fii)=0.\end{cases}\]
In other words, we replace top-most occurrences of subcircuits $\Box\psi$ such that $e(\psi)=0$ with~$\bot$. Notice
that $\lh{\fii^e}\le\lh\fii$.

Assume we are given an $\CF L$ proof $\pi=\p{\theta_0,\dots,\theta_z}$, where $\theta_z$ is positive. Let $\nu$ be the
substitution such that $\nu(p_i)=\top$ for all variables~$p_i$, and put $\Phi=\{\nu(\fii):\Box\fii\in\Sub(\pi)\}$. Let
$e\colon\Phi\to\two$ satisfy the conclusion of Lemma~\ref{lem:fix-var-free}. Notice that $\fii^{e\circ\nu}$ is defined
for all $\fii\in\Sub(\pi)$, where $e\circ\nu$ denotes the composite assignment $(e\circ\nu)(\fii)=e(\nu(\fii))$.

Since $\theta_z$ is positive, $\vdash_L\nu(\fii)$ for all $\fii\in\Sub(\theta_z)$, thus we must have $e(\nu(\fii))=1$
whenever $\Box\fii\in\Sub(\theta_z)$. It follows that $\theta_z^{e\circ\nu}=\theta_z$, hence it suffices to show that
the sequence
\[\theta_0^{e\circ\nu},\dots,\theta_z^{e\circ\nu}\]
can be extended to a polynomially larger $\CF{\ob L}$ proof.

By Corollary~\ref{cor:frege-sim}, we may assume the $\CF L$ system is axiomatized by axioms and rules of~$\CPC$ (which are
trivially preserved by the $(-)^{e\circ\nu}$ translation), \eqref{eq:91}, and a single axiom schema consisting of
substitution instances of a formula~$\alpha$. For~\eqref{eq:91}, notice that $\vdash_L\nu(\theta_i)$, hence
$e(\nu(\theta_i))=1$, i.e., $\theta_i^{e\circ\nu}\ru(\Box\theta_i)^{e\circ\nu}$ is again an instance of~\eqref{eq:91}.

Concerning instances of~$\alpha$, let $X=\{\beta:\Box\beta\in\Sub(\alpha)\}$, and if $\sigma$ is a substitution such
that $\sigma(\alpha)\in\pi$, define $e_\sigma\colon X\to\two$ by $e_\sigma=e\circ\nu\circ\sigma$. Let
$\sigma^{e\circ\nu}$ be the substitution such that $\sigma^{e\circ\nu}(p_i)=\bigl(\sigma(p_i)\bigr)^{e\circ\nu}$.
Unwinding the definition of the translation, we find
\[\bigl(\sigma(\alpha)\bigr)^{e\circ\nu}=\sigma^{e\circ\nu}(\alpha^{e_\sigma}).\]
Since there is only a constant number of choices for~$e_\sigma$, the translations of all instances
of~$\alpha$ in the proof are instances of a constant number of axiom schemata, and as such have linear-size $\CF{\ob
L}$ proofs by Observation~\ref{obs:frege-schema}, as long as these schemata are valid in~$\ob L$. Thus, it remains to show that
\[\vdash_{\ob L}\alpha^{e_\sigma}\]
for all $\sigma$ such that $\sigma(\alpha)\in\pi$.

Let $M=\p{V,{<},v_M}$ be a finite Kripke $\ob L$-model, which we may assume to be a (binary) tree; we will show
$M\model\alpha^{e_\sigma}$. We embed the underlying frame $\p{V,{<}}$ as a dense subframe in~$\bintr_{h,\RI}$ for
some~$h$, in such a way that the root of $\p{V,{<}}$ is the root of $\bintr_{h,\RI}$, and all leaves
of~$\bintr_{h,\RI}$ are outside~$V$, i.e., every point of~$V$ sees an element of $\bintr_{h,\RI}\bez V$. Using
Lemma~\ref{lem:fix-var-free}, let us fix an $L$-frame~$F=\p{W,{<},A}$ and a dense subreduction $f$ from~$F$
to~$\bintr_{h,\RI}$ that satisfies~\eqref{eq:78}. We may assume that $F$ is rooted and its root~$r$ is mapped to the
root of~$\p{V,{<}}$ by~$f$, hence $f^{-1}[V]$ is a lower subset of~$W$. We endow $F$ with an admissible valuation as
follows:
\[F,u\model p_i\iff
\left\{\begin{aligned}
  M,f(u)&\model p_i,&&\text{if $u\in f^{-1}[V]$,}\\
  F,u&\model\nu(\sigma(p_i)),&&\text{otherwise.}
\end{aligned}\right.
\]
Since $W\bez f^{-1}[V]$ is an upper subset of~$W$, we obtain
\begin{equation}\label{eq:79}
F,u\model\fii\iff F,u\model\nu(\sigma(\fii))
\end{equation}
for all $\fii$ and $u\notin f^{-1}[V]$. We claim that
\begin{equation}\label{eq:80}
F,u\model\beta\iff M,f(u)\model\beta^{e_\sigma}
\end{equation}
for all $u\in f^{-1}[V]$ and $\beta\in\Sub(\alpha)$. Since $F\model\alpha$, this implies $M\model\alpha^{e_\sigma}$,
finishing the proof.

We prove~\eqref{eq:80} by induction on the complexity of~$\beta$. It holds for variables by definition, and the
induction steps for Boolean connectives follow immediately as they commute with~$(-)^{e_\sigma}$.

Assume that \eqref{eq:80} holds for~$\beta\in X$, we will prove it for~$\Box\beta$.

If $e_\sigma(\beta)=1$, we have $F,r\model\Box\nu(\sigma(\beta))$ by~\eqref{eq:78}, thus $F,v\model\beta$ for all
$v\notin f^{-1}[V]$ by~\eqref{eq:79}. It follows that for any $u\in f^{-1}[V]$, we have
\begin{align*}
F,u\model\Box\beta
&\iff\forall v>u\:\bigl(v\in f^{-1}[V]\implies F,v\model\beta\bigr)\\
&\iff\forall v>u\:\bigl(v\in f^{-1}[V]\implies M,f(v)\model\beta^{e_\sigma}\bigr)\\
&\iff\forall y>f(u)\:M,y\model\beta^{e_\sigma}\\
&\iff M,f(u)\model(\Box\beta)^{e_\sigma},
\end{align*}
using the induction hypothesis and $f$'s being a subreduction.

If $e_\sigma(\beta)=0$, $(\Box\beta)^{e_\sigma}=\bot$ is false in $f(u)$. On the other hand, there exists $v>u$ such
that $v\in f^{-1}[\bintr_{h,\RI}\bez V]$, and $F,v\nmodel\Box\nu(\sigma(\beta))$ by~\eqref{eq:78}, hence there exists
$w>v$ such that $F,w\nmodel\beta$ by~\eqref{eq:79}. Thus, $F,u\nmodel\Box\beta$.
\end{Pf}

In order to apply Theorem~\ref{thm:posit-sim}, we need a convenient supply of positive tautologies. In fact, there is a
simple general method of converting any tautology to a positive one:
\begin{Def}\label{def:posit-tr}
Given a formula or circuit $\fii(\vec p)$, we define a positive formula or circuit $\fii^+(\vec p,r)$ using a new
variable~$r$ as follows. We first rewrite all negations $\neg\psi$ inside~$\fii$ as $\psi\to\bot$, so that w.l.o.g.\
$\fii$ uses only the connectives $\{\land,\lor,\to,\top,\bot,\Box\}$. Let $\fii'(\vec p,r)$ be the circuit obtained
from~$\fii$ by replacing $\bot$ with~$r$, thus $\fii'$ is positive and $\fii(\vec p)=\fii'(\vec p,\bot)$. Then we put
\[\fii^+(\vec p,r)=\ET_i\boxdot(r\to p_i)\to\fii'(\vec p,r).\]
\end{Def}
\begin{Lem}\label{lem:posit-proof}
Let $L$ be an extension of~$\kiv$ by positive axiom schemata, and $\fii$ a circuit.
\begin{enumerate}
\item\label{item:7}
There is a poly-time constructible $\CF L$ proof of $\sigma(\fii^+)\to\fii$, where $\sigma$ is the substitution
$\sigma(r)=\bot$.
\item\label{item:8}
Given an $\CF L$ or $\SCF L$ proof of~$\fii$, we can construct in polynomial time an $\CF L$ or $\SCF L$ proof
\brak{respectively} of~$\fii^+$.
\end{enumerate}
\end{Lem}
\begin{Pf}
\ref{item:7} is obvious. Observe that $L$ can be axiomatized by \eqref{eq:90}, \eqref{eq:91}, positive
axiom schemata, and the schema $\bot\to\psi$. With this in mind, \ref{item:8} can be shown by virtually the same proof
as \cite[Thm.~3.8]{ej:implic}; we leave the details to the reader.
\end{Pf}
\begin{Thm}\label{thm:sf-ef-posit}
Let $L\Sset\kiv$ be a logic such that for every $h\ge0$, there exists a dense subreduction from an $L$-frame to a
Kripke frame with skeleton~$\bintr_h$.

Then $\SF L$ has superpolynomial speed-up over $\EF L$, unless $\psp=\np=\conp$, and unless the conclusion of
Theorem~\ref{thm:lb-hru-style} holds.
\end{Thm}
\begin{Pf}
Let $\RI\in\{\I,\R\}$ be as in Lemma~\ref{lem:refl-irrefl}, and put $\ob L=\lgc{GLBB_2}$ if $\RI=\I$, and
$\ob L=\lgc{S4GrzBB_2}$ if $\RI=\R$. By the proofs of Theorems \ref{thm:ef-sf-sep} and~\ref{thm:lb-hru-style}, there exists a sequence
of tautologies $\{\fii_n:n<\omega\}$ that have polynomial-size $\SCF\kiv$ proofs, while the conclusion of the
theorem holds if they have polynomial-size $\CF{\ob L}$ proofs. Now, by Lemma \ref{lem:posit-proof}~\ref{item:8}, the
tautologies $\fii_n^+$ also have polynomial-size $\SCF\kiv$ proofs, and if we assume they have
polynomial-size $\CF L$ proofs, then they have polynomial-size $\CF{\ob L}$ proofs by Theorem~\ref{thm:posit-sim}, thus
$\fii_n$ have polynomial-size $\CF{\ob L}$ proofs by Lemma \ref{lem:posit-proof}~\ref{item:7}.
\end{Pf}
\begin{Exm}\label{exm:dense}
Theorem~\ref{thm:sf-ef-posit} applies to all transitive logics included in $\lgc{S4.2GrzBB_2}$ or
in $\lgc{GL.2BB_2}$: indeed, $\bintr_{h,\I}$ with an extra irreflexive point on top is a
$\lgc{GL.2BB_2}$-frame for any~$h$, and similarly in the reflexive case.
\end{Exm}
\begin{Rem}\label{rem:psp-hard}
Logics $L$ satisfying the assumption of Theorem~\ref{thm:sf-ef-posit} are $\psp$-hard by Theorem~\ref{thm:psp-hard}, hence
$\psp\ne\np$ implies superpolynomial lower bounds on all Cook--Reckhow proof systems for~$L$, in particular on $\SF L$.
\end{Rem}

\section{Superintuitionistic logics}\label{sec:super-logics}

Intuitionistic logic ($\IPC$) and its extensions (superintuitionistic logics) behave in many respects analogously to
transitive modal logics; in particular, many interesting properties are preserved by the Blok--Esakia isomorphism
between extensions of~$\IPC$ and extensions of $\lgc{S4Grz}$. We will now indicate how to transfer our results to the
case of superintuitionistic logics. Our basic tool will be an efficient transformation of proofs from
superintuitionistic logics to modal logics by means of the G\"odel--Tarski--McKinsey translation, which reduces the
decision problems associated with $\DP$ and similar rules to the modal case; in this way, we will obtain analogues of
the extension rule complexity estimates from Theorem~\ref{thm:dec-ext} and the first equivalence in Theorem~\ref{thm:intern}, and
of the conditional separations from Theorem~\ref{thm:ef-sf-sep}, Corollary~\ref{cor:ef-sf-sep-itp}, and (a monotone form of)
Remark~\ref{rem:ef-sf-sep-itp}.

There is not much point in formally introducing an intuitionistic analogue of the class of $\RI$-extensible
logics, as the only such logic is~$\IPC$ itself (being complete w.r.t.\ finite trees). The intuitionistic equivalent of
the bounded branching logics are the \emph{Gabbay--de Jongh logics%
\footnote{Introduced as $\lgc D_{k-1}$ in Gabbay and de~Jongh~\cite{gab-dj:bb}. We find the off-by-one error in the
subscript too distressing, hence we follow the notation of~\cite{cha-zax} instead.}%
}~$\lgc T_k$, axiomatized by
\begin{align*}
\lgc T_k
&=\IPC+\ET_{i\le k}\Bigl[\Bigl(\fii_i\to\LOR_{j\ne i}\fii_j\Bigr)\to\LOR_{j\ne i}\fii_i\Bigr]\to\LOR_{i\le k}\fii_j\\
&=\IPC+\Bigl[\LOR_{i\le k}\Bigl(\fii_i\to\LOR_{j\ne i}\fii_j\Bigr)\to\LOR_{i\le k}\fii_i\Bigr]\to\LOR_{i\le k}\fii_j.
\end{align*}
As in Lemma~\ref{lem:bb}, the logic~$\lgc T_k$ is complete w.r.t.\ finite trees (or more general finite intuitionistic
Kripke frames) of branching~$\le k$, and a frame~$F$ validates~$\lgc T_k$ iff there is no dense subreduction from~$F$
to~$\Psi_{k+1}$.

The disjunction property for superintuitionistic logics is defined by $L$-admissibility of the multi-conclusion rules
\[\tag{$\DP_n$}\fii_0\lor\dots\lor\fii_{n-1}\ru\fii_0,\dots,\fii_{n-1}.\]
The intuitionistic analogue of the extension rules are \emph{Visser's rules}
\[\tag{$\VR_n$}\ET_{i<n}(\fii_i\to\psi_i)\to\LOR_{i<n}\fii_i
  \Ru\ET_{i<n}(\fii_i\to\psi_i)\to\fii_0,\dots,\ET_{i<n}(\fii_i\to\psi_i)\to\fii_{n-1}.\]
We mention that similarly to Theorem~\ref{thm:feas-ext}, Visser's rules are constructively feasible for $\CF\IPC$
\cite{min-koj,ej:modfrege} by an argument using an efficient version of Kleene's slash in place of Boolean assignments.

We assume $\IPC$ is formulated in a language using connectives $\{\land,\lor,\to,\bot\}$, with $\neg\fii$ being defined
as $\fii\to\bot$, and $\top$ as~$\neg\bot$. The \emph{G\"odel--McKinsey--Tarski translation} of intuitionistic formulas
(or circuits) to modal formulas (circuits, resp.) is defined such that $\T(p_i)=\Box p_i$ for propositional
variables~$p_i$, $\T$ commutes with the monotone connectives $\land$, $\lor$, and~$\bot$, and
\[\T(\fii\to\psi)=\Box\bigl(\T(\fii)\to\T(\psi)\bigr).\]
A modal logic $L'\Sset\lgc{S4}$ is a \emph{modal companion} of a superintuitionistic logic~$L$ if
\begin{equation}\label{eq:81}
{}\vdash_L\fii\iff{}\vdash_{L'}\T(\fii)
\end{equation}
for all formulas~$\fii$. If $L=\IPC+\{\fii_i:i\in I\}$, then $\tau L=\lgc{S4}\oplus\{\T(\fii_i):i\in I\}$ is the
smallest modal companion of~$L$, while $\sigma L=\tau L\oplus\lgc{Grz}$ is the largest modal companion of~$L$. (See
\cite[\S9.6]{cha-zax} for details.) We have $\tau\lgc T_k=\lgc{S4BB}_k$ and $\sigma\lgc T_k=\lgc{S4GrzBB}_k$.
\begin{Lem}\label{lem:t-box}
Given a formula or circuit~$\fii$, we can construct in polynomial time an $\CF{\lgc{S4}}$ proof of
$\T(\fii)\eq\Box\T(\fii)$.
\end{Lem}
\begin{Pf}
By induction on the complexity of~$\fii$.
\end{Pf}
\begin{Lem}\label{lem:mod-comp}
Let $L'$ be a modal companion of a superintuitionistic logic~$L$. Given an $\CF L$ proof \brak{or $\SCF L$ proof} of~$\fii$,
we can construct in polynomial time an $\CF{L'}$ proof \brak{$\SCF{L'}$ proof, resp.} of\/ $\T(\fii)$.
\end{Lem}
\begin{Pf}
Using Lemma~\ref{lem:t-box}, the $\T$~translation commutes with substitution up to shortly provable
equivalence. This means we can just apply $\T$ to the whole proof line by line, and fix it up to make a valid proof; we
leave the details to the reader.
\end{Pf}

\begin{Lem}\label{lem:tk-paral}
Let $k\ge2$. Given $n$, there are $\poly(n)$-time constructible $\Fr{\lgc T_k}$ proofs of
\[\Bigl[\ET_{l<n}\LOR_{i\le k}\Bigl(q_{l,i}\to\LOR_{j\ne i}q_{l,j}\Bigr)\to\ET_{l<n}\LOR_{i\le k}q_{l,i}\Bigr]
  \to\ET_{l<n}\LOR_{i\le k}q_{l,i}.\]
\end{Lem}
\begin{Pf}
Put $\beta_{l,i}=q_{l,i}\to\LOR_{j\ne i}q_{l,j}$. We prove
\begin{equation}\label{eq:88}
\Bigl(\ET_{l<m}\LOR_{i\le k}\beta_{l,i}\to\ET_{l<n}\LOR_{i\le k}q_{l,i}\Bigr)
  \to\ET_{l<n}\LOR_{i\le k}q_{l,i}
\end{equation}
by induction on $m\le n$. The base case $m=0$ is trivial. Assuming we have a proof of \eqref{eq:88} for~$m$, we derive
it for $m+1$ by
\begin{align*}
\Bigl(\ET_{l\le m}\LOR_{i\le k}\beta_{l,i}\to\ET_{l<n}\LOR_{i\le k}q_{l,i}\Bigr)
&\to\Bigl[\LOR_{i\le k}\beta_{m,i}\to\Bigl(\ET_{l<m}\LOR_{i\le k}\beta_{l,i}\to\ET_{l<n}\LOR_{i\le k}q_{l,i}\Bigr)\Bigr]\\
&\to\Bigl(\LOR_{i\le k}\beta_{m,i}\to\ET_{l<n}\LOR_{i\le k}q_{l,i}\Bigr)\\
&\to\Bigl(\LOR_{i\le k}\beta_{m,i}\to\LOR_{i\le k}q_{m,i}\Bigr)\\
&\to\LOR_{i\le k}q_{m,i}\\
&\to\LOR_{i\le k}\beta_{m,i}\\
&\to\ET_{l<n}\LOR_{i\le k}q_{l,i}
\end{align*}
using an instance of~$\lgc T_k$.
\end{Pf}
\begin{Lem}\label{lem:rkt-hardness-int}
For any $k\ge t\ge2$, $\decp(\rrule_{k,t},\CF\CPC)\le_s\decp(\VR_t,\CF{\lgc T_k})$.
\end{Lem}
\begin{Pf}
Assume we are given a $\CF\CPC$ proof of
\[\ET_{l<n}\LOR_{i\le k}p_{l,i}\to\LOR_{u<t}\fii_u(\vec p),\]
where $\fii_u$ are monotone circuits. We can make it an $\CF\IPC$ proof by \cite[Thm.~3.9]{ej:sfef}, hence we can
construct an $\CF\IPC$ proof of the substitution instance
\begin{equation}\label{eq:89}
\ET_{l<n}\LOR_{i\le k}\beta_{l,i}\to\LOR_{u<t}\fii_u(\dots,\beta_{l,i},\dots),
\end{equation}
where $\beta_{l,i}=q_{l,i}\to\LOR_{j\ne i}q_{l,j}$. Using~\eqref{eq:89} and Lemma~\ref{lem:tk-paral}, we can construct a
$\CF{\lgc T_k}$ proof of
\begin{align*}
\ET_{u<t}\Bigl(\fii_u(\dots,\beta_{l,i},\dots)\to\ET_{l<n}\LOR_{i\le k}q_{l,i}\Bigr)
&\to\Bigl(\ET_{l<n}\LOR_{i\le k}\beta_{l,i}\to\ET_{l<n}\LOR_{i\le k}q_{l,i}\Bigr)\\
&\to\ET_{l<n}\LOR_{i\le k}q_{l,i}\\
&\to\ET_{l<n}\LOR_{i\le k}\beta_{l,i}\\
&\to\LOR_{u<t}\fii_u(\dots,\beta_{l,i},\dots),
\end{align*}
which gives a reduction to~$\decp(\VR_t,\CF{\lgc T_k})$. In order to see that it is sound, if $u<t$ is such that
\[\vdash_{\lgc T_k}\ET_{v<t}\Bigl(\fii_v(\dots,\beta_{l,i},\dots)\to\ET_{l<n}\LOR_{i\le k}q_{l,i}\Bigr)
    \to\fii_u(\dots,\beta_{l,i},\dots),\]
then
\[\vdash_{\lgc T_k}\ET_{l<n}\LOR_{i\le k}q_{l,i}\to\fii_u\Bigl(\dots,\LOR_{j\ne i}q_{l,j},\dots\Bigr).\]
By substituting $\ET_{j\ne i}p_{l,j}$ for $q_{l,i}$, we obtain
\[\vdash_{\lgc T_k}\ET_{l<n}\ET_{i<j\le k}(p_{l,i}\lor p_{l,j})\to\fii_u(\vec p)\]
as in the proof of Lemma~\ref{lem:rkt-hardness}.
\end{Pf}

We note that the same argument also shows $\consp(\rrule_{k,t},\CF\CPC)\le\consp(\VR_t,\CF{\lgc T_k})$. However, we will not
obtain any upper bound on the complexity of $\consp(\VR_t,\CF{\lgc T_k})$.
\begin{Thm}\label{thm:dec-vis}
If $k\ge t\ge2$, then $\decp(\VR_t,\CF{\lgc T_k})$, and therefore $\decp(\DP_t,\CF{\lgc T_k})$, is subsumed by a
total $\conp$~search problem. Specifically, $\decp(\VR_t,\CF{\lgc T_k})\equiv_s\decp(\rrule_{k,t},\CF\CPC)$.
\end{Thm}
\begin{Pf}
In view of Theorems \ref{thm:dec-ext} and~\ref{thm:intern} and Lemma~\ref{lem:rkt-hardness-int}, it suffices to construct a reduction of
$\decp(\VR_t,\CF{\lgc T_k})$ to $\decp(\Ext^\R_t,\CF{\lgc{S4BB}_k})$. Given a $\CF{\lgc T_k}$ proof of
\[\ET_{u<t}(\fii_u\to\psi_u)\to\LOR_{u<t}\fii_u,\]
we can construct an $\CF{\lgc{S4BB}_k}$ proof of
\[\ET_{u<t}\Box\bigl(\Box\T(\fii_u)\to\Box\T(\psi_u)\bigr)\to\LOR_{u<t}\Box\T(\fii_u)\]
by Lemmas \ref{lem:mod-comp} and~\ref{lem:t-box}. Using
\[\bigl[\bigl(\Box\T(\fii_u)\to\Box\T(\psi_u)\bigr)\to\Box\bigl(\Box\T(\fii_u)\to\Box\T(\psi_u)\bigr)\bigr]
   \to\Box\bigl(\Box\T(\fii_u)\to\Box\T(\psi_u)\bigr)\lor\Box\T(\fii_u),\]
we obtain an $\CF{\lgc{S4BB}_k}$ proof of
\[\ET_{u<t}B^\R\bigl(\Box\T(\fii_u)\to\Box\T(\psi_u)\bigr)\to\LOR_{u<t}\Box\T(\fii_u).\]
This is a sound reduction, as
\[{}\vdash_{\lgc{S4BB}_k}\ET_{u<t}\Box\bigl(\Box\T(\fii_u)\to\Box\T(\psi_u)\bigr)\to\T(\fii_v)
   \implies{}\vdash_{\lgc T_k}\ET_{u<t}(\fii_u\to\psi_u)\to\fii_v\]
by \eqref{eq:81} and~Lemma~\ref{lem:t-box}.
\end{Pf}
\begin{Rem}\label{rem:vis-k-m}
The logics~$\lgc T_k$ in fact admit Visser's rules in a more general form
\[\tag{$\VR_{t,m}$}\ET_{i<t}(\fii_i\to\psi_i)\to\LOR_{i<t+m}\fii_i
  \Ru\ET_{i<t}(\fii_i\to\psi_i)\to\fii_0,\dots,\ET_{i<t}(\fii_i\to\psi_i)\to\fii_{t+m-1}\]
for $t\le k$ and all $m\ge0$; it is possible to derive $\VR_{t,m}$ by iteration of $\VR_{t,0}=\VR_t$. However, as in
Remark~\ref{rem:dp-gt-k}, we do not get any nontrivial bounds on the complexity of $\decp(\VR_{t,m},\CF{\lgc T_k})$ for
$t+m>k$.
\end{Rem}
\begin{Rem}\label{rem:intern-int}
We do not know if a full analogue of Theorem~\ref{thm:intern} holds for~$\lgc T_k$. Instead of using translation to modal
logic as in our proof of Theorem~\ref{thm:dec-vis}, it is straightforward to give a self-contained argument with efficient
Kleene's slash taking the role of Boolean assignments as in \cite[4.11--4.13]{ej:modfrege}. This in turn can be
internalized along the lines of Section~\ref{sec:intern}, and we can prove analogues of Lemmas \ref{lem:back} and~\ref{lem:sound}
with no particular difficulty. Unfortunately, this does not seem to lead anywhere, as $\lgc T_k$ does not prove the
crucial tautology~\eqref{eq:58}, i.e.,
\[\ET_{\substack{l<m\\i_0<i_1\le k}}\Bigl[\Bigl(\psi_{l,i_0}\to\LOR_{j\ne i_0}\psi_{l,j}\Bigr)
    \lor\Bigl(\psi_{l,i_1}\to\LOR_{j\ne i_1}\psi_{l,j}\Bigr)\Bigr],\]
just like $\lgc{S4BB}_k$ does not prove the boxed version of~\eqref{eq:58}:
\[\ET_{\substack{l<m\\i_0<i_1\le k}}\Bigl[\Box\Bigl(\Box\psi_{l,i_0}\to\LOR_{j\ne i_0}\Box\psi_{l,j}\Bigr)
    \lor\Box\Bigl(\Box\psi_{l,i_1}\to\LOR_{j\ne i_1}\Box\psi_{l,j}\Bigr)\Bigr].\]
Our inability to circumvent this problem is directly related to our failure to solve Remark~\ref{rem:single}.
\end{Rem}

We now turn to lower bounds. We define the intuitionistic versions~$A_\Phi^I$ of the $A_\Phi$~circuits by
dropping all boxes from Definition~\ref{def:qbf-mod}. It is straightforward to adapt the proofs of
Lemmas \ref{lem:bool-dec}, \ref{lem:qbf-dec-sf}, and~\ref{lem:aphi-phi} (again, by essentially dropping all boxes) to show the following:
\begin{Lem}\label{lem:qbf-dec-sf-int}
Given a QBF $\Phi(p_0,\dots,p_{n-1})$, there are poly-time constructible $\SCF\IPC$ proofs of
\[\ET_{i<n}(p_i\lor\neg p_i)\to A^I_\Phi\lor A^I_{\ob\Phi}.\noproof\]
\end{Lem}
\begin{Lem}\label{lem:aphi-phi-int}
Let $\Phi$ be a QBF in free variables~$\vec p$, let $\vec a$ be a Boolean assignment to~$\vec p$, and $\vec p/\vec a$
denote the corresponding substitution. If $L$ is a superintuitionistic logic with DP, and
\[\vdash_LA^I_\Phi(\vec p/\vec a),\]
then $\Phi(\vec a)$ is true.
\noproof\end{Lem}

As with the notion of extensible logics, in the superintuitionistic case there is not much point in considering a
complicated condition on logics as in Theorem~\ref{thm:sf-ef-posit}: one can check that a superintuitionistic logic $L$ has
the property that for each~$h$ there exists a subreduction from an $L$-frame to~$B_h$ if and only if
$L\sset\lgc T_2+\lgc{KC}$, where $\lgc{KC}$ is the logic of weak excluded middle
\[\lgc{KC}=\IPC+\neg\fii\lor\neg\neg\fii,\]
hence we may as well just directly state the results for sublogics of $\lgc T_2+\lgc{KC}$.

The superintuitionistic analogues of Lemma~\ref{lem:posit-proof} and Theorem~\ref{thm:posit-sim} were already proved in
Je\v r\'abek~\cite{ej:implic}. Given a formula or circuit~$\fii(\vec p)$, let $\fii'(\vec p,r)$ be the positive
circuit obtained by replacing all occurrences of~$\bot$ with~$r$, so that $\fii(\vec p)=\fii'(\vec p,\bot)$. Then we
put $\fii^+(\vec p,r)=\ET_i(r\to p_i)\to\fii'(\vec p,r)$. The following is Theorem~3.8 in~\cite{ej:implic}.
\begin{Lem}\label{lem:posit-proof-int}
Let $L$ be an extension of $\IPC$ by positive axioms, and $\fii$ a circuit.
\begin{enumerate}
\item\label{item:9}
There is a poly-time constructible $\CF\IPC$ proof of $\sigma(\fii^+)\to\fii$, where $\sigma$ is the substitution
$\sigma(r)=\bot$.
\item\label{item:10}
Given an $\CF L$ or $\SCF L$ proof of~$\fii$, we can construct in polynomial time an $\CF L$ or $\SCF L$ proof
\brak{respectively} of $\fii^+$.\noproof
\end{enumerate}
\end{Lem}
The next lemma is a special case of Theorem~4.5 in~\cite{ej:implic}.
\begin{Lem}\label{lem:posit-sim-int}
Given a $\CF{(\lgc T_2+\lgc{KC})}$ proof of a positive formula or circuit~$\fii$, we can construct in polynomial time a
$\CF{\lgc T_2}$ proof of~$\fii$.
\noproof\end{Lem}
\begin{Thm}\label{thm:ef-sf-sep-int}
If\/ $\IPC\sset L\sset\lgc T_2+\lgc{KC}$, then $\SF L$ has superpolynomial speed-up over $\EF L$ unless
$\psp=\np=\conp$, and unless the disjoint $\np$~pair version of\/ $\decp(\rrule_{2,2},\CF\CPC)$ is a complete disjoint
$\psp$ pair under nonuniform poly-time reductions.
\end{Thm}
\begin{Pf}
As before, it suffices to show a conditional separation between $\CF L$ and $\SCF L$ proofs of circuits using
intuitionistic variants of Lemmas \ref{lem:cf-ef} and~\ref{lem:scf-sf}.

For any QBF $\Phi$, the circuits $(A_\Phi^I)^+$ have polynomial-time constructible $\SCF\IPC$ proofs by
Lemmas \ref{lem:qbf-dec-sf-int} and~\ref{lem:posit-proof-int}. Thus, if $\CF L$ weakly simulates $\SCF L$, then the circuits
$A^I_\Phi$ have polynomial-size $\CF{\lgc T_2}$ proofs~$\pi_\Phi$ by Lemmas \ref{lem:posit-sim-int} and~\ref{lem:posit-proof-int}. In
view of Theorem~\ref{thm:dec-vis} amd Lemma~\ref{lem:aphi-phi-int}, this implies that $\psp=\np$ by guessing $\pi_\Phi$ nondeterministically
as in the proof of Theorem~\ref{thm:ef-sf-sep}, and that all disjoint $\psp$ pairs nonuniformly reduce to
$\decp(\rrule_{2,2},\CF\CPC)$ by using the $\pi_\Phi$ as advice as in the proof of Corollary~\ref{cor:ef-sf-sep-itp}.
\end{Pf}

We will also show a monotone lower bound. We are not able to extend the full statement of Theorem~\ref{thm:lb-hru-style}
to $\lgc T_2+\lgc{KC}$, but we will prove a monotone version of Remark~\ref{rem:ef-sf-sep-itp}.
\begin{Def}\label{def:dual}
If $\Phi$ is a QBF in negation normal form, its \emph{dual $\Phi^\dual$} is constructed by replacing each $\land$
with~$\lor$, $\top$ with~$\bot$, $\forall$ with~$\exists$, and vice versa.
\end{Def}
\pagebreak[2]
\begin{Lem}\label{lem:dual}
\ \begin{enumerate}
\item\label{item:11}
Given a monotone formula or circuit~$\fii(p_0,\dots,p_{n-1})$, we can construct in polynomial time an
$\CF\IPC$ proof of
\[\ET_{i<n}(p_i\lor q_i)\to\fii(\vec p)\lor\fii^\dual(\vec q).\]
\item\label{item:12}
Given a QBF~$\Phi(p_0,\dots,p_{n-1})$ in negation normal form which is monotone in $\vec p$, and
uses quantified variables~$\{r_i:i<d\}$, we can construct in polynomial time an $\SCF\IPC$ proof of
\[\ET_{i<n}(p_i\lor q_i)\to A^I_\Phi(\vec p,\vec r)\lor A^I_{\Phi^\dual}(\vec q,\vec r).\]
\end{enumerate}
\end{Lem}
\begin{Pf}
\ref{item:11}: By straightforward induction on the complexity of~$\fii$.

\ref{item:12}: By induction on~$d$. The base case $d=0$ is~\ref{item:11}. For the induction step from $d$ to~$d+1$,
assume w.l.o.g.\ that $\Phi$ is existentially quantified. We can write $\Phi(\vec p)=\exists r_d\,\Phi_0(\vec p,r_d,\neg r_d)$,
where $\Phi_0(\vec p,r,r')$ is monotone in $r$ and~$r'$. It is easy to check that
\[A^I_{\Phi_0(\vec p,r_d,\neg r_d)}(\vec p,r_d,\vec r)=A^I_{\Phi_0(\vec p,r,r')}(\vec p,r_d,\neg r_d,\vec r),\]
hence
\begin{equation}\label{eq:82}
A^I_\Phi(\vec p,\vec r,r_d)=\bigl[\bigl(r_d\to A^I_{\Phi_0(\vec p,r,r')}(\vec p,r_d,\neg r_d,\vec r)\bigr)
      \lor\bigl(\neg r_d\to A^I_{\Phi_0(\vec p,r,r')}(\vec p,r_d,\neg r_d,\vec r)\bigr)\bigr],
\end{equation}
and likewise,
\begin{equation}\label{eq:83}
A^I_{\Phi^\dual}(\vec p,\vec r,r_d)=\bigl(r_d\lor\neg r_d\to A^I_{\Phi_0^\dual(\vec p,r,r')}(\vec p,r_d,\neg r_d,\vec r)\bigr).
\end{equation}
By the induction hypothesis, we have an $\SCF\IPC$ proof of
\[\ET_{i<n}(p_i\lor q_i)\land(r\lor s)\land(r'\lor s')
     \to A^I_{\Phi_0}(\vec p,r,r',\vec r)\lor A^I_{\Phi_0^\dual}(\vec q,s,s',\vec r).\]
Using the substitution rule, we obtain
\begin{align*}
\ET_{i<n}(p_i\lor q_i)&\to\bigl(A^I_{\Phi_0}(\vec p,\top,\bot,\vec r)\lor A^I_{\Phi_0^\dual}(\vec q,\bot,\top,\vec r)\Bigr),\\
\ET_{i<n}(p_i\lor q_i)&\to\bigl(A^I_{\Phi_0}(\vec p,\bot,\top,\vec r)\lor A^I_{\Phi_0^\dual}(\vec q,\top,\bot,\vec r)\Bigr),
\end{align*}
hence (suppressing the variables $\vec p,\vec r$ in~$A_{\Phi_0}^I$ and $\vec q,\vec r$ in~$A_{\Phi_0^\dual}^I$ for
readability)
\begin{align*}
\ET_{i<n}(p_i\lor q_i)
&\to\bigl(A^I_{\Phi_0}(\top,\bot)\lor A^I_{\Phi_0}(\bot,\top)\bigr)
     \lor\bigl(A^I_{\Phi_0^\dual}(\top,\bot)\land A^I_{\Phi_0^\dual}(\bot,\top)\bigr)\\
&\to\bigl[\bigl(r_d\to A^I_{\Phi_0}(r_d,\neg r_d)\bigr)\lor\bigl(\neg r_d\to A^I_{\Phi_0}(r_d,\neg r_d)\bigr)\bigr]
     \lor\bigl(r_d\lor\neg r_d\to A^I_{\Phi_0^\dual}(r_d,\neg r_d)\bigr)\\
&\to A^I_\Phi(\vec p,\vec r,r_d)\lor A^I_{\Phi^\dual}(\vec p,\vec r,r_d)
\end{align*}
by \eqref{eq:82} and~\eqref{eq:83}.
\end{Pf}
\begin{Thm}\label{thm:lb-hru-int}
Let $\IPC\sset L\sset\lgc T_2+\lgc{KC}$, and assume that $\EF L$ weakly simulates $\SF L$.

Then for every monotone $\psp$ language~$P$, there exists a sequence of polynomial-size monotone Boolean circuits
$\{C_n^\forall,C_n^\exists:n\in\omega\}$ such that $C_n^\forall$ and~$C_n^\exists$ use variables $\{p_i:i<n\}$ and
$\{s_{l,r}:l<m_n,r<3\}$, and for every $w\in\two^n$, we have
\begin{align}
w\in P
\label{eq:84}&\iff\forall\vec s\:\Bigl(\ET_{l<m_n}\ET_{i<j<3}(s_{l,i}\lor s_{l,j})\to C_n^\forall(w,\vec s)\Bigr)\\
\label{eq:85}&\iff\exists\vec s\:\Bigl(\ET_{l<m_n}\ET_{i<j<3}(s_{l,i}\lor s_{l,j})\land C_n^\exists(w,\neg\vec s)\Bigr),
\end{align}
while the circuits
\begin{equation}\label{eq:86}
\ET_{l<m_n}\LOR_{r<3}s_{l,r}\land C_n^\exists(\vec p,\neg\vec s)\to C_n^\forall(\vec p,\vec s)
\end{equation}
have polynomial-size $\CF\CPC$ proofs.

If $P\in\psp$ is not necessarily monotone, the above holds with $C_n^\forall$ and~$C_n^\exists$ monotone in~$\vec s$.
\end{Thm}
\begin{Pf}
Using Lemmas \ref{lem:posit-proof-int} and~\ref{lem:posit-sim-int} and intuitionistic versions of
Lemmas~\ref{lem:cf-ef} and~\ref{lem:scf-sf}, we
may assume that $\CF{\lgc T_2}$ weakly simulates $\SCF\IPC$ on circuits. Let $P\in\psp$ be monotone. There exists a
polynomial-time constructible sequence of QBF $\{\Phi_n(p_0,\dots,p_{n-1}):n\in\omega\}$ in negation normal form such
that $\Phi_n$ is monotone in~$\vec p$, and
\[w\in P\iff\Phi_n(w)\]
for all $w\in\two^n$. By Lemma~\ref{lem:dual} and the assumption, there are polynomial-size $\CF{\lgc T_2}$ proofs of
\[\ET_{i<n}(p_i\lor q_i)\to A^I_{\Phi_n}(\vec p,\vec r)\lor A^I_{\Phi_n^\dual}(\vec q,\vec r),\]
hence using Lemmas \ref{lem:mod-comp} and~\ref{lem:t-box}, there are polynomial-size $\CF{\lgc{S4BB_2}}$ proofs of
\[\ET_{i<n}(\Box p_i\lor\Box q_i)\to\Box\T(A^I_{\Phi_n})(\vec p,\vec r)\lor\Box\T(A^I_{\Phi_n^\dual})(\vec q,\vec r).\]
By Theorem~\ref{thm:hru-style}, there exist polynomial-size monotone circuits $C_n^u(\vec p,\vec q,\vec s)$, $u=0,1$,
polynomial-size $\CF\CPC$ proofs of
\begin{equation}\label{eq:87}
\ET_{i<n}(p_i\lor q_i)\land\ET_{l<m_n}\LOR_{r<3}s_{l,r}\to\LOR_{u<2}C^u_n(\vec p,\vec q,\vec s),
\end{equation}
and polynomial-size $\CF{\lgc{S4BB_2}}$ proofs of
\begin{align*}
\ET_{\substack{l<m\\r<3}}\Bigl(s_{l,r}\land\Box\psi_{l,r}\to\LOR_{i\ne r}\Box\psi_{l,i}\Bigr)
           \land\ET_{i<n}(p_i\to\Box p_i)\land\ET_{i<n}(q_i\to\Box q_i)\land C_n^1(\vec p,\vec q,\vec s)
&\to\Box\T(A^I_{\Phi_n})(\vec p,\vec r),\\
\ET_{\substack{l<m\\r<3}}\Bigl(s_{l,r}\land\Box\psi_{l,r}\to\LOR_{i\ne r}\Box\psi_{l,i}\Bigr)
           \land\ET_{i<n}(p_i\to\Box p_i)\land\ET_{i<n}(q_i\to\Box q_i)\land C_n^0(\vec p,\vec q,\vec s)
&\to\Box\T(A^I_{\Phi_n^\dual})(\vec q,\vec r),
\end{align*}
for some formulas $\{\psi_{l,i}:l<m_n,\,i<3\}$. Using the same argument as in the proof of
Theorem~\ref{thm:lb-hru-style}, this implies the validity of the QBF
\begin{align*}
\forall\vec s\:\Bigl(\ET_{l<m_n}\ET_{i<j<3}(s_{l,i}\lor s_{l,j})\to C_n^1(\vec p,\vec q,\vec s)\Bigr)
&\to\Phi_n(\vec p),\\  
\forall\vec s\:\Bigl(\ET_{l<m_n}\ET_{i<j<3}(s_{l,i}\lor s_{l,j})\to C_n^0(\vec p,\vec q,\vec s)\Bigr)
&\to\Phi^\dual_n(\vec q).
\end{align*}
Observe $\Phi^\dual(\vec p)\equiv\neg\Phi(\neg\vec p)$. Thus, putting
$C_n^\forall(\vec p,\vec s)=C_n^1(\vec p,\vec\top,\vec s)$,
$C_n^\exists(\vec p,\vec s)=(C_n^0)^\dual(\vec\bot,\vec p,\vec s)\equiv\neg C_n^0(\vec\top,\neg\vec p,\neg\vec s)$, and
using the monotonicity of $C_n^u$, we have
\begin{align*}
\forall\vec s\:\Bigl(\ET_{l<m_n}\ET_{i<j<3}(s_{l,i}\lor s_{l,j})\to C_n^\forall(\vec p,\vec s)\Bigr)
&\to\Phi_n(\vec p),\\  
\forall\vec s\:\Bigl(\ET_{l<m_n}\ET_{i<j<3}(s_{l,i}\lor s_{l,j})\to\neg C_n^\exists(\vec p,\neg\vec s)\Bigr)
&\to\neg\Phi_n(\vec p),
\end{align*}
i.e.,
\[\Phi_n(\vec p)\to
  \exists\vec s\:\Bigl(\ET_{l<m_n}\ET_{i<j<3}(s_{l,i}\lor s_{l,j})\land C_n^\exists(\vec p,\neg\vec s)\Bigr).\]
Using the monotonicity of~$C_n^u$, substitution of $\neg p_i$ for~$q_i$ in~\eqref{eq:87} yields~\eqref{eq:86}. This in turn
implies
\[\exists\vec s\:\Bigl(\ET_{l<m_n}\ET_{i<j<3}(s_{l,i}\lor s_{l,j})\land C_n^\exists(\vec p,\neg\vec s)\Bigr)
\to\forall\vec s\:\Bigl(\ET_{l<m_n}\ET_{i<j<3}(s_{l,i}\lor s_{l,j})\to C_n^\forall(\vec p,\vec s)\Bigr),\]
hence \eqref{eq:84} and~\eqref{eq:85}: indeed,
\begin{align*}
\ET_{l<m_n}\ET_{i<j<3}(t_{l,i}\lor t_{l,j})&\land C_n^\exists(\vec p,\neg\vec t)
  \land\ET_{l<m_n}\ET_{i<j<3}(s_{l,i}\lor s_{l,j})\\
&\to\ET_{l<m_n}\LOR_{r<3}(s_{l,r}\land t_{l,r})\land C_n^\exists\bigl(\vec p,\neg(\vec t\land\vec s)\bigr)\\
&\to C_n^\forall(\vec p,\vec t\land\vec s)\\
&\to C_n^\forall(\vec p,\vec s),
\end{align*}
using once again the monotonicity of $C_n^\exists$ and $C_n^\forall$.

For nonmonotone~$P\in\psp$, we proceed as in Theorem~\ref{thm:lb-hru-style}.
\end{Pf}
\begin{Rem}\label{rem:psp-np}
That \eqref{eq:86} has short proofs, and in particular, is a tautology, is a crucial part of Theorem~\ref{thm:lb-hru-int}:
the existence of $C_n^\forall$ and~$C_n^\exists$ satisfying \eqref{eq:84} and~\eqref{eq:85} already follows from
$\psp=\np$. Indeed, if $P\in\conp$ is monotone, there exists a polynomial-time constructible sequence of monotone
formulas $\alpha_n(p_0,\dots,p_{n-1},q_0,\dots,q_{m-1},q'_0,\dots,q'_{m-1})$ such that
\[w\in P\iff\forall\vec q\:\alpha_n(w,\vec q,\neg\vec q)\]
for all $w\in\two^n$. (Note that $\alpha_n$ can be made monotone in~$\vec p$ as in the beginning of the proof of
Theorem~\ref{thm:lb-hru-style}.) Then
\[w\in P\iff\forall\vec s\:\Bigl(\ET_{l<m}\ET_{i<j<3}(s_{l,i}\lor s_{l,j})\to C_n^\forall(w,\vec s)\Bigr),\]
where $C_n^\forall(\vec p,\vec s)$ is the monotone formula
\[\alpha_n(\vec p,s_{0,0},\dots,s_{m-1,0},s_{0,1},\dots,s_{m-1,1})\lor\LOR_{l<m}(s_{l,0}\land s_{l,1}).\]
\end{Rem}

\section{Conclusion}\label{sec:conclusion}

We have characterized the decision complexity of extension rules in basic transitive modal logics of bounded branching
and the corresponding superintuitionistic logics, and as a consequence, we proved superpolynomial separation of
$\EF{}$ and $\SF{}$ systems for these logics under plausible hypotheses, solving Problem~7.1 from~\cite{ej:sfef}. Our work raises a few questions. First, we did
not manage to obtain \emph{unconditional} separations or lower bounds, but it is not clear if this is a result of
insufficiency of our methods, or if the problems are fundamentally hard (say, as hard as lower bounds on classical
Frege-like systems). Several additional problems were mentioned in Remark~\ref{rem:single}:
\begin{Que}\label{que:improve}
Let $\RI\in\{\I,\R\}$, $k\ge t\ge2$, and $L=L_0\oplus\lgc{BB}_k$, where $L_0$ is a $\RI$-extensible logic.
\begin{enumerate}
\item What is the complexity of\/ $\decp(\DP_t,\CF L)$? Is it equivalent to $\decp(\Ext^\RI_t,\CF L)$? Is it feasible?
\item Are the single-conclusion extension rules $\Ext^{\RI,\lor}_t$ feasible for $\CF L$? Are all $\EF{}$ \brak{or
$\CF{}$} systems for~$L$ p-equivalent even if allowed to use non-derivable admissible rules?
\end{enumerate}
\end{Que}
Similar questions also concern the superintuitionistic logics~$\lgc T_k$.

\begin{figure}
\centering
\includegraphics{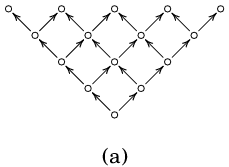}
\hskip 10em
\includegraphics{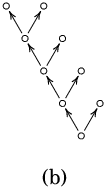}
\caption{Some frames of branching two: (a) clipped grid, (b) binary caterpillar.}
\label{fig:frames}
\end{figure}
On a more general note, our results only apply to $\RI$-extensible logics augmented with the $\lgc{BB}_k$ axioms, which
are among the weakest logics of bounded branching. They do not show much light on other logics of bounded branching and
unbounded width, especially strong logics such as the logic of square grids
$\p{\{0,\dots,n\}\times\{0,\dots,n\},{\le}}$ (or the similar logic of ``clipped'' grids as in
Fig.~\ref{fig:frames}~(a), which even has the disjunction property) and the logic of binary caterpillars
(Fig.~\ref{fig:frames}~(b)).

The results of~\cite{ej:sfef} were consistent with the mental picture of a clear dividing line between weak logics for
which we can prove unconditional exponential separations between $\EF{}$ and~$\SF{}$ using some forms of feasible
disjunction properties, and strong logics for which---at least if they are sufficiently well-behaved---$\SF{}$
and~$\EF{}$ are p-equivalent, and up to a translation, p-equivalent to $\EF\CPC$.

The results here rather seem to suggest a more complicated landscape where, as logics get stronger, the complexity of
disjunction properties goes up until it perhaps becomes irrelevant for separation of proof systems, while perhaps
the gap between $\EF{}$ and $\SF{}$ gradually becomes smaller, or perhaps it becomes dominated by tautologies of a
completely different nature than seen here. In any case, there seems to be a law of diminishing returns at play, as it
took us quite a lot of effort to get a modest improvement over~\cite{ej:sfef}, and it appears even more effort would be
needed for further progress; at the same time, we are moving into a territory where the number of natural modal logics
is quite underwhelming.

We now have a decent understanding of the relationship between $\EF{}$ and~$\SF{}$, but we know nothing much about what
happens below or above these proof systems. These might be currently the most important problems in the proof
complexity of nonclassical logics:
\begin{Que}\label{que:f-ef}
Can we separate $\Fr L$ from $\EF L$ for some modal or superintuitionistic logics~$L$?
\end{Que}
\begin{Que}\label{que:sf}
Can we unconditionally \brak{or at least, less trivially than by assuming $\psp\ne\np$} prove superpolynomial
lower bounds on the lengths of $\SF L$ proofs for some modal or superintuitionistic logics~$L$?
\end{Que}

\subsection*{Acknowledgements}

I want to thank Pavel Pudl\'ak and Pavel Hrube\v s for clarifying discussion, and the anonymous reviewer for helpful
suggestions to improve the presentation.

Supported by grant 19-05497S of GA~\v CR. The Institute of Mathematics of the Czech Academy of Sciences is supported by
RVO:~67985840.

\bibliographystyle{mybib}
\bibliography{mybib}
\end{document}